\documentclass[doublespace,a4paper,12pt]{thesis}
\usepackage{amsmath,amsfonts,amsthm,amssymb,euscript,eufrak,mathrsfs,epsf,epsfig}
\usepackage{array,verbatim}
\usepackage[sort&compress]{natbib}
\makeatletter
\makeatletter
\@addtoreset{equation}{section}
\makeatother

\newcounter{multieqs}

\newcommand{\nn}{\nonumber}

\newcommand{\Gt}{\tilde{\Gamma}}
\newcommand{\tr}{\textrm{tr}}
\newcommand{\Tr}{\textrm{Tr}}
\newcommand{\Str}{\textrm{STr}}

\newcommand{\tD}{\tilde{D}}
\newcommand{\D}{\tilde{D}}

\newcommand{\ra}{\rightarrow}
\newcommand{\ba}{\bar{a}}
\newcommand{\bbb}{\bar{b}}
\newcommand{\bc}{\bar{c}}
\newcommand{\bd}{\bar{d}}

\newcommand{\bg}{\bar{g}}

\newcommand{\p}{\partial}
\newcommand{\bz}{\bar{Z}}
\newcommand{\bp}{\bar{\psi}}
\newcommand{\e}{\epsilon}
\newcommand{\be}{\bar{\epsilon}}
\newcommand{\g}{\Gamma}
\newcommand{\G}{\Gamma}

\newcommand{\mud}{\dot{\mu}}
\newcommand{\nud}{\dot{\nu}}
\newcommand{\rhod}{\dot{\rho}}
\newcommand{\sigmad}{\dot{\sigma}}
\newcommand{\lambdad}{\dot{\lambda}}
\newcommand{\omegad}{\dot{\omega}}
\newcommand{\deltad}{\dot{\delta}}
\newcommand{\alphad}{\dot{\alpha}}
\newcommand{\betad}{\dot{\beta}}
\newcommand{\gammad}{\dot{\gamma}}

\newcommand{\mH}{\mathcal{H}}
 
\newcommand{\la}{\langle}

\newcommand{\tF}{\tilde{F}}

\begin{document}

\vspace{1truecm}

\centerline{\LARGE \bf Aspects of Supersymmetry  \\}
\centerline{\LARGE \bf in Multiple Membrane Theories \\}
\vspace{1truecm}
\thispagestyle{empty} \centerline{
    {\large \bf Andrew M. Low ${}^{\dag}$}}

\vspace{.4cm}
\centerline{{\it A thesis submitted for the degree of Doctor of Philosophy}}
\vspace{2cm}

\centerline{{\it ${}^\dag$ Centre for Research in String Theory, Department of Physics}}
\centerline{{ \it Queen Mary University of London}} \centerline{{\it Mile End Road, London E1 4NS, UK}}
\centerline{{ a.m.low@qmul.ac.uk}}

\vspace{.4cm}

\topmatter{Summary}
This thesis consists of two parts. In the first part we investigate the worldvolume supersymmetry algebra of multiple membrane theories. We begin with a description of M-theory branes and their intersections from the perspective of spacetime and worldvolume supersymmetry algebras. We then provide an overview of the recent work on multiple M2-branes focusing on the Bagger-Lambert theory and its relation to the Nambu-Poisson M5-brane and the ABJM theory. The worldvolume supersymmetry algebras of these theories are explicitly calculated and the charges interpreted in terms of spacetime intersections of M-branes. 

The second part of the thesis looks at $l_p^3$ corrections to the supersymmetry transformations of the Bagger-Lambert theory. We begin with a review of the dNS duality transformation which allows a gauge field to be dualised to a scalar field in 2+1 dimensions. Applying this duality to $\alpha'^2$ terms of the non-abelian D2-brane theory gives rise to the $l_p^3$ corrections of the Lorentzian Bagger-Lambert theory. We then apply this duality transformation to the ${\alpha'^2}$ corrections of the D2-brane supersymmetry transformations. For the `abelian' Bagger-Lambert theory we are able to uniquely determine the $l_p^3$ corrections to the supersymmetry transformations of the scalar and fermion fields. Generalising to the `non-abelian' Bagger-Lambert theory we are able to determine the $l_p^3$ correction to the supersymmetry transformation of the fermion field. Along the way make a number of observations relating to the implementation of the dNS duality transformation at the level of supersymmetry transformations.

\topmatter{Declaration}

This thesis is my own work. It is based on the sole-authored research presented in the papers \cite{Low:2009kv}, \cite{Low:2009de} and \cite{Low:2010ie}. A plan of the structure of the thesis, including details of major references for each chapter is included at the end of Chapter 1. This dissertation has not previously been submitted in whole or in part for a degree at this or any other institution.
\\
\\
\\
\\
\\
\\
\\
\\
\\
\\
\\
Andrew M. Low

\topmatter{Acknowledgements}
I would like to thank my supervisor Bill Spence for his continued support throughout my PhD. I would like to thank David Berman for providing context to the whole experience. I am most grateful. I would like to thank Daniel Thompson for his time and patience spent discussing M-theory with me. I would like to thank Moritz McGarrie for his enthusiasm and for providing me with a floor to sleep on. I would like to thank Vincenzo Calo for being my first Italian friend. I would like to thank Gianni Tallarita for being my second. I would like to thank all the PhD students who made me smile. I would like to thank David Birtles for smoking cigarettes with me. I would like to acknowledge the influence of Arthur Schopenhauer's `Die Welt als Wille und Vorstellung', Rachmaninov's second piano concerto and Chopin's first Ballade. I would like to thank the selflessness of Loganayagam and all those at the Tata Institute in Mumbai. I would like to thank B-Boy for his services in East London. I would like to acknowledge the hospitality of Nomad bookshop on Fulham Road. I would like to thank the coffee bean industry. I would like to thank my wonderful parents for their unconditional love and support. I would like to thank my Father for providing me with intellectual curiosity. I would like to thank my Mother for providing me with an aesthetic sensibility. Let us not forget that these acknowledgements relate to this thesis. For that reason I would like to thank Margarita Gelepithis above all.

\tableofcontents

\mainmatter

\chapter{Introduction}
M-theory is a non-perturbative description of string theory. It offers a way of relating the five superstring theories and eleven-dimensional supergravity through an intricate web of dualities. 
In a sense it represents the ultimate goal of unification: the ability to have a unique theory capable of describing all physical phenomena. Furthermore it highlights the significance of the reductionist mindset in the development of physics, from both a theoretical and ontological perspective: theoretical in the sense that the history of physics is the history of one theoretical framework being subsumed by another; ontological because the theory is based on the idea that there exists `fundamental' objects in nature. In M-theory, these fundamental objects are the M2-brane and the M5-brane. 

The existence of such objects is implied by the presence of a three-form gauge field in 11-dimensional supergravity, which is the low energy limit of M-theory. 
The M2-brane couples electrically to the three-form whereas the M5-brane  couples magnetically. This is analogous to the situation in four-dimensional Maxwell theory where a charged point particle couples to a one-form gauge field. The coupling of p-dimensional objects to (p+1)-form gauge fields appears naturally in string theory 
where the Ramond-Ramond fields of type II string theory are sourced by extended objects called D-branes. What is truly remarkable is that fundamental strings and D-branes can be obtained from the M-branes by considering certain compactifications. For example, wrapping one of the transverse directions of the M2-brane worldvolume around a circle and taking the radius to zero gives rise to the type IIA fundamental string. The circle radius is related to the type IIA string coupling in such a way that small radius corresponds to weak coupling. Turning this around, one sees that at strong coupling the IIA theory grows an extra dimension. When the extra dimension is large, one is outside the regime of perturbative string theory, and new techniques are required. 

A complete understanding of the fundamental degrees of freedom of this eleven-dimensional theory is likely to involve an understanding of the M-branes and how they interact with one-another. However this task is made difficult by the fact that M-theory contains no natural dimensionless parameter with which to define a perturbative expansion. Instead, one must rely on intuition gained from D-brane physics, duality relations and supersymmetry to probe the theory. Often this can involve asking questions about how string theory and D-brane phenomena are uplifted to M-theory, and then interpreting the result. For example, in string theory it is well known that fundamental strings end on D-branes. Dimensionally uplifting this result implies that an M2-brane can end on an M5-brane. From the supergravity perspective, the M2-brane and M5-brane appear as solitonic solutions of the supergravity field equations preserving half the supersymmetry of the vacuum, whereas the M2-M5 intersection appears as a quarter supersymmetric spacetime configuration. 

The existence of the M2-brane and M5-brane can also be seen at the level of the M-theory superalgebra, which is a modification of the 11-dimensional supertranslation algebra to include 2-form and 5-form central charges. The spatial components of these charges couple to the M2-brane and M5-brane worldvolumes. However it also possible to consider the time components of the 2-form and 5-form central charges. In this case it can be shown that these charges couple to the M-KK-monopole (uplift of D6-brane) and M9-brane (boundary of spacetime in Horava-Witten heterotic string theory). The M2-brane, M5-brane, M9-brane, M-KK and M-wave are collectively known as `M-theory objects'. These states all preserve half the supersymmetry of the vacuum. 

States which preserve a fraction of the supersymmetry of the vacuum are called BPS states. Such states play a vital role in string theory as they offer the possibility of extrapolating information from weak to strong coupling (and therefore the possibility of testing proposed duality relations). The reason this extrapolation is possible is related to the idea of a BPS bound. The concept of a BPS bound and its saturation can be illustrated by looking at extended supersymmetry algebras. The central charges appearing in such algebras can be thought of as electric and magnetic charges which couple to the gauge fields belonging to the supergravity multiplet. The structure of the algebra implies that the mass of a state preserving some fraction of supersymmetry is bounded from below by the magnitude of the central charges. States in which the mass is equal to the charge are said to saturate the BPS bound. In this case, zeroes appearing in the superalgebra lead to a shortening of the supermultiplet. Since the dimension of a multiplet cannot change by varying any parameters in the theory, it is possible to follow BPS states from weak coupling to strong coupling with precise control.\footnote{The only way this could fail is if another representation becomes degenerate with the BPS multiplet so that they can pair up to produce the degrees of freedom necessary to fill out a long representation.} 

Studying the superalgebra of a theory can provide valuable information about the types of BPS objects which exist in the theory, as well as their interactions. It is a remarkable feature of intersecting brane configurations that the intersection appears as a half supersymmetric soliton solution of the worldvolume theory of one of the constituent branes \cite{Callan:1997kz, Gibbons:1997xz}. For example, the fact that an M2-brane can end on an M5-brane can be seen from the worldvolume perspective of the M5-brane where the boundary of the M2-brane appears as a soliton of the M5-brane worldvolume theory. Likewise, the M5-M5 intersection can be described as a 3-brane vortex on the worldvolume of one of the M5-branes. Astonishingly, the spacetime interpretation is already implicit at the level of the worldvolume supersymmetry algebra \cite{Bergshoeff:1997bh}. In the case of the M5-brane, the worldvolume superalgebra includes a worldvolume 1-form charge carried by the self-dual string soliton living on the M5-brane (corresponding to the M2-M5 intersection) as well as a worldvolume 3-form charge carried by the 3-brane soliton (corresponding to an M5-M5 intersection). 

Although the worldvolume theories of a single M2-brane and M5-brane are quite well understood, it is only recently that effort has been directed toward a Lagrangian description of the worldvolume theory for multiple M2-branes. This work began with the efforts of Bagger and Lambert\footnote{See also Gustavsson \cite{Gustavsson:2007vu}.} \cite{Bagger:2006sk, Bagger:2007jr} who proposed a theory based on a novel algebraic structure called a Lie 3-algebra. Their original motivation was to write down a theory capable of reproducing the Basu-Harvey equation \cite{Basu:2004ed} as a BPS equation of the theory. The Basu-Harvey equation has a `fuzzy funnel' solution in which the worldvolume of the multiple M2-branes opens up into an M5-brane. This is the M-theory analogue of the Nahm equation \cite{Nahm:1979yw}, which is a BPS equation of the non-abelian D1-brane theory describing multiple D1-branes opening into a D3-brane in type IIB string theory \cite{Constable:1999ac}. In the case of IIB string theory, the charge representing the energy bound of the D1-D3 configuration appears in the superalgebra of the coincident D1-brane theory. In a similar way, the charge corresponding to the M2-M5 configuration should appear in the worldvolume superalgebra of the multiple M2-brane theory. In the first part of this thesis we investigate the worldvolume superalgebra of the recently proposed M2-brane theories of Bagger and Lambert. We explicitly calculate the central charges and provide interpretations in terms of spacetime intersections of the M2-brane with other M-theory objects. In particular, we identify the central charge corresponding to the energy bound of the M2-M5 configuration. 

Remarkably, it is possible to make contact with the worldvolume description of a single M5-brane (in a three-form background) by defining Bagger-Lambert theory with an infinite dimensional Lie 3-algebra based on the Nambu-Poisson bracket \cite{Ho:2008nn, Ho:2008ve}. This can be seen by rewriting the Bagger-Lambert field theory on the three-dimensional worldvolume $\mathcal{M}$ as a field theory on a six-dimensional manifold $\mathcal{M} \times \mathcal{N}$ where $\mathcal{N}$ is an `internal' three-manifold with Nambu-Poisson structure. In this case, the `non-abelian' gauge symmetry of the original Bagger-Lambert theory is promoted to an infinite-dimensional local symmetry of volume preserving diffeomorphisms of the manifold $\mathcal{N}$. The bosonic content of this model includes a self-dual gauge field on $\mathcal{M} \times \mathcal{N}$, as well as five scalar fields parameterising directions transverse to the six-dimensional worldvolume (the expected field content of an M5-brane). In the second part of this thesis we investigate the worldvolume superalgebra of the Nambu-Poisson M5-brane. Given the interpretation of the Nambu-Poisson model as a description of an M5-brane in 3-form background, one might expect to find central charges corresponding to M5-M2 and M5-M5 intersections in the presence of background flux. Indeed we will see that the central charges appearing in the superalgebra may be interpreted as energy bounds of M5-brane worldvolume solitons in the presence of a background three-form gauge field.

The M2-brane Lagrangian proposed by Bagger and Lambert can be thought of as the leading order term in an $l_p$ expansion of a (not yet determined) non-linear multiple M2-brane theory. This is analogous to the situation in string theory where super Yang-Mills theory represents the leading order terms of the non-abelian Born-Infeld action, which describes the dynamics of coincident D-branes. Ultimately one is interested in determining the full non-linear theory, of which the leading order terms are those of the Bagger-Lambert Lagrangian. Toward this end it is worth considering non-linear corrections to 3-algebra theories. In the third and final part of this thesis we investigate $l_p^3$ corrections to the supersymmetry transformations of the Bagger-Lambert theory. For the `abelian' Bagger-Lambert theory we are able to uniquely determine the $l_p^3$ corrections to the supersymmetry transformations of the scalar and fermion fields. Generalising to the `non-abelian' Bagger-Lambert theory we are able to determine the $l_p^3$ correction to the supersymmetry transformation of the fermion field.

The thesis is structured as follows: In Chapter 2 we provide an overview of branes and their interactions with an emphasis on the role played by superalgebras. This includes a review of the D1-D3 system in type IIB string theory and a discussion of the Basu-Harvey equation. In Chapter 3 we review the Bagger-Lambert theory and its relation to the ABJM theory of multiple membranes. In Chapter 4 we calculate the worldvolume superalgebra of the $\mathcal{N} = 6$ Bagger-Lambert theory, as reported in \cite{Low:2009kv}. For a particular choice of three-algebra we derive the superalgebra of the ABJM theory. We interpret the associated central charges in terms of BPS brane configurations. In particular we find the charge corresponding to the energy bound of the BPS fuzzy-funnel configuration of the ABJM theory. In Chapter 5 we investigate the worldvolume superalgebra of the Nambu-Poisson M5-brane theory as reported in \cite{Low:2009de}. We derive the central charges corresponding to M5-brane solitons in 3-form backgrounds. We also show that the double dimensional reduction of the superalgebra gives rise to the Poisson bracket terms of a non-commutative D4-brane superalgebra. We provide interpretations of the D4-brane charges in terms of spacetime intersections. In Chapter 6 we focus on ${l}_p^3$ corrections to the supersymmetry transformations of the $\mathcal{N}=8$ Bagger-Lambert theory as reported in \cite{Low:2010ie}. We begin with a review of the dNS duality transformation which relates a gauge field and scalar field in (2+1) dimensions \cite{Nicolai:2003bp, deWit:2003ja, deWit:2004yr}. We apply this duality transformation to the ${\alpha'^2}$ corrections of the non-abelian D2-brane supersymmetry transformations. Along the way make a number of observations relating to the implementation of the dNS duality transformation at the level of supersymmetry transformations. Finally, in Chapter 7 we offer some concluding remarks and discuss directions for future research. An Appendix summarises conventions and provides calculational details pertaining to results presented in the thesis. 

\chapter{Branes and their Interactions}

In this chapter we provide a brief overview of M-theory branes and their interactions.\footnote{For an alternative review see  \cite{Berman:2007bv}.} A large portion of this thesis will be concerned with extended supersymmetry algebras. Therefore it seems natural to introduce M-branes from the perspective of the M-theory superalgebra. We will begin with a review of the role played by supersymmetry algebras in providing information about M-theory objects and their spacetime intersections. In particular we will see that M-brane intersections are encoded in the p-form charges of the respective worldvolume supersymmetry algebras. 


The quest for a Lagrangian description of the worldvolume theory of coincident M2-branes in M-theory has been a problem of longstanding interest. This has been partly motivated by the fact that theories of coincident D-branes have provided interesting ideas about symmetry enhancement and `dual' descriptions in string theory. One example of a system exhibiting dual descriptions is the type IIB configuration in which $N$ coincident D1-branes end on a D3-brane. From the D3-brane perspective, the endpoint of the D1-branes appear as a BPS monopole on the D3-brane worldvolume \cite{Callan:1997kz, Gibbons:1997xz, Howe:1997ue}. From the perspective of the multiple D1-brane theory, the configuration gives rise to a `fuzzy funnel' soliton, which is a fuzzy 2-sphere whose radius grows without bound as the D3-brane is reached \cite{Constable:1999ac}. These two descriptions of the same physical state provide a stringy realisation of the Nahm construction \cite{Nahm:1979yw}. This leads one to hope that the multiple M2-brane worldvolume theory admits a fuzzy-funnel solution that satisfies a `generalised' Nahm equation. 

In the second half of this chapter will focus on non-abelian D-brane intersections using the example of the D1-D3 system. In particular we will see how the BPS equation of the D1-brane worldvolume theory corresponds to the Nahm equation describing the D1-brane growing into a D3 brane. Furthermore we will see that the charge corresponding to the energy bound of this configuration appears in the D1-brane supersymmetry algebra. Using the intuition gained from the D1-D3 system, Basu and Harvey proposed a `generalised' Nahm equation capabale of describing multiple M2-branes ending on an M5-brane \cite{Basu:2004ed}. In the last part of this chapter we will briefly review their work and derive an energy bound corresponding to the M2-M5 configuration. In Chapter 4 we will see how this energy bound appears as a charge in the worldvolume supersymmetry algebra of the M2-brane.

\section{M-theory Superalgebra and M-theory Objects}
A remarkable amount of information is contained within spacetime and worldvolume supersymmetry algebras. We begin by studying the 11-dimensional superpoincare algebra. The existence of 2-brane and 5-brane solutions in 11-dimensional supergravity \cite{Cremmer:1978km} motivates the inclusion of additional `charges' in the algebra. Furthermore these additional charges imply the existence of other half-supersymmetric objects in M-theory. This subsection is based on the work presented in \cite{Townsend:1997wg}.

\subsection{11-dimensional superpoincare algebra} \label{lolol}
Let us begin by considering the 11-dimensional superpoincare algebra. Eleven dimensions is the maximum dimension in which one can expect to find an interacting supersymmetric field theory \cite{Nahm:1977tg}. Translation invariance implies that $P$ and $Q$ commute leaving the anicommutator
\begin{equation}
\{ Q_\alpha , Q_\beta \} = (C \g^M )_{\alpha \beta} P_M  \label{poincare1}
\end{equation}
where $M = 0, \ldots 10$ is a spacetime Lorentz index and $Q$ is a 32-component Majorana spinor. Now consider a state $|S \rangle$ which preserves some fraction, $\alpha$, of the supersymmetry of the vacuum. In other words this state will be annihilated by some combination of supersymmetry charges. The expectation value of $\{Q_\alpha , Q_\beta \}$ acting on $|S \rangle$ will be a real symmetric matrix with $32 \alpha$ zero eigenvalues. A matrix with zero eigenvalues has a vanishing determinant. This implies
\begin{equation}
0 = \textrm{det} (\g. P) = (P^2)^{16}
\end{equation}
which follows from the fact that
\begin{align}
[\textrm{det} (\g. P)]^2 &=  \textrm{det} (\g^M P_M)\textrm{det} (\g^N P_N) \nn \\
&= \textrm{det}( \g^M \g^N P_M P_N ) \nn \\
&= \textrm{det} ( P^2.I_{32} ) \nn \\
&= (P^2)^{32}
\end{align}
So for the particular algebra \eqref{poincare1} we see that states preserving a fraction of the supersymmetry of the vacuum have null momenta. To determine the fraction of supersymmetry preserved by these massless states we can pick a frame in which
\begin{equation}
P_M = \frac{1}{2} (-1; \pm 1, 0 , \ldots 0).
\end{equation}
Furthermore if we choose a Majorana representation then $C = \g^0$ and \eqref{poincare1} becomes
\begin{equation}
\{ Q_\alpha , Q_\beta \} = \frac{1}{2} (1 \mp \g_{01})_{\alpha \beta}.
\end{equation}
It then follows that eigenspinors of $\{Q_\alpha , Q_\beta \}$ with zero eigenvalue satisfy
\begin{equation}
\g_{01} \e = \e. 
\end{equation}
The fact that $\g_{01}$ is traceless and $(\g_{01})^2 = 1$ implies that the space of solutions is 16-dimensional and therefore $\alpha = \frac{1}{2}$. In other words there exists a massless state which preserves half the supersymmetry of the vacuum. There are no other possibilities allowed by the supersymmetry algebra \eqref{poincare1}. This state is identified as a gravitational wave, or in M-theory parlance, an M-wave. 

As it stands, the 11-dimensional superpoincare algebra \eqref{poincare1} would seem to suggest that 11-dimensional supergravity  admits only one type of state which preserves a nonzero fraction of the supersymmetry of the vacuum (other than the vacuum itself). However it is well known that an analysis of the eleven-dimensional supergravity equations of motion reveals the existence a membrane solution \cite{Bergshoeff:1987cm, Duff:1990xz} as well as a five-brane solution \cite{Gueven:1992hh}, both of which preserve half the supersymmetry of the vacuum. With the advent of M-theory, these states have since been identified as the M2-brane and M5-brane. If the 11-dimensional superpoincare algebra is meant to encode information about 11-dimensional supergravity, which is the low energy limit of M-theory, then it appears as if we have a slight problem, because the algebra \eqref{poincare1} only admits one type of state, the M-wave state. So how do we solve this problem? The answer is to modify the supersymmetry algebra \cite{deAzcarraga:1989gm, Nicolai:1982gi, Zizzi:1984cq, Bergshoeff:1995hm, Sezgin:1996cj} such that
\begin{equation}
\{ Q_\alpha , Q_\beta \} = (C \g^M )_{\alpha \beta} P_M + \frac{1}{2} (C \g_{MN})_{\alpha \beta} Z^{MN} + \frac{1}{5!} (C \g_{MNPQR})_{\alpha \beta} Y^{MNPQR}. \label{fullalg}
\end{equation}
The spatial components of the two form charge $Z^{MN}$ relate to the M2-brane whereas the spatial components of the 5-form charge $Y^{MNPQR}$ relate to the M5-brane. Let us quickly see how this works for the M2-brane. $Z^{MN}$ can be expressed as an integral over the two-cycle occupied by the M2-brane in spacetime
\begin{equation}
Z^{MN} = Q_2 \int dX^M \wedge dX^N
\end{equation}
with $Q_2$ the unit of charge carried by a single membrane.\footnote{The charge $Q_2$ is analogous to the string winding number in string theory; it vanishes unless the two-cycle is non-contractible. The case of an infinite planar membrane can be dealt with by considering it as a limit of one wrapped on a large torus. In this case
$T_2$ and the charge $Q_2$ remain finite even though $P^0$ and components of $Z^{MN}$ may be infinite.} We see that for an M2-brane lying in  the $(12)$ plane $Z_{12}$ is non-zero. In the case of a static membrane, choosing a suitable Majorana basis in which $C= \g^0$ results in
\begin{equation}
\{ Q , Q \} = P^0 + \g^{012}Z_{12}. \label{M2alg}
\end{equation}
In the Majorana representation, the supercharge $Q$ is real and so the left-hand side of \eqref{M2alg} is manifestly positive. The sign of the charge $Z_{12}$ depends on whether it is an M2-brane or anti-M2-brane. Therefore it must be true that $P^0 \ge 0$. For $P^0 = 0$ we have the vacuum. For $P > 0$ we derive the bound $P^0 > |Z_{12}|$ which is equivalent to the statement
\begin{equation}
T_2 > |Q_2|. \label{bound1}
\end{equation}
For the case in which the bound is saturated and $T_2=Q_2$ the algebra can be written as
\begin{equation}
\{ Q, Q \} = P^0 [1 \mp \g^{012}].
\end{equation}
It then follows that spinors $\e$ satisfying
\begin{equation}
\g^{012} \e = \pm \e
\end{equation} 
are eigenspinors of $\{ Q, Q \}$ with zero eigenvalue. Again, since $(\g^{012})^2 = 1$ and $\g^{012}$ is traceless, the dimension of the zero-eigenvalue eigenspace of $\{ Q, Q \}$ is $16$. Thus we can conclude that a membrane saturating the bound \eqref{bound1} preserves half the supersymmetry of the vacuum, as expected. A similar analysis shows that for a static five-brane in the $(12345)$ plane with non-zero $q_5 = Y_{12345}/P^0$ the anticommutator of supercharges takes the form
\begin{equation}
\{ Q, Q \} = P^0 [1 + \g^{012345} q_5].
\end{equation}
Identifying $|q_5|$ as the ratio of the fivebrane's charge $Q_5$ to its tension $T_5$ we see that positivity implies
that $T_5 \ge Q_5$. When the five-brane saturates this bound it preserves half the supersymmetry of the vacuum. Note that with the central charges of the M2-brane and M5-brane there is no room for any additional charges in the superalgebra \eqref{fullalg}. We can see this from the fact that $\{Q ,Q \}$ is a real symmetric $32 \times 32$ matrix with $528$ entries. This is the same as the total number of components of the momenta and two central charges ($11 + 55 + 462$). 

We have seen that the spatial components of the charges appearing in \eqref{fullalg} correspond to the M2-brane and M5-brane. Furthermore we see that each state corresponds to a spinor constraint of the form $\g \e = \e$ where $\g$ a product of gamma matrices. The number of gamma matrices appearing in the product is related to the dimension of the worldvolume of the object. Based on this intuition, one may ask what the time components of the charges correspond to. The time components of the 2-form charge leads to a spinor constraint of the form
\begin{equation}
\g^{0123456789} \e = \pm \e
\end{equation}
suggesting a 9-brane. This may be associated with a boundary of eleven-dimensional spacetime  \cite{Sorokin:1997ps, Hull:1997kt}, as in the Horava-Witten construction of the heterotic string \cite{Horava:1995qa}. This is often referred to as the M-9-brane. 
The time component of the 5-form charge has a spinor constraint of the form
\begin{equation}
\g^{0123456}\e = \pm \e
\end{equation}
suggesting a six dimensional object. This has an M-theory interpretation \cite{Townsend:1995kk} as a Kaluza-Klein monopole  \cite{Gross:1983hb, Sorkin:1983ns}. We shall refer to this as the M-KK-monopole. The M9-brane and M-KK monopole can also be shown to preserve half the supersymmetry of the vacuum. Thus by a simple consideration of the extended 11-dimensional superpoincare algebra we have seen that there are in fact 5 basic half-supersymmetric M-theory objects: The M-wave, M2-brane, M5-brane, M-KK-monopole and M9-brane.

\subsection{Intersecting M-branes}

Each of the basic half supersymmetric objects of M-theory are associated with a constraint of the form $\g \e = \e$ for some traceless product of gamma matrices $\g$. Given two such products we have two such objects with these properties, calling them $\g$ and $\g'$ respectively. Let $q$ and $q'$ be the respective charge/tension ratios. Then when $P^i =0$
\begin{equation}
\{ Q, Q \} = P^0 [1 + q \g + q' \g'] \label{kaka}
\end{equation}
Positivity imposes a bound on $q$ and $q'$. The structure of the bound depends on whether $\g , \g'$ commute or anticommute. For $\{ \g , \g' \} = 0$
\begin{equation}
T \ge \sqrt{Z^2 + Z^{'2}}
\end{equation}
where $Z$ and $Z'$ are the charges of the two branes. This corresponds to a `bound state' which still preserves half the supersymmetry of the vacuum. When $[\g , \g'] = 0$ the gamma matrices may be simultaneously diagonalised and positivity implies the bound
\begin{equation}
T \ge |Z| + |Z'|.
\end{equation}
When this bound is saturated one can define the projectors  $\mathcal{P} = 1/2 (1 - \g)$ and ${\mathcal{P}}' = 1/2 (1 - \g')$ and rewrite \eqref{kaka} as
\begin{equation}
\{ Q, Q \} = 2 P^0 [q \mathcal{P} + q' {\mathcal{P}}']
\end{equation}
Since $[\mathcal{P}, {\mathcal{P}}'] = 0$, a zero eigenvalue eigenspinor of $\{ Q, Q \}$ must be annihilated by both of them and therefore must satisfy the joint condition
\begin{equation}
\g \e = \e \qquad \g' \e = \e
\end{equation}
which implies that the dimension of the zero eigenvalue eigenspace of $\{ Q, Q \}$ is 8. In other words the configuration is quarter-supersymmetric.  The simplest example of a quarter-supersymmeric configuration is when two M2-branes orthogonally intersect over a point \cite{Papadopoulos:1996uq}. This configuration can be written as $(0| M2, M2)$ where the first entry indicates the dimension of the subspace over which the intersection occurs. We will adopt this notation for the remainder of the thesis. This particular configuration may be realised by picking, for example, $\g = \g_{012}$ and $\g' = \g_{034}$, corresponding to two M2-branes aligned along the (12) and (34) spatial planes respectively.\footnote{Discussions relating to various intersecting brane configurations and projection rules can be found in \cite{Bergshoeff:1996rn, Bergshoeff:1997tt, Gauntlett:1997cv, Argurio:1997gt, Ohta:1997gw}.}
Most other quarter-supersymmetric configurations can be obtained from this one from duality transformations. For example, compactifying on the M-theory circle in a direction transverse to both the M2-branes will give rise to two D2-branes intersecting on a point in ten dimensions. Performing a T-duality along the (56) directions will result in two D4-branes intersecting over a 2-plane. Decompactifying the M-theory circle then leads to an M5-M5 intersection over a 3-plane, which is known to be a quarter-supersymmetric state \cite{Green:1996vh, Townsend:1996em, Tseytlin:1996bh}:
\begin{equation}
 (0 | M2 , M2) \rightarrow (0| D2 , D2) \rightarrow (2| D4 , D4) \rightarrow (3 | M5, M5 ). \label{dual}
\end{equation}
Alternatively, one could begin by compactifying along one of the transverse directions of one of the M2-branes. This will result in a fundamental string intersecting a D2-brane in ten dimensions. T-dualising in the (45) directions maps this to a fundamental string intersecting a D4-brane. Decompactifying then gives rise to an M2-M5 intersection in eleven dimensions: 
\begin{equation}
 (0 | M2 , M2) \rightarrow (1| F1 , D2) \rightarrow (1 | F1 , D4) \rightarrow (1 | M2, M5 ). 
\end{equation}
In the next section we will discover that these quarter-supersymmetric spacetime configurations are encoded in the worldvolume supersymmetry algebras of the M2-brane and M5-brane. To this we now turn.

\subsection{Worldvolume superalgebra's}

A wonderful aspect of worldvolume theories of branes is that BPS-saturated states, some of which correspond to half-supersymmetric classical solutions of the theory, can be interpreted from the spacetime perspective as intersections with other branes. The most obvious example of this is an electric charge on a Type II D-brane, which can be interpreted as the endpoint of a fundamental Type II string \cite{Polchinski:1995mt}. The M-theory analogue of this is the self-dual string soliton of the M5-brane worldvolume theory which can be interpreted as the boundary of an M2-brane \cite{Strominger:1995ac, Townsend:1995af}. It is also possible to find vortices on the M2 brane or M5 brane, which acquire an interpretation as 0-brane or 3-brane intersections with a second M2-brane or M5-brane \cite{Green:1996vh, Townsend:1996em, Tseytlin:1996bh}. In all these cases, the half-supersymmetric solutions of the worldvolume field theory of a single D-brane or M-brane have been found \cite{Papadopoulos:1996uq, Callan:1997kz, Gibbons:1997xz, Howe:1997et}. A remarkable feature of these worldvolume solitons is that they suggest their own spacetime interpretation. This arises from the fact that the world-volume scalars determine the spacetime embedding. Furthermore, in \cite{Bergshoeff:1997bh}, it was shown that that the spacetime interpretation of intersections is already implicit in the central charge structure of the worldvolume supersymmetry algebra. 

\subsubsection{M5-brane worldvolume superalgebra}
The first point to appreciate is that a worldvolume p-brane soliton is associated with a p-form charge in the worldvolume supersymmetry algebra. As an example let us consider the M5-brane worldvolume supersymmetry algebra. Studies of the worldvolume theory of the M5-brane reveal that there exist half-supersymmetric soliton solutions which describe the self-dual string and the 3-brane vortex. The existence of these solitons should be reflected by the existence of 1-form and 3-form charges in the worldvolume superymmetry algebra of the M5-brane. Indeed, in \cite{Howe:1997et}, the M5-brane worldvolume supersymmetry algebra was shown to take the form
\begin{align}
\{Q_\alpha^I , Q_\beta^J\} &= \Omega^{IJ} (\g^m )_{\alpha \beta} P_m + (\g^m )_{\alpha \beta} Y^{IJ}_m + (\g^{mnp})_{\alpha \beta} Z^{IJ}_{mnp} \nn \\
&= \Omega^{IJ} P_{[\alpha \beta ]} + {\mathcal{Y}}^{(IJ)}_{(\alpha \beta )} + {\mathcal{Z}}^{[IJ]}_{[\alpha \beta ]}. 
\end{align}
Here $\alpha, \beta = 1, \ldots , 4$ is an index of $SU(4) \cong \textrm{Spin} (5,1)$ and $I = 1, \ldots, 4$ is an index of $Sp(2)$, with $\Omega^{IJ}$ its invariant antisymmetric tensor. Lowercase Roman indices $m,n = 0, \ldots 5$ refer to worldvolume cordinates of the M5-brane. Thus we see that $Y^{IJ}_m$ is a worldvolume one-form and $Z^{IJ}_{mnp}$ is a worldvolume three-form. In this sense one can think of the original use of the terminology `p-form charge' as characterising p-directions in the worldvolume along which the soliton is aligned. The representations of the R-symmetry group (in this case $Sp(2)$) encode all information regarding the possible interpretations of the solitons as intersections of the M5-brane with other M-theory objects. For example, the self dual string charge transforms in the irreducible \textbf{5} of $Sp(2)$ which can be viewed as a vector in the space transverse to the M5-brane worldvolume. It can be naturally interpreted as defining the direction in which the M2-brane `leaves' the M5-brane wordvolume. In other words, viewing the string charge as a vector of $Sp(2)$ leads to the spacetime interpretation $(1 | M2, M5 )$. In a similar way, the 3-brane vortex charge, which transforms in the irreducible \textbf{10} of Sp(2), can be thought of as defining a 2-form in space transverse to the M5-brane. This 2-form is naturally interpreted as defining a 2-plane along which a second M5-brane is aligned. This corresponds to the spacetime picture $(3 | M5 , M5 )$.

Each of the worldvolume charges is expressible as an integral over the subspace $\omega_q$ defined by directions transverse to the p-brane soliton on the worldvolume of the brane. The $q$-form which is integrated over $\omega_q$ can be thought of as being defined by the representation of the R-symmetry group describing rotational invariance in directions transverse to the brane. So for example, from the perspective of the transverse space, the charge corresponding to the spacetime configuration $(3 | M5 , M5)$ should be expressible as the integral of a 2-form. The integral should be taken over the susbspace of the 5-dimensional worldspace transverse to the soliton on the M5-brane worldvolume. This is a two-dimensional space and so we should integrate a two-form over this space. The two-form should include two scalar fields, consistent with its interpretation as a 2-form in the transverse space. A natural choice is
\begin{equation}
Z =  \int dX \wedge d{Y} \label{xx21}
\end{equation}
where $X$ and $Y$ represent two scalar fields of the M5-brane worldvolume theory. 
What about the self-dual string charge? As we have seen, the string charge defines a 5-vector of $Sp(2)$ which can be viewed as a 1-form in the 5-space transverse to the M5-brane worldvolume in spacetime. This tells us that the corresponding integral charge should contain one scalar field. Furthermore we know that the integral should be taken over the subspace of the 5-dimensional worldspace transverse to the soliton, which in this case is 4-dimensional. This implies that we should integrate a 4-form. A natural choice is
\begin{equation}
Y = \int dX \wedge H \label{hhh21}
\end{equation}
where $X$ is a worldvolume scalar and $H$ is the self-dual three-form field strength associated with the 2-form gauge field living on the M5-brane worldvolume. Note that $Y$ is not given simply by an integral of $H$ over a 3-sphere surrounding the string in the M5-brane, as one might naively expect. It includes dependence on one scalar field, as required by its identification with the magnitude of a 1-form in the transverse space to the M5-brane.  

Note that the \textbf{5} representation of $Sp(2)$ could equally have been interpreted as a transverse 4-form in the transverse 5-space. In this case the 4-form can be seen as parameterising the directions in which an object with at least four spatial coordinates leaves the M5-brane wordvolume, with one coincident coordinate parameterised by the original 1-form of the 1-brane soliton. Thus we have the natural interpretation of this being the intersection of two M5-branes over a string $(1| M5 , M5 )$, which is a known quarter-supersymmetric spacetime configuration \cite{Gauntlett:1996pb}. In a similar way, the \textbf{10} of $Sp(2)$ can be thought of as defining a transverse 3-form which may be interpreted as the intersection of the M5-brane with a 6-dimensional object over a 3-brane. This six dimensional object we interpret as the M-KK monopole and therefore  $(3 | M5, KK)$ is another spacetime intersection allowed by the algebra \cite{Bergshoeff:1997tt}.\footnote{It is possible to view the time component of the 1-form charge as the spatial components of a dual 5-form \cite{Howe:1997et}. This leads to the spacetime intersections $(5|M5, KK)$ and $(5|M5, M9)$ (see \cite{Tseytlin:1996hi, Bergshoeff:1997tt, Costa:1996yb}). Note that the 3-form is self-dual and therefore the time component provides no new information. One can also consider the $Sp(2)$ 5-vector $P$ which can be interpreted as a 0-form or 5-form in the transverse space leading to spacetime interpretations $(1 | M5, MW)$ and $(1 | M5, KK)$ respectively.}

It is possible to verify  that \eqref{xx21} and \eqref{hhh21} are the correct expressions for the worldvolume charges by explicitly constructing the supersymmetry generators as Noether charges and determining their algebra directly \cite{deAzcarraga:1989gm, Sorokin:1997ps}. Alternatively, one can show that the p-volume tensions of worldvolume p-brane solitons are bounded from below by expressions that are precisely of this form \cite{Gauntlett:1997ss}. Both these approaches have been investigated in the literature and both confirm the results outlined above.
 
\subsubsection{M2-brane worldvolume superalgebra}
For the case of the M2-brane the worldvolume supersymmetry algebra is given by the maximal central extension
of the 3-dimensional $\mathcal{N} = 8$ super-Poincare algebra \cite{Bergshoeff:1997bh}. The anticommutator of supersymmetry charges is given by
\begin{align}
\{ Q^i_\alpha , Q^j_\beta \} = \delta^{ij} P_{(\alpha \beta)} +  W^{(ij)}_{(\alpha \beta)} + \varepsilon_{\alpha \beta} X^{[ij]}, \qquad (\delta_{ij} W^{(ij)} = 0),
\end{align}
where $Q^i$ are the eight Majorana spinor supercharges (i = 1, \ldots ,8) and $P$ is the 3-momentum. This supersymmetry algebra has an $SO(8)$ automorphism group, which we interpret as the rotation group in the transverse 8-space. $W$ represents a worldvolume 0-form transforming in the \textbf{28} of SO(8) whereas $X$ represents a worldvolume 1-form transforming in the ${\textbf{35}}^+$ of SO(8). The \textbf{28} can be interpreted as defining either a 2-form or a 6-form in the transverse space. When defining a 2-form, the natural spacetime interpretation is the intersection of two M2-branes over a point $(0 | M2, M2)$. The 2-form defines the 2-plane along which the second M2-brane is aligned. When defining a 6-form, the natural spacetime interpretation is an M2-brane intersecting a M-KK-monopole over a point \cite{deRoo:1997gq}, namely $(0 | M2 , KK)$. 

The 1-form worldvolume charge transforms under ${\textbf{35}}^+$ which defines a transverse self-dual 4-form. From the spacetime perspective this can be viewed as an M2-brane intersecting an M5-brane $(1 | M2 , M5)$. This is consistent with the fact that the 4-form parameterises a 4-plane transverse to the M2-brane along which four of the spatial dimensions of the M5-brane are aligned. The time component of the 1-form worldvolume charge is dual to the spatial components of a 2-form. In this case, the transverse 4-form suggests that the M2-brane is coincident with two of the spatial coordinates of a 6-dimensional object. In other words the M2-brane is `inside' an M-KK-monopole intersecting over a 2-plane, $(2 | M2, KK)$. However the story doesn't end here. One must also consider configurations in which only the worldvolume 3-vector $P$ is non-zero. Since $P$ is a singlet of SO(8) it may be interpreted as representing either a 0-form or 8-form in the transverse space. Since $P$ is a vector from the worldvolume perspective, the corresponding spacetime intersections will be over a one-dimensional string. The obvious interpretation of the transverse 0-form is the intersection of M2-brane with an M-wave, $(1 | M2, MW)$ \cite{Bergshoeff:1997tt}. The 8-form  describes the intersection of the M2-brane with a nine-dimensional object which we interpret as being the M9-brane, $(1 | M2 , M9)$. This exhausts all possible quarter-supersymmetric spacetime configurations involving the M2-brane. 

As with the M5-brane, one might wonder whether its possible to express the 0-form and 1-form worldvolume charges as integrals over the subspaces transverse to the worldvolume solitons on the M2-brane. Let us begin by considering the spacetime configuration $(0| M2, M2)$. The worldvolume soliton is zero dimensional meaning that the subspace over which we must integrate is two-dimensional. This implies that we must integrate a 2-form. Furthermore, the 2-form should include two scalar fields which parameterise the 2-plane of the second M2-brane. Thus we are naturally led to an expression of the form
\begin{equation}
W = \int dX \wedge dY
\end{equation} 
where $X$ and $Y$ are two scalar fields of the M2-brane worldvolume theory. Note that this has exactly the same form as the charge corresponding to $(3| M5, M5)$. This is to be expected from the equivalence under spacetime duality of the intersecting brane configurations associated with the M2-brane vortex and the M5-brane 3-brane (recall \eqref{dual}).

What about the spacetime configuration $(1 | M2, M5)$? From the perspective of the M2-brane, the subspace transverse to the 1-brane soliton is clearly 1-dimensional. Thus we expect the corresponding charge to be expressed as the integral of a 1-form. Furthermore, the integral charge should contain four scalar fields consistent with its interpretation as a transverse 4-form. However a problem arises when trying to construct this type of charge from the worldvolume theory of a single M2-brane. This is related to the fact that the worldvolme theory of a single M2-brane contains no soliton solution of its equations of motion which allows for a spacetime interpretation corresponding to an M2-M5 intersection. Instead, if one wishes to uncover knowledge of the fact that a single M2-brane can end on an M5-brane, one must investigate the boundary theory of the M2-brane. For the case of a single membrane this has been investigated in \cite{Strominger:1995ac, Townsend:1995af, Townsend:1996em} . 

The situation is different if one considers the possibility of multiple M2-branes. This is related to the fact that in string theory, given an intersection between D-branes with different dimensions, the description of the system using the worldvolume of lower dimensionality is usually based on a non-abelian worldvolume theory. Using the intuition gained from the multiple D1-D3 system, Basu and Harvey were able to construct an equation capable of describing multiple M2-branes growing into an M5-brane. The energy bound corresponding to this configuration should appear in the worldvolume supersymmetry algebra of the M2-brane as a central charge with the properties outlined above. In Chapter 4 we will see that this is indeed the case. In order to pave the way we will now review the D1-D3 system (including the D1-brane supersymmetry algebra), as well as the work of Basu and Harvey. This provides an opportunity to outline the approach that will be adopted in Chapters 4 and 5.

\section{Non-abelian D-brane phenomena and the Basu-Harvey equation}


A single Dp-brane contains a $U(1)$ gauge field living on its worldvolume which couples to the endpoint of a fundamental string.
Scalar fields on the worldvolume describe transverse displacements of the brane. One can think of the scalars as the Goldstone bosons associated with spontaneously broken translation symmetry in the transverse directions. 
When $N$ Dp-branes coincide, the worldvolume theory is promoted to a $U(N)$ gauge theory \cite{Witten:1995im}. This is due to new massless states arising as a result of stretched strings between the Dp-branes achieving vanishing length. Thus while the number of light degrees of freedom is proportional to $N$ for $N$ widely separated D-branes, this number grows like $N^2$ for $N$ coincident D-branes. In this case the transverse coordinates described by the worldvolume scalars are also promoted to $N \times N$ matrices since they are T-dual to the nonabelian gauge fields and so inherit their behaviour. 

The matrix-valued scalar fields provide a natural framework in which to study non-commutative geometry. Detailed studies of nonabelian worldvolume theories reveal that noncommutative geometries appear dynamically in many physical situations. For example, one finds that D-branes of one dimension metamorphose into D-branes of a higher dimension through noncommutative configurations \cite{Diaconescu:1996rk, Gauntlett:1997ss, Brecher:1998tv, Giveon:1998sr, Kapustin:1998pb, Tsimpis:1998zh}. The prototype example is the nonabelian D1-brane theory in which the Nahm equation appears as a BPS equation of the theory. We will now review this system.


\subsection{Review of the D1-D3 system}

An interesting aspect of the abelian Born-Infeld action is that it supports solitonic configurations describing lower-dimensional branes protruding from the original D-brane. For example, in the case of a D3 brane, one finds spike solutions, known as `bions', corresponding to fundamental strings and/or D1-branes extending out of the D3-brane \cite{Callan:1997kz, Gibbons:1997xz, Howe:1997ue}. Importantly, in these configurations both the worldvolume gauge fields and transverse scalar fields are excited. The gauge field can be thought of as coupling to the point charge arising from the end-point of the attached string. The scalar field describes the deformation of the D3-brane geometry caused by attaching the strings. We can see this explicitly by considering the low energy dynamics of a single D3-brane in Minkowski space. This system can be described by the Born-Infeld action \cite{Leigh:1989jq}, 
\begin{equation}
S = - T_3 \int d^4 \sigma \sqrt{- \textrm{det} (\eta_{\mu \nu} + \lambda^2 \partial_\mu X^i \partial_\nu X^i + \lambda F_{\mu \nu})}
\end{equation}
where $\lambda = 2 \pi \alpha' = 2 \pi l^2_s$. Note that a static gauge choice has been made such that $\sigma^\mu$ denotes the worldvolume coordinates of the D3 brane with $\mu = 0 , \ldots 3$ and $X^i$ denotes the scalar fields describing transverse fluctuations of the brane with $i = 4, \ldots , 9$ . $F_{\mu \nu}$ describes the $U(1)$ field strength living on the worldvolume of the D3-brane. D1-branes appear as BPS magnetic monopoles of the D3-brane worldvolume theory. Thus we choose a configuration of fields (consistent with the equations of motion) in which only one of the scalar fields $X = X^9$ is excited along with the magnetic field $B^a = \frac{1}{2} \varepsilon^{abc} F_{cd}$ where $a = 1, \ldots 3$ represents the spatial coordinates of the D3-brane. For static configurations, the energy of the system is
\begin{align}
E = - \mathcal{L} &= T_3 \int d^3 \sigma \sqrt{\lambda^2 |\vec{\nabla} X \mp \vec{B}|^2 + (1 \pm \lambda^2 \vec{B}. \vec{\nabla} X )^2 } \nn \\
&\ge T_3 \int d^3 \sigma (1 \pm \lambda^2 \vec{B}. \vec{\nabla} X). \label{E}
\end{align}
The first term in this lower bound is simply the energy of the D3-brane. The second term can be written as a total derivative  $\vec{B}. \vec{\nabla} X = \vec{\nabla} . (\vec{B} X)$ by using the Bianchi identity $\vec{\nabla}.\vec{B} = 0$. This term is therefore topological and only depends on the boundary values of the fields. Thus the last line in \eqref{E} provides a true minimum of the energy for a given set of boundary conditions. We see that the lower bound is achieved when
\begin{equation}
\vec{\nabla} X = \pm \vec{B}. \label{mif}
\end{equation}
This coincides with the BPS condition for magnetic monopoles \cite{Callan:1997kz, Gibbons:1997xz, Howe:1997ue}. Furthermore, using the Bianchi identity, \eqref{mif} implies that $\nabla^2 X = 0$ and therefore the functions describing the monopole on the D3-brane worldvolume are harmonic in form. The simplest solution, which corresponds to the bion spike,  can be expressed as
\begin{equation}
X(r) = \frac{N}{2r}, \qquad \vec{B} (\vec{r}) = \mp \frac{N}{2r^3} \vec{r}
\end{equation}
where $r^2 = (\sigma^1)^2 + (\sigma^2)^2 + (\sigma^3)^2$ and $N$ is an integer resulting from the quantisation of magnetic charge. The energy of this configuration is easily calculated to be
\begin{equation}
E = T_3 \int d^3 \sigma + N T_1 \int^{\infty}_0 d (\lambda X) \label{energy1}
\end{equation}
where $T_1 = (2 \pi l_s)^2 T_3$. The physical distance in the transverse direction is represented by $\lambda X$. Therefore this expression can be thought of as the energy of a BPS configuration consisting of $N$ semi-infinite D1-branes ending on an orthogonal D3-brane.

In order to describe the D1-D3 intersection from the perspective of the worldvolume theory of the D1-brane requires knowledge of the non-abelian D1-brane worldvolume theory. A particular soliton solution of this theory describes a fuzzy 2-sphere whose radius grows without bound into the D3-brane worldvolume. The low energy dynamics of $N$ D1-branes in a flat background is well described by the non-abelian Born-Infeld action \cite{Tseytlin:1999dj, Myers:1999ps}. However, if one is only interested in the leading nontrivial terms in a weak field expansion, the result is exactly describable in terms of super Yang-Mills theory with $U(N)$ gauge group, dimensionally reduced to 1+1 dimensions. Therefore, for a configuration in which all gauge fields and fermions have been set to zero, it is sufficient (for our purpose) to consider
 

\begin{equation}
\mathcal{L} = -T_1 \lambda^2 \Tr (\frac{1}{2} \partial_\mu X^i \partial^\mu X^i + \frac{1}{4} [X^i , X^j]^2)
\end{equation}
where we have assumed static gauge such that the worldvolume coordinates are identified with those of spacetime as $\sigma^0 = t$ and $\sigma^1 = x^9 = s$. For the case in which only three of the scalar fields are non-zero it is possible to write the energy of this system as
\begin{align}
E &= T_1 \lambda^2 \int dx^9 \Tr ( \frac{1}{2}\partial_s X^i \partial^s X^i + \frac{1}{4}[X^i, X^j] [X^i , X^j] ) \nn \\
&= T_1 \lambda^2 \int dx^9 \left[ \frac{1}{2} \Tr (\partial_s X^i - \frac{1}{2} \varepsilon_{ijk} [X^j , X^k])^2 + \frac{1}{2}\e_{ijk} \partial_s X^i [X^j , X^k] \right ]\nn \\
&\ge \frac{T_1 \lambda^2}{2}\int dx^9\e^{ijk} \partial_s X^i [X^j , X^k]. \label{b1}
\end{align}
We see see that the minimum energy condition 
\begin{equation}
\partial_s X^i =  \frac{1}{2}\e_{ijk} [X^j , X^k] \label{Na}
\end{equation}
can be identified as the Nahm equation \cite{Nahm:1979yw, Diaconescu:1996rk}. The desired solution to this equation is given by
\begin{equation}
X^i (s) = \pm \frac{\alpha^i}{2 s}
\end{equation}
where the $\alpha^i$ are an ${N} \times N$ representation of the SU(2) algebra $[\alpha^i , \alpha^j] = 2i \e^{ijk} \alpha^k$. The Casimir $C$ for this algebra is defined by $\sum (\alpha^i)^2 = C {\textbf{1}}_N$ where ${\textbf{1}}_N$ is the $N \times N$ identity matrix. If we focus on the irreducible $N \times N$ representation for which $C = N^2 - 1$ then this noncommutative scalar field configuration describes a fuzzy two-sphere with physical radius
\begin{equation}
R (s) = \lambda \sqrt{\Tr [X^i (s)]^2/ N} = \frac{N \pi l_s^2}{s} \sqrt{1 - 1/N^2}.
\end{equation}
Hence the solution describes a fuzzy funnel in which the D1-brane expands to cover the $X^{1,2,3}$ hyperplane at $s = 0$.\footnote{Note that more general configurations involving more than three scalar fields describe D1-branes expanding into intersecting D3-branes. Supersymmetry dictates that the intersecting D3-branes must lie on a calibrated 3-surface of spacetime. See for example \cite{Constable:2002yn, Harvey:1982xk, Becker:1995kb, Becker:1996ay, Gibbons:1998hm, Gauntlett:1998vk, Acharya:1998en, Berman:2005re}.} This geometry can be compared to the D3-brane solution after re-labelling $s \rightarrow \lambda X$ and $R \rightarrow r$. We see that both descriptions yield the same geometry in the limit of large $N$, up to $1/N^2$ corrections. The energy can be calculated from the boundary term appearing in \eqref{b1} and is found to be
\begin{equation}
E = N T_1 \int^\infty_0 ds + \frac{1}{\sqrt{1 - \frac{1}{N}}} T_3 \int 4 \pi R^2 dR
\end{equation} 
which matches the result arising from the D3-brane analysis \eqref{energy1} in the large $N$ limit. It is possible to show that this configuration preserves half the supersymmery of the D1-brane worldvolume theory. The linearised supersymmetry variation of the D1-brane theory is simply the dimensional reduction of the fermion variation appearing in ten-dimensional super Yang-Mills. Demanding that this variation vanishes requires
\begin{equation}
(2 \g^{s i} \partial_s X^i + \g^{ij} [X^i , X^j]) \e = 0. \label{s2}
\end{equation}
We see that if the scalar fields satisfy the Nahm equation \eqref{Na} then the supersymmetry constraint \eqref{s2} is satisfied provided the spinor $\e$ satisfies
\begin{equation}
\g^{s123} \e = \pm \e.
\end{equation}
This constraint tells us that solutions satisfying the Nahm equation preserve half the supersymmetry of the D1-brane worldvolume theory. The BPS energy bound corresponding to the `fuzzy funnel' configuration should appear in the extended worldvolume supersymmetry algebra of the D1-brane theory. The structure of the charge should allow for a spacetime interpretation as the energy bound corresponding to the D1-D3 configuration. In the next section we will show how this works.

\subsection{D1-brane superalgebra}
A neat way of deriving the worldvolume supersymmetry algebra of the multiple D1-brane theory is to consider the 10-dimensional supersymmetry algebra of super Yang-Mills theory and then dimensionally reduce the result to 1+1 dimensions\footnote{Parts of this section are influenced by the lecture notes `Electromagnetic Duality for Children' by J. M. Figueroa-O`Farrill.}. This is an illustrative exercise as we will see explicitly how the charge corresponding to the Nahm configuration arises in the supersymmetry algebra. Furthermore it provides us with an opportunity to outline the method of deriving a superalgebra. This methodology will be put to use in Chapters 4 and 5. We begin by considering ten dimensional super Yang-Mills theory defined by the Lagrangian
\begin{equation}
\mathcal{L} = -\frac{1}{4}\Tr (F_{MN},  F^{MN}) + \frac{i}{2} \Tr (\bp,  \g^M D_M \psi)  \label{YM}
\end{equation}
and supersymmetry transformations
\begin{align}
\delta A_M &= i\be \g_M \psi \nn \\
\delta \psi &= \g^{MN}F_{MN} \e \label{s1}
\end{align}
where $M, N$ are ten dimensional Lorentz indices and $\psi$ is a complex Majorana-Weyl spinor. Varying the Lagrangian with respect to these supersymmetry transformations one finds
\begin{equation}
\delta \mathcal{L} = \partial_M ( - i \be F^{NP}  \g_P \psi - \frac{i}{4} \be \g_{NP} \g^M F^{NP} \psi ) = \partial_M V^M \label{bzq}
\end{equation}
In order to calculate the superalgebra of this theory we need to first determine the supersymmetry charge, which is the spatial integral of the time-like component of the supersymmetry current. The supersymmetry current is the Noether current associated with global supersymmetry transformations. Noethers theorem asserts that corresponding to every global symmetry there exists a corresponding conserved current. Consider an infinitesimal transformation that takes a ten-dimensional field $\Phi$ to $\Phi + \delta \Phi$ and changes the Lagrangian by a total derivative
\begin{equation}
\delta \mathcal{L} = \mathcal{L} (\Phi + \delta \Phi ) - \mathcal{L} (\Phi ) = \partial_M V^M.
\end{equation}
For a Lagrangian of the form $\mathcal{L} (\Phi , \p_M \Phi)$ it is possible to write the variation of the action as
\begin{align}
\delta S &= \int d^{10} x \left(\frac{\p \mathcal{L}}{\p \Phi} \delta \Phi + \frac{\p \mathcal{L}}{\p (\p_M \Phi)} \p_M \delta \Phi \right) \nn \\
 &=\int d^{10} x \left(
 \left[ \frac{\p \mathcal{L}}{\p \Phi}  - \left( \p_M \frac{\p \mathcal{L}}{\p (\p_M \Phi)} \right) \right ] \delta \Phi + \p_M \left(\frac{\p \mathcal{L}}{\p (\p_M \Phi)} \delta \Phi \right)  \right). \nn \\
\end{align}
The terms in square brackets represent the Euler-Lagrange equations. When these equations are satisfied (i.e when the equations of motion are satisfied) we can always define a conserved current
\begin{equation}
\partial_M J^M = \p_M \left(\frac{\p \mathcal{L}}{\p (\p_M \Phi)} \delta \Phi - V^M \right) = 0. \label{nun}
\end{equation}
Therefore, applying this to super Yang-Mills theory, the Noether supersymmetry current takes the form
\begin{align}
\be J^M &= \frac{\partial \mathcal{L}}{\partial (\partial_M \Lambda )} \delta \Lambda - V^M \nn \\
&= \frac{i}{2} \be F_{NP} \g^{NP} \g^M \psi
\end{align}
where in the first line, summation over $\Lambda$ is implied, where $\Lambda$ represents all the fields in the super Yang-Mills Lagrangian \eqref{YM} and we have substituted \eqref{bzq}. The supersymmetry algebra can be derived by varying the supersymmetry current. This follows from the fact that the supercharge is the generator of supersymmetry transformations and the infinitesimal variation of an anticommuting field is given by $\delta \Phi \propto \{ Q, \Phi \} $. Therefore we have
\begin{equation}
\be \{ Q , \bar{Q} \} \propto  \int_{\textrm{space}} \delta {\bar{J}}^0. 
\end{equation}
Ignoring fermion terms, one finds that
\begin{align}
- i \delta {\bar{J}}^M &= - \frac{1}{4} \be \g^{NP} \g^M \g^{QR} F_{NP} F_{QR} \nn \\
&= \be [- \frac{1}{4} \g^{NPQRM} F_{NP} F_{QR} + 2 F^{MN} F_{NP} \g^P + \frac{1}{2} F^{NP} F_{NP} \g^M ] \nn \\
&= 2 \be [\frac{1}{8. 5!} \varepsilon^{NPQRMSTUVW} F_{NP} F_{QR} \g_{STUVW} + (F^{MN} F_{NP} + \frac{1}{4} F^{QR} F_{QR} \delta ^M_P) \g^P ].
\end{align}
Defining 
\begin{align}
T^{MN} &= F^{MP}F_{P}^{\ N} + \frac{1}{4} \eta^{MN} F^{PQ}F_{PQ} \nn \\
Z^{MNPQRS} &= \frac{1}{8} \varepsilon^{MNPQRSTUVW} F_{TU} F_{VW}
\end{align}
we recognise $T$ as the bosonic part of the improved energy-momentum tensor of the super Yang-Mills theory. The momentum, as usual, is given by the spatial integral of $T^{0M}$
\begin{equation}
P^M = \int_{\textrm{space}} T^{0M}.
\end{equation}
On the other hand it is possible to define the topological charge
\begin{equation}
\mathcal{Z}^{MNPQR} = \int_{\textrm{space}} Z^{0MNPQR}
\end{equation}
With these definitions the 10-dimensional super Yang-Mills supersymmetry algebra takes the compact form
\begin{equation}
\{ Q, \bar{Q} \} = 2 P^M \g_M + \frac{2}{5!} \mathcal{Z}^{MNPQR} \g_{MNPQR}. \label{a11}
\end{equation}
In order to write down the supersymmetry algebra in 1+1 dimensions we need to dimensionally reduce both the momenta and topological charge. In order to perform the dimensional reduction we split the ten dimensional indices into worldvolume and transverse indices such that $F_{\mu \nu}$ is the field strength on the D1-brane, $F_{\mu i} = D_\mu X^i$ and $F_{ij}= [X^i , X^j]$, where $X^i$ represent the transverse coordinates. We assume a configuration in which only three of the scalar fields are active and label them $X^a$ with $a=2,3,4$. Performing the reduction one finds charges in 1+1 dimensions of the form
\begin{align}
T^{0a} &\in  E. \partial_s X^a   \\
Z^{0MNPQR} &\in \frac{1}{2}\e^{abc}  \partial_s X^a [X^b , X^c]. \label{c1}
\end{align}
The first term represents an electric charge corresponding to the endpoint of a fundamental string. The second term we identify as the charge corresponding to the fuzzy 2-sphere soliton associated with the Nahm equation which we found in the D1-brane worldvolume theory. The form of this charge could have been motivated along the lines considered in Chapter 2. The D1 brane intersects the D3-brane on a point. This may be expressed as $(0| D1, D3)$. Thus from the worldvolume perspective this corresponds to a zero-form central charge. However the integral expression representing this charge should be expressible as an integral over the subspace transverse to the soliton on the worldvolume of the D1-brane. In this case the subspace is 1-dimensional indicating that a one-form should be integrated. Furthermore, the zero-form charge 
can be interpreted as a 3-form in the transverse space. The integral charge should reflect this fact and contain three scalar fields. Based on this line of reasoning we see that the charge had to have the form given in \eqref{c1}.  

Now that we have seen how to describe the D1-D3 system from the worldvolume and superalgbera perspective, the question arises whether it is possible to extend this analysis to the M2-M5 system in M-theory. Indeed, if one investigates the worldvolume theory of a single M5-brane it is possible to find a self-dual string soliton solution which is interpreted as the boundary of an M2-brane. Furthermore, the charge corresponding to the self-dual string has been found explicitly in the M5-brane superalgebra. Using the intuition gained from the D1-brane system, one expects the multiple M2-brane worldvolume theory to contain a BPS equation which describes multiple M2-branes ending on an M5-brane. The cross section of this solution should describe a fuzzy 3-sphere. In the next section we will review the work of Basu and Harvey who proposed a `generalised' Nahm equation as a candidate for the BPS equation of a multiple M2-brane theory.

\subsection{The Basu-Harvey equation}
In the previous section we found that the D1-D3 system exhibits dual descriptions: The D1-brane appears as a spike on the worldvolume of the D3-brane and is described by a monopole equation. From the D1-brane perspective we found that the Nahm equation describes a fuzzy 2-sphere with a radius that grows as one approaches the D3-brane. Importantly, there was a region in which the two descriptions overlap. It is reasonable to expect a similar picture in M-theory regarding the M2-M5 system. From the perspective of the M5-brane, the M2-brane boundary appears as the self-dual string soliton. In order to see this explicitly from the M5-brane perspective one can look for half-BPS solutions. The simplest approach is to look for solutions to $\delta \psi = 0$ where $\psi$ is the worldvolume fermion living on the M5-brane. Explicitly, for a single M5-brane the worldvolume fermion variation takes the form \cite{Howe:1997fb, Bandos:1997ui, Aganagic:1997zq, Pasti:1997gx}
\begin{equation}
\delta \psi = \g^\mu \g^I \partial_\mu X^I \e + \frac{1}{12} \g^{\mu \nu \lambda} H_{\mu \nu \lambda}
\end{equation}
where $\mu = 0, \ldots, 5$ is a worldvolume coordinate and $I = 6, \ldots 10$ labels transverse directions to the brane.
As outlined at the beginning of this chapter, half-BPS solutions correspond to certain constraints on the supersymmetry parameters of the theory. Thus one expects the constraint corresponding to the string on the M5-brane to be of the form
\begin{equation}
\g^{016} \e = \e \label{bug}
\end{equation} 
where $\g^0, \g^1$ are M5-brane worldvolume gamma matrices and $\g^6$ is a gamma matrix in the transverse space. This represents a configuration in which the M2-brane is aligned along the (16) spatial plane with the (1) direction being coincident with the M5-brane. Since we are looking for a string solution we consider the situation where the M5-brane worldvolume fields only depend on the radial coordinate tangent to the M5-brane which we label $R$, where  
\begin{equation}
R^2 = X^2_2 + X^2_3 + X^2_4 + X^2_5.
\end{equation}
Setting $\delta \psi = 0$ and using the spinor constraint \eqref{bug} one arrives at
\begin{equation}
H = * d s(R) \label{sqj}
\end{equation}
where $s(R)$ parameterises the direction along which the M2-brane `leaves' the M5-brane, $H$ is the 3-form worldvolume field strength and $*$ denotes Hodge duality along the four wordvolume coordinates transverse to the string. The Bianchi identity allows us to write \eqref{sqj} as
\begin{equation}
d * d s(R) = 0
\end{equation}
which implies that $s(R)$ is harmonic suggesting a solution of the form
\begin{equation}
s(R) \sim \frac{Q }{R^2} \label{tyty}
\end{equation} 
where Q refers to the self-dual string charge. We now want to consider the  
M2-M5 system from the perspective of the M2-brane. Recall that in the D1-brane system, the Nahm equation involved three active scalar fields which represented directions transverse to the D1-brane worldvolume which ended up forming the worldvolume of the D3-brane. In this case, solutions to the Nahm equation described a fuzzy 2-sphere with a radius that grew as one approached the D3-brane. Following a similar line of reasoning, in order to describe the M2-brane growing into the M5-brane from the M2 perspective will require a generalised Nahm equation involving 4 scalar fields. The equation should describe a fuzzy 3-sphere whose radius diverges to fill out the remaining worldvolume coordinates of the M5-brane. Furthermore, the radial profile should match that of the self-dual string \eqref{tyty}. Based on these requirements, Basu and Harvey proposed the following equation to describe membranes ending on an M5-brane\footnote{The generalisation of this equation describing M2-branes ending on intersecting M5-branes was considered in \cite{Berman:2005re}.}
\begin{equation}
\frac{d X^i}{ds} + \frac{M^3_{11}}{64 \pi} \e_{ijkl} \frac{1}{4!} [G , X^j , X^k , X^l] = 0 \label{BH1}
\end{equation}
where $i,j, k = 3, 4, 5, 6$ and $G$ is a fixed matrix with the property $G^2 = 1$. The 4-bracket is defined by
\begin{equation}
[X^1, X^2, X^3, X^4] = \sum_{\textrm{perms} (\sigma)} \textrm{sign} (\sigma ) X^{\sigma (1)} X^{ \sigma (2)} X^{\sigma (3)} X^{\sigma (4)}.
\end{equation}
A solution to \eqref{BH1} was found in \cite{Basu:2004ed} and takes the form
\begin{equation}
X^i (s) = \frac{i \sqrt{2 \pi}}{M^{3/2}_{11}} \frac{1}{\sqrt{s}} G^i
\end{equation}
From this it is possible to define the physical radius as
\begin{equation}
R = \sqrt{\frac{\Tr (X^i)^2}{\Tr (\textbf{1}) }}
\end{equation}
which implies
\begin{equation}
s(R) \sim \frac{N}{R^2}
\end{equation}
which matches \eqref{tyty} when $N$ is identified with the number of membranes. Looking at \eqref{BH1} we see that it's possible to use a Bogomol'nyi argument to write the energy of this configuration as 
\begin{align}
E &= T_2 \int d^2 \sigma \Tr \left[ (\partial_s X^i + \frac{1}{4!}\e_{ijkl} [G , X^j , X^k , X^l] )^2 + ( \{ \partial_s X^i , \frac{1}{2.4!} [G , X^j , X^k , X^l ]\})^2\right ]^{\frac{1}{2}}  \\
&= T_2 \int d^2 \sigma  \Tr \left[ \frac{1}{2}(\partial_s X^i - \frac{1}{3!} \e^{ijkl} [X^j , X^k , X^l])^2 + \frac{1}{3} \e^{ijkl} \partial_s X^i [X^j , X^k , X^l ] \right ] \nn \\
&\ge \frac{1}{3} \int d^2 \sigma \e^{ijkl} \partial_s X^i [X^j , X^k , X^l ] \label{woopw}
\end{align}
where we have made use of the Basu-Harvey equation and the fact that $G^2 =1$. The Nambu 3-bracket $[X^i, X^j , X^k]$ is defined as 
\begin{equation}
[X^1 , X^2, X^3] = \sum_{\textrm{perms} (\sigma)} \textrm{sign} (\sigma ) X^{\sigma (1)} X^{ \sigma (2)} X^{\sigma (3)}.
\end{equation}
We see from \eqref{woopw} that when the Basu-Harvey equation is satisfied, the energy is bounded by a term involving four scalar fields and a single derivative. We recall from the beginning of this chapter that the M2-brane 1-form charge transforms in the ${\textbf{35}}^+$ of SO(8) and can be interpreted as a 4-form in the transverse space. Given this, plus the fact that the subspace transverse to the worldvolume 1-brane soliton on the M2-brane is 1-dimensional, we argued that the charge should be a 1-form containing 4 scalar fields. We see that the energy bound appearing in \eqref{woopw} is exactly of this form. The question now arises whether it is possible to `derive' this charge from a multiple M2-brane Lagrangian theory. In other words, is it possible to write down a supersymmetric worldvolume theory of multiple M2-branes which reproduces \eqref{woopw} for a configuration in which all fermions and gauge fields have been set to zero, and only half the scalar fields are active. Given such a supersymmetric worldvolume theory of the M2-brane, it should be possible to explicitly calculate the worldvolume supersymmetry algebra by looking at the anticommutator of supercharges. This is the task we will undertake in Chapter 4. In the next chapter we will review some recent work on Lagrangian descriptions of multiple M2-branes. In particular we will focus on the work of Bagger and Lambert who proposed a description of multiple M2-branes based on a novel algebraic structure  called a 3-algebra. 


\chapter{Bagger-Lambert Theory}
There exist a number of interesting problems relating to the dynamics of multiple M2-branes. Firstly, there is no dilaton in the theory to enable a weak coupling limit.\footnote{Interestingly the coupling between the string world sheet Euler character and the dilaton in string theory can be shown to arise from a careful treatment of the M2-brane partition
function measure \cite{Berman:2006vg}.} Secondly it is known that the degrees of freedom of $N$ coincident M2-branes scales like $N^{3/2}$ as compared to $N^2$ for $N$ coincident D-branes \cite{Klebanov:1997kc, Klebanov:1996un, Berman:2006eu}. Thirdly, for a long time it was believed that a Lagrangian description of multiple M2-branes was not possible \cite{Schwarz:2004yj}. There \textit{is} however a way of defining an M2-brane Lagrangian which is correct by definition and possesses manifest $\mathcal{N} = 8$ supersymmetry (although not manifest conformal symmetry), namely 
\begin{equation}
\lim_{g_{YM} \rightarrow \infty} \frac{1}{g^2_{YM}} {\mathcal{L}}_{SYM}. \label{pollop}
\end{equation}
One can think of this as the infra-red limit of Yang-Mills theory on D2-branes. The question is whether this conformal IR fixed point has an explicit Lagrangian description where all of the symmetries of the theory are manifest. Furthermore, we know that the Basu Harvey equation should correspond to the BPS equation of a multiple M2-brane theory in much the same way that the Nahm equation is the BPS equation of the non-abelian D1-brane theory. This in fact was the original motivation that led Bagger and Lambert to propose a supersymmetric Lagrangian theory for multiple M2-branes in which the scalar fields take values in an algebra that admits a totally antisymmetric tri-linear product. It was conjectured that this model could be made maximally supersymmetric by including a non-propagating gauge field. The correpsonding supersymmetry algebra was shown to close onto equations of motion. These were `integrated' to arrive at a Lagrangian expression. The theory is consistent with all the symmetries expected from multiple M2-branes: in other words a conformal and gauge invariant action with 16 supersymmetries. The theory has an SO(8) R-symmetry that acts on the eight transverse scalars, a nonpropagating gauge field, and no free parameters, modulo a rescaling of the structure constants. 

In this chapter we will review the $\mathcal{N}=8$ Bagger-Lambert-Gustavsson theory. We will see how the construction of an M2-brane theory with the correct symmetries naturally leads to a 3-algebra structure. In the second part of this chapter we will review the $\mathcal{N} = 6$ Bagger-Lambert theory and its relation to the ABJM theory. In the final part of this chapter we will review the `novel Higgs mechanism' which relates the M2-brane theory to D2-branes. This will provide us with the material necessary to calculate and interpret the worldvolume supersymmetry algebras of these theories in Chapter 5. The content of this chapter is largely based on the following papers \cite{Bagger:2006sk, Bagger:2007jr, Bagger:2007vi, Bagger:2008se, Mukhi:2008ux, Gustavsson:2007vu}. When the content of the chapter refers to material not contained in these papers, explicit reference will be made in the text.

\section{$\mathcal{N}= 8$ Bagger-Lambert Theory}
\subsection{Constructing the theory}
A theory describing multiple M2-branes should have 8 scalar fields $X^I$, parameterising directions transverse to the worldvolume, as well as their fermionic superpartners, the Goldstinos, which correspond to broken supersymmetries. The fermionic field $\Psi$ is a Majorana spinor in $10+1$ dimensions and as a result $\Psi$ has 16 real fermionic components, equivalent to 8 bosonic degrees of freedom. Preserved supersymmetries should be reflected in the projection constraint of the supersymmetry parameters $\e$. Thus we have
\begin{align}
\g^{\mu \nu \lambda} \e &= \varepsilon^{\mu \nu \lambda} \e \\
\g^{\mu \nu \lambda} \Psi &= - \varepsilon^{\mu \nu \lambda}\Psi  \label{byep}
\end{align}
where the first constraint corresponds to the preserved supersymmetries and the second to the broken supersymmetries. In order to construct a membrane theory involving these scalar and fermion fields, one can begin by assuming that they live in some vector space $\mathcal{A}$. This is familiar from D-brane physics where we know that the worldvolume scalar fields are valued in a Lie 2-algebra. In 2+1 dimensions a scalar field has mass dimension $1/2$ whereas a fermion has mass dimension $1$. A little thought reveals that the supersymmetry transformations (to leading order in $l_p$) must be of the form
\begin{align}
\delta X^I &= i\be \g^I \psi \label{cxh}\\
\delta \psi &= \partial_\mu X^I \g^\mu \g^I \e +  \kappa [ X^I, X^J , X^K ] \g^{IJK} \e. \label{cxi}
\end{align}
where $\mu = 0,1,2$ are the M2-brane worldvolume coordinates and $I, J , K = 3, \ldots 10$ label the eight scalar fields which define directions transverse to the brane. The triple product $[X^I , X^J , X^K]$ is antisymmetric and linear in each of the fields. We note that the scalar variation \eqref{cxh} has the expected form for a supersymmetric theory. As for the fermion, the first term in the variation is the leading order free field term for a single M2-brane \cite{Bergshoeff:1987cm}. Symmetry pretty much dictates what else you can add to $\delta \psi$ in addition to the free-field term. The M2-brane theory is known to be superconformal. The mass dimension of the scalar implies that the potential in 2+1 dimensions must be sextic \cite{Schwarz:2004yj} and therefore whatever additional term appears in $\delta \psi$ should be cubic in scalar fields (due to the relation between $\delta \psi$ and $V$). Furthermore, we know that $\psi$ and $\e$ have opposite chirality and the supersymmetry transformation should respect this. This limits the gamma matrix structure of the second term to be either $\g^I$ or $\g^{IJK}$. The fact that $\delta \psi = 0$ should give rise to a BPS equation with a solution that displayed the correct divergence to reproduce the profile of the self-dual string soliton on the M5-brane limits this choice to $\g^{IJK}$. The next thing to do is check the closure of the proposed supersymmetry transformations. The scalar fields close onto translations plus a term that looks like
\begin{equation}
\delta X^I \propto i \be_2 \g_{JK} \e_1 [X^J , X^K , X^I] \label{gauget}
\end{equation} 
which can be viewed as a local version of the global symmetry transformation
\begin{equation}
\delta X = [a, b , X],
\end{equation}
where $a, b \in \mathcal{A}$. It proves convenient to introduce a basis for the algebra $\mathcal{A}$ involving the generators $T^a$ where $a = 1, \ldots N$ where $N$ is the dimension of $\mathcal{A}$. The structure constants are defined by
\begin{equation}
[T^a, T^b , T^c] = f^{abc}_{\ \ \ d} T^d \label{quibit}
\end{equation}
which immediately implies $f^{abc}_{\ \ \ d} = f^{[abc]}_{\ \ \ d}$. In this case, the symmetry transformation \eqref{gauget} can be expressed generally as
\begin{equation}
\delta X_d = f^{abc}_{\ \ \ d} \Lambda_{ab} X_c \equiv {\tilde{\Lambda}}^b_{\ a} X_b 
\end{equation}
In order to promote this global symmetry to a local symmetry a covariant derivative is defined such that $\delta (D_\mu X) = \delta (D_\mu) X + D_\mu (\delta X)$. A natural choice is
\begin{equation}
(D_\mu X)_a = \partial_\mu X_a - {\tilde{A}}_{\mu \ a}^{\ b} X_b \label{cov4}
\end{equation}
with ${\tilde{A}}_{\mu \ a}^{\ b} = f^{cdb}_{\ \ \ a } A_{\mu cd}$. One can think of ${\tilde{A}}_{\mu \ a}^{\ b}$ as living in the space of linear maps from $\mathcal{A}$ to itself. The field strength can be calculated in the normal way by considering the commutator of covariant derivatives $([D_\mu , D_\nu] X)_a = {\tilde{F}}_{\mu \nu \ a}^{\ \ b} X_b$. As a result of the gauging, Bagger and Lambert (and independently Gustavsson) showed that the full set of supersymmetry transformations, including gauge field, take the form
\begin{align}
\delta X^I_a &= i \be \g^I \psi_a  \nn \\
\delta \psi_a &= D_\mu X^I_a \g^\mu \g^I \e -\frac{1}{6} X^I_b X^J_c X^K_d f^{bcd}_{\ \ \ a} \g^{IJK} \e  \nn \\
\delta {\tilde{A}}_{\mu \ a}^{\ b} &= i \be \g_\mu \g^I X^I_c \psi_d f^{cdb}_{\ \ \ a} \label{gyg}.
\end{align}
where the factor of $1/6$ appearing in the fermion variation is fixed by closure of the algebra. We note that the form of the gauge-field transformation is essentially determined by dimensional analysis. In \cite{Bagger:2007jr} it was shown that this algebra closes on-shell provided that the structure constants satisfy the `fundamental identity'
\begin{equation}
f^{efg}_{\ \ \ d} f^{abc}_{\ \ \ g} = f^{efa}_{\ \ \ g} f^{bcg}_{\ \ \ d} + f^{efb}_{\ \ \ g} f^{cag}_{\ \ \ d} + f^{efc}_{\ \ \ g} f^{abg}_{\ \ \ d}. \label{vidi}
\end{equation}
This identity ensures that the gauge symmetry acts as a derivation
\begin{equation}
\delta ([X, Y, Z]) = [\delta X , Y , Z] + [X , \delta Y , Z] + [X, Y , \delta Z].
\end{equation}
This is analogous to the Jacobi identity in ordinary Lie algebra where the Jacobi identity arises from demanding that the transformation $\delta X = [a, X]$ acts as a derivation.\footnote{Note that if one takes the fields $X^I$ to be valued in the Lie algebra $U(N)$ (as with the D2-brane theory) then the $[X^I , X^J , X^K]$ would be given by a double commutator $[X^I , X^J , X^K] = \frac{1}{3!} [[X^I , X^J], X^K] \pm \textrm{cyclic}$ and this
would vanish by the Jacobi identity.} It is possible to construct an invariant Lagrangian by defining a trace-form on the algebra $\mathcal{A}$ which acts as a bi-linear map $\Tr : \mathcal{A} \times \mathcal{A} \rightarrow \mathcal{C}$ that satisfies
\begin{equation}
\Tr (A, B) = \Tr (B, A) , \qquad \Tr (A.B, C) = \Tr (A, B.C).
\end{equation} 
The second of these relations implies
\begin{equation}
\Tr([A, B, C], D) = - \Tr (A, [B, C, D]) \label{invariance}
\end{equation}
The trace form provides a notion of metric
\begin{equation}
h^{ab} = \Tr (T^a, T^b) \label{met}
\end{equation}
which can be used to raise indices. The relation \eqref{invariance} on the trace-form together with antisymmetry of the triple-product implies
\begin{equation}
f^{abcd} = f^{[abcd]}.
\end{equation}
The fermion variation appearing in \eqref{gyg} closes onto the fermion equation of motion. The super-variation of this then gives the bosonic equations of motion. These equations of motion can be obtained from the following Lagrangian 
\begin{align}
\mathcal{L} = -\frac{1}{2} (D_\mu X^{aI})(D^\mu X^I_a) + \frac{i}{2} \bp^a \g^\mu D_\mu \psi_a + \frac{i}{4} \bp_b \g^{IJ} X^I_c X^J_d \psi_a f^{abcd} \nn \\
- V + \frac{1}{2} \e^{\mu \nu \lambda} (f^{abcd} A_{\mu ab} \partial_\nu A_{\lambda cd} + \frac{2}{3} f^{cda}_{\ \ \ g} f^{efgb} A_{\mu ab} A_{\nu cd} A_{\lambda ef}) \label{act}
\end{align}
where
\begin{equation}
V = \frac{1}{12} f^{abcd} f^{efg}_{\ \ \ d} X^I_a X^J_b X^K_c X^I_e X^J_f X^K_g.
\end{equation}
The gauge potenial has no canonical kinetic term, but only a Chern-Simons term, and hence it has no propagating degree of freedom. In \cite{Bandres:2008vf} it was verified that this theory possesses $OSp(8|4)$ superconformal symmetry and that it is parity conserving despite the fact that it contains a Chern-Simons term. The Lagrangian is invariant under the supersymmetry transformations \eqref{gyg} up to a surface term $\delta \mathcal{L} = \partial_\mu V^\mu$ where
\begin{align}
V^\mu = -i \be \g^I \psi^a D^\mu X^I_a - \frac{i}{2} \be \g^\nu \g^I \g^\mu \psi^a D_\nu X^I_a - \frac{i}{12} \be \g^\mu\g^{IJK} \psi^a X^I_b X^J_c X^K_d f^{abcd}. \label{sfc}
\end{align}
In the next chapter we will see that this surface term contributes to the supercurrent of the $\mathcal{N}=8$ Bagger-Lambert theory and therefore plays a role in defining the superalgebra of this theory.

\subsection{Interpreting the theory}
When considering finite-dimensional representations of the three-algebra with positive-definite metric there is essentially one unique theory\footnote{One point worth mentioning is that the equations of motion for the Bagger-Lambert theory, which arise through closure of the supersymmetry algebra, do not require that the structure constant $f$ be totally antisymmetric and in this case one may in fact construct infinitely many examples \cite{Gran:2008vi}. However it is worth noting that without a metric, there is no gauge-invariant trace, and so it would appear that one is unable to construct observables. } \cite{Papadopoulos:2008sk, Gauntlett:2008uf}, this is the so-called Euclidean ${\mathcal{A}}_4$ theory in which
\begin{equation}
f^{abcd} = \frac{2 \pi}{k} \e^{abcd}.  
\end{equation}
In this case the theory is simply an $SU(2) \times SU(2)$ Chern-Simons gauge theory coupled to matter transforming in the bi-fundamental representation \cite{VanRaamsdonk:2008ft, Berman:2008be}. The Chern-Simons quantisation condition demands that $k$, the Chern-Simons level, is integer valued $k \in \mathbb{Z}$. So what is this theory describing? An investigation of the vacuum moduli space of the theory goes some way to answering this question (as does the novel Higgs mechanism discussed later in this chapter). In \cite{Lambert:2008et, Distler:2008mk} it was shown that the vacuum moduli space of this theory is 
\begin{equation}
\frac{{\mathbb{R}}^8 \times {\mathbb{R}}^8}{{\mathbb{D}}_{2k}}
\end{equation}
where ${\mathbb{D}}_{2k}$ is the Dihedral group. Specifically for $k=1$ the moduli space is ${\mathbb{R}}^8/ {\mathbb{Z}}_2 \times {\mathbb{R}}^8/ {\mathbb{Z}}_2$ which describes the moduli space of an $SO(4)$ gauge theory. This can be thought of as the strong coupling limit of two D2-branes on an $O2^-$ orientifold, whose worldvolume theory is the maximally supersymmetric $SO(4)$ gauge theory. The $k=2$ moduli space describes an $SO(5)$ gauge theory. Both these examples would seem to suggest that the theory is describing two objects living on ${\mathbb{R}}^8/ {\mathbb{Z}}_2$. Although two membranes are better than one, the dream still remains. In the search for a theory of $N$ M2-branes one might consider ways of generalising the Lagrangian construction outlined above. One such possibility is to look for theories with a reduced number of supersymmetries. In \cite{Aharony:2008ug} Aharony, Bergman, Jafferis and Maldacena (ABJM) constructed an infinite class of brane configurations whose low energy effective Lagrangian is a Chern-Simons theory with SO(6) R-symmetry and $\mathcal{N}= 6$ supersymmetry (12 supercharges) with $U(N)\times U(N)$ gauge groups for any $N$ and level $k$. \footnote{In \cite{Fuji:2008yj, Gaiotto:2008sd}  a class of Chern-Simons Lagrangians with $\mathcal{N}=4$ supersymmetry (8 supercharges) was constructed.} These theories were shown to describe $N$ M2-branes in an ${\mathbb{R}}^8 / {\mathbb{Z}}_k$ orbifold background. One advantages of this construction is that the limit in which the number of branes, $N$, and the Chern-Simons level $k$ are large, with $\lambda = N/k$ fixed, the theory admits a dual geometric description given by $AdS4 \times \mathbb{C}{\mathbb{P}}^3$.  Motivated by the work of ABJM, Bagger and Lambert derived the general form for a three-dimensional scale-invariant field theory with $\mathcal{N} = 6$ supersymetry, $SU(4)$ $R$-symmetry and a $U(1)$ global symmetry \cite{Bagger:2008se}. This was achieved by relaxing the constraint on the structure constants. In the next section we will review the $\mathcal{N}=6$ Bagger-Lambert theory and its relation to the ABJM model.

\section{$\mathcal{N}=6$ Bagger-Lambert Theory }
In the previous section we observed that when $f^{abc}_{\ \ \ d}$ is real and antisymmetric in $a,b,c$ then, for any such triple product, one finds equations of motion that are invariant under 16 supersymmetries and SO(8) R-symmetry. Here we follow \cite{Bagger:2008se} and take the $3$-algebra to be a complex vector space and only demand that the triple product be antisymmetric in the first two indices. Thus one defines
\begin{equation}
[T^a , T^b ; T^{\bc}] = f^{ab \bc}_{\phantom{1} \phantom{1} \phantom{1} d} T^d .
\end{equation} 
with  $a = 1 \ldots  N$ and $f^{a b \bc \bd} = - f^{ba \bc \bd}$. A complex notation is used in which the SO(8) R-symmetry of the $\mathcal{N}=8$ theory is broken to $SU(4) \times U(1)$. Next one introduces four complex $3$-algebra valued scalar fields $Z^A_a$ with $A=1,2,3,4$. Their complex conjugates are written as $\bz_{A \ba} = (Z^A_a)^*$. The fermionic super-partners are written as $\psi_{Aa}$ and their complex conjugates as $\psi^A_{\ba} = (\psi_{Aa})^*$. Note that the act of complex conjugation raises and lowers the $A$ index and interchanges $a \leftrightarrow \ba$. When the $A$ index is raised it means that the corresponding field transforms in the $\textbf{4}$ of $SU(4)$ and a lowered index field transforms in the $\bar{\textbf{4}}$. Both the fermions and scalars are assigned a U(1) charge of $1$. The supersymmetry parameters $\e_{AB}$ are in the $\textbf{6}$ of $SU(4)$. They satisfy the reality condition $(\e_{AB})^*= \epsilon^{AB} = \frac{1}{2} \varepsilon^{ABCD} \epsilon_{CD}$ and therefore carry no $U(1)$ charge. In order to construct a Lagrangian for this theory one uses a similar approach to the $\mathcal{N}=8$ theory. In other words one writes the most general form for the supersymmetry transformations and checks that the algebra closes. This then determines the fundamental identity as well as the equations of motion for the theory. A Lagrangian is then constructed which gives rise to these equations of motion. In order to construct a Lagrangian it is necessary to define a trace form on the $3$-algebra which provides a notion of an inner product, namely
\begin{equation}
h^{\ba b} = \Tr (T^{\ba} , T^b ). \label{gauge}
\end{equation}
Gauge-invariance of the Lagrangian requires that the metric defined by \eqref{gauge} be gauge invariant. In order for this to be true it can be shown \cite{Bagger:2008se} that the structure constants $f^{ab \bc \bd}$ must satisfy
\begin{equation}
f^{ab \bc \bd} = f^{* \bc \bd a b}.
\end{equation}
In other words complex conjugation acts on $f^{ab \bc \bd}$ as
\begin{equation}
(f^{ab\bc\bd})^* = f^{*\ba \bbb c d} = f^{cd \ba \bbb}. \label{com}
\end{equation}
Given this information Bagger and Lambert were able to construct the following Lagrangian
\begin{align}
\mathcal{L} =& -D^\mu \bz^a_A D_\mu Z^A_a - i \bp^{Aa} \gamma^\mu D_\mu \psi_{Aa} - V + \mathcal{L}_{CS} \nonumber\\
& -i f^{ab \bc \bd} \bp^A_{\bd} \psi_{Aa} Z^B_b \bz_{B \bc} + 2i f^{ab \bc \bd} \bp^A_{\bd} \psi_{Ba} Z^B_b \bz_{A \bc} \label{zop} \\
& + \frac{i}{2} \varepsilon_{ABCD} f^{ab \bc \bd} \bp^A_{\bd} \psi^B_{\bc} Z^C_a Z^D_b - \frac{i}{2} \varepsilon^{ABCD} f^{cd \ba \bbb} \bp_{Ac} \psi_{Bd} \bz_{C \ba} \bz_{D \bbb} \label{vfl}, 
\end{align}
with the potential given by
\begin{equation}
V = \frac{2}{3} \Upsilon^{CD}_{Bd} {\bar{\Upsilon}}^{Bd}_{CD} \label{potential}
\end{equation}
where
\begin{align}
\Upsilon^{CD}_{Bd} &= f^{ab\bc}_{\phantom{1}\phantom{1}\phantom{1} d} Z^C_a Z^D_b \bz_{B \bc} - \frac{1}{2} \delta^C_B f^{ab\bc}_{\phantom{1}\phantom{1}\phantom{1} d} Z^E_a Z^D_b \bz_{E \bc} + \frac{1}{2} \delta^D_B f^{ab\bc}_{\phantom{1}\phantom{1}\phantom{1} d} Z^E_a Z^C_b \bz_{E\bc} \label{upsi}
\end{align}
and the Chern-Simons term $\mathcal{L}_{CS}$ is given by
\begin{equation}
\mathcal{L}_{CS} = \frac{1}{2} \varepsilon^{\mu \nu \lambda} \left( f^{ab \bc \bd} A_{\mu \bc b} \partial_\nu A_{\lambda \bd a} + \frac{2}{3} f^{ac\bd}_{\phantom{1}\phantom{1}\phantom{1} g} f^{ge \bar{f} \bbb} A_{\mu \bbb a} A_{\nu \bd c} A_{\lambda \bar{f} e} \right).
\end{equation}
The covariant derivative is defined by $D_\mu Z^A_d = \partial_\mu Z^A_d - {\tilde{A}}^{\phantom{1}c}_{\mu \phantom{1} d} Z^A_c$. It follows that $D_\mu \bz_{A \bd} = \partial_\mu \bz_{A \bd} - {\tilde{A}}^{* \bc}_{\mu \phantom{1} \bd} \bz_{A \bc}$. Supersymmetry requires that $D_\mu \psi^A_{\bd} = \partial_\mu \psi^A_{\bd} - {\tilde{A}}^{* \bc}_{\mu \phantom{1} \bd} \psi^A_{\bc}$ and $D_\mu \psi_{Ad} - {\tilde{A}}^{\phantom{1}c}_{\mu \phantom{1} d} \psi_{Ac}$. The gauge field kinetic term is of Chern-Simons type and thus does not lead to propagating degrees of freedom. The above Lagrangian is invariant under the following supersymmetry transformations 
\begin{align}
\delta Z^A_a &= i {\bar{\epsilon}}^{AB} \psi_{Ba} \nn \\
\delta \psi_{Aa} &= - \gamma^\mu D_\mu Z^B_a \e_{AB} - f^{db \bar{c}}_{\phantom{1}\phantom{1}\phantom{1}a} Z^C_d Z^B_b \bz_{C \bar{c}} \e_{AB} + f^{db \bar{c}}_{\phantom{1}\phantom{1}\phantom{1}a} Z^C_d Z^D_b \bz_{A \bar{c}} \e_{CD} \label{susy1}\\
\delta {\tilde{A}}^{\phantom{1}c}_{\mu \phantom{1} d} &= -i \be_{AB} \gamma_\mu Z^A_a \psi^B_{\bbb} f^{ca\bbb}_{\phantom{1}\phantom{1}\phantom{1}d} + i \be^{AB} \gamma_\mu \bz_{A \bbb} \psi_{Ba} f^{ca\bbb}_{\phantom{1}\phantom{1}\phantom{1}d} \nn 
\end{align}
up to a surface term (See Appendix). The supersymmetry algebra closes into a translation plus a gauge transformation provided that the $f^{ab \bc}_{\phantom{1} \phantom{1} \phantom{1} d}$  satisfy the following fundamental identity,
\begin{equation}
f^{ef\bg}_{\ \ \ b} f^{cb \ba}_{\ \ \ d} + f^{fe\ba}_{\ \ \ b} f^{cb\bg}_{\ \ \ d} + f^{*\bg \ba f}_{\ \ \ \bbb} f^{ce \bbb}_{\ \ \ d} + f^{*\ba \bg e}_{\ \ \ \ \bbb} f^{cf \bbb}_{\ \ \ d} = 0 .
\end{equation}
the $f^{ab\bc \bd}$ generate the Lie algebra $\mathcal{G}$ of gauge transformations. In particular if the Lie algebra $\mathcal{G}$ is of the form
\begin{equation}
\mathcal{G} = \otimes_\lambda \mathcal{G}_\lambda
\end{equation}
where $\mathcal{G}_\lambda$ are commuting subalgebras of $\mathcal{G}$, then 
\begin{equation}
f^{ab \bc \bd} = \sum_\lambda \omega_\lambda \sum_\alpha (t^\alpha_\lambda)^{a \bd} (t^\alpha_\lambda)^{b \bc} , 
\end{equation}
where the $t^\alpha_\lambda$ span a $u(N)$ representation of the generators of $\mathcal{G}_\lambda$ and the $\omega_\lambda$ are arbitrary constants. This form of $f^{ab \bc \bd}$ allows one to rewrite the Lagrangian \eqref{zop} as
\begin{align}
\mathcal{L} = &-\Tr (D^\mu \bz_A , D_\mu Z^A) - i \Tr (\bp^A , \gamma^\mu D_\mu \psi_A) - V + \mathcal{L}_{CS}  \nn\\
&- i \Tr (\bp^A , [\psi_A , Z^B ; \bz_B ]) + 2i \Tr (\bp^A , [\psi_B , Z^B ; \bz_A ])\label{zip} \\ 
&+ \frac{i}{2} \varepsilon_{ABCD} \Tr (\bp^A , [Z^C , Z^D ; \psi^B ]) - \frac{i}{2} \varepsilon^{ABCD} \Tr (\bz_D , [\bp_A , \psi_B ; \bz_C ]), \nn
\end{align}
where now
\begin{align}
V = \frac{2}{3} \Tr (\Upsilon^{CD}_B , {\bar{\Upsilon}}^B_{CD}),
\end{align}
with
\begin{align}
\Upsilon^{CD}_B &= [Z^C, Z^D ; \bz_B ] - \frac{1}{2} \delta^C_B [Z^E, Z^D ; \bz_E] + \frac{1}{2} \delta^D_B [Z^E , Z^C ; \bz_E ].
\end{align}
The equivalence of \eqref{zip} and \eqref{zop} can be verified by expanding the fields $Z^A , \psi_A$ in terms of the generators $T^a$ and defining the trace form as in \eqref{gauge}. For example
\begin{align}
\Tr (\bp^A , [\psi_A , Z^B ; \bz_B ]) &= \Tr ( \bp^A_{\bd} T^{\bd} , [\psi_{A a} T^a , Z^B_b T^b ; \bz_{B \bc} T^{\bc}] ) \nn\\
&= \bp^A_{\bd} \psi_{A a} Z^B_b \bz_{B \bc} \Tr (T^{\bd} , [T^a , T^b , T^{\bc}]) \nn \\
&= \bp^A_{\bd} \psi_{A a} Z^B_b \bz_{B \bc} f^{ab \bc \bd}.
\end{align}
In \cite{Bagger:2008se} it was shown that for a particular choice of triple product one is able to recover the $\mathcal{N}=6$ Lagrangian of ABJM written in component form \cite{Aharony:2008ug, Benna:2008zy}. Given two complex vector spaces $V_1$ and $V_2$ of dimension $N_1$ and $N_2$ respectively one may consider the vector space $\mathcal{A}$ of linear maps $X: V_1 \rightarrow V_2$. A triple product may be defined on $\mathcal{A}$ as
\begin{equation}
[X, Y ; Z] = \lambda (X Z^\dagger Y - Y Z^\dagger X) \label{bracket}
\end{equation}
where $\dagger$ denotes the transpose conjugate and $\lambda$ is an arbitrary constant. The inner product acting on this space may be written as
\begin{equation}
\Tr( X, Y) = \tr (X^\dagger Y). \label{trace}
\end{equation}
With this choice of 3-algebra, the Lagrangian \eqref{zip} takes the form of the ABJM theory Lagrangian presented in \cite{Benna:2008zy} for $N_1 = N_2$. For $N_1 \ne N_2$ one obtains the $U(N_1) \times U(N_2)$ models of ABJ proposed in \cite{Aharony:2008gk}. In the next Chapter we will calculate the superalgebra for the $\mathcal{N}=6$ Bagger-Lambert theory and express the central charges in terms of 3-brackets. We can then make use of \eqref{bracket} and \eqref{trace} to derive the ABJM central charges. Before doing so we will briefly review the novel Higgs mechanism \cite{Mukhi:2008ux} which relates both the BLG and ABJM theories to D2-brane theories.

\section{The Novel Higgs Mechanism}
We can summarise both the Bagger-Lambert and ABJM theories as being $G \times G$ Chern-Simons matter theories where for Bagger-Lambert theory $G = SU(2)$ and for ABJM theory $G = U(N)$ or $SU(N)$. Given a theory of multiple M2-branes one might expect that compactification in a direction transverse to the M2-brane would result in a theory of multiple D2-branes, which to leading order in $\alpha'$ would be describable in terms of super Yang-Mills theory reduced to (2+1) dimensions. In this section we will review the `Novel Higgs Mechanism', first discussed in the context of $SU(2) \times SU(2)$ Bagger-Lambert theory \cite{Mukhi:2008ux}, which relates the M2-brane and D2-brane theories.\footnote{The structure of this sub-section was inspired by discussions with Sunil Mukhi. } It was later applied to ABJM theory in \cite{Pang:2008hw, Li:2008ya}. It is possible to summarise the Higgs procedure by stating that, for all $G \times G$ Chern-Simons theories, the following holds true: If we give a VEV to one component of the bi-fundamental scalar fields, then at energies below this VEV, the Lagrangian becomes
\begin{equation}
{\mathcal{L}}^{G \times G}_{CS} |_{vev = v} = \frac{1}{v^2} {\mathcal{L}}^{(G)}_{SYM} + \mathcal{O} \left( \frac{1}{v^3}\right) \label{xxq}
\end{equation}
where on the right-hand side the gauge field has become dynamical. So in other words, giving a VEV to one of the scalars makes one of the gauge fields, the diagonal gauge field of $G \times G$, into a dynamical gauge field. We note that this is not what happens with Yang-Mills theory. If one starts with super Yang-Mills theory and gives a VEV to one of the scalars then one simply retains super Yang-Mills theory for the subgroup $G'$ of the original group $G$ which commutes with the VEV. Lets see how this works for the Euclidean ${\mathcal{A}}_4$ theory when $k=1$. In this case we can write the Chern-Simons term appearing in the Bagger-Lambert Lagrangian as
\begin{align}
{\mathcal{L}}_{CS} &= \Tr (A \wedge dA + \frac{2}{3} A \wedge A \wedge A \wedge A - \tilde{A} \wedge d \tilde{A} - \frac{2}{3} \tilde{A} \wedge \tilde{A} \wedge \tilde{A}) \nn \\
&= \Tr (B_- \wedge F_+ + \frac{1}{6} B_- \wedge B_- \wedge B_-) \label{kog} 
\end{align}
where
\begin{equation}
B_{\pm} = A \pm \tilde{A} \qquad F_+ = d B_+ + \frac{1}{2} B_+ \wedge B_+.
\end{equation}
We see that \eqref{kog} has the form of a $B \wedge F$ type theory. Note that because $X$ is a bi-fundamental field, the covariant derivative contains the gauge field $A$ multiplying $X$ from the right and $\tilde{A}$ multiplying from the left
\begin{equation}
D_\mu X = \partial_\mu X - A_\mu X + X {\tilde{A}}_\mu.
\end{equation}
Now if one considers giving $X$ a VEV $\langle X \rangle  = v$ then the scalar kinetic term will become
\begin{equation}
- (D_\mu X)^2 \sim - v^2 (B_-)_\mu (B_-)^\mu + \ldots
\end{equation}
which resembles a mass term for the field $B_-$. Because $B_-$ is non-dynamical in \eqref{kog} it can be integrated out resulting in a kinetic term for the gauge field $B_+$
\begin{equation}
-\frac{1}{4v^2} (F_+)_{\mu \nu} (F^+)^{\mu \nu} + \mathcal{O} (\frac{1}{v^3}).
\end{equation}
In other words $B_+$ has become dynamical. In this way the Chern-Simons theory has transmuted into Yang-Mills theory. So how does one interpret \eqref{xxq}? At first sight it would appear as if the M2-brane theory has been compactified. But this cannot be the case since all that has happened is that one of the scalars has acquired a VEV. However looking more closely at \eqref{xxq} we see that this isn't \textit{exactly} super Yang-Mills theory because there are additional terms $\mathcal{O} (1/v^3)$. Thus, at best one can say
\begin{equation}
{\mathcal{L}}^{G \times G}_{CS} |_{v \rightarrow \infty} =  \lim_{v \rightarrow \infty} \frac{1}{v^2} {\mathcal{L}}_{SYM}
\end{equation}
which is by definition the theory of M2-branes as noted earlier in \eqref{pollop}. In other words there is one point on the moduli space, corresponding to very large $v$, in which the theory is equivalent to a multiple M2-brane theory. Note that this discussion was for level $k=1$. It is possible to extend the analysis for general $k$ in which case one finds
\begin{equation}
{\mathcal{L}}^{G \times G}_{CS} |_{vev = v} = \frac{k}{v^2} {\mathcal{L}}^G_{SYM} + \mathcal{O} (\frac{k}{v^3}).
\end{equation}
We see now that if one takes $k \rightarrow \infty$ and $v \rightarrow \infty$ with $v^2/k = g_{YM}$ fixed then in this limit the extra terms on the right-hand side disappear from the theory and the Lagrangian for D2-branes is recovered at finite coupling. So it would seem in this case that the theory really has been compactified. No longer have we just gone on the moduli space because we have also varied $k$ and as we have seen this labels the rank of the orbifold of the moduli space. This mechanism of compactification has been understood in terms of the deconstruction of the orbifold ${\mathbb{C}}^4/ {\mathbb{Z}}_k$ \cite{Distler:2008mk}. The angle of the orbifold cone is $2 \pi/k$. If you send $k \rightarrow \infty$ while pulling the branes away from the orbifold point (sending $v \rightarrow \infty$) then the cone turns into a cylinder of radius $\sqrt{v^2 / k}$. Therefore the M2-branes are effectively experiencing one compact direction transverse to their worldvolume and as a result one recovers the D2-brane theory.\footnote{Note that the orbifold ${\mathbb{C}}^4/ {\mathbb{Z}}_k$ has $\mathcal{N}=6$ supersymmetry and SU(4) R-symmetry. Therefore this deconstruction picture strictly speaking applies to the $\mathcal{N}=6$ models.}    

Before closing this section let us mention that the theories we have been discussing can be thought of as having coupling constant $1/k$ and therefore large $k$ corresponds to weak coupling. As with several discussions of the leading-order Bagger-Lambert ${\mathcal{A}}_4$ and ABJM theories, the classical action is most meaningful for large $k$ where the theory is weakly coupled and loop corrections can be ignored. Nevertheless, it is usually written down and studied as a function of $k$ and one hopes it has some significance even for small $k$.

\chapter{M2-brane superalgebra}
In this chapter we will calculate the extended worldvolume supersymmetry algebra of the $\mathcal{N}= 8$ and $\mathcal{N}=6$ Bagger-Lambert theories and interpret the charges from the spacetime perspective as intersections of the M2-brane with other M-theory objects. The majority of this chapter is based on the work presented in \cite{Low:2009kv}. Let us quickly remind ourselves of the method outlined in Chapter 2 for calculating the superalgebra of a theory: Firstly one derives the conserved supercurrent, which is the Noether current associated with supersymmetry transformations. The supercharge is then defined by the spatial integral of the zeroth component of the supercurrent. Using the fact that the supercharge is the generator of supersymmetry transformations and that the infinitesimal variation of an anticommuting field is given by $\delta \Phi \propto \{ Q, \Phi \} $, it follows that $\int  \delta J^0 \propto  \{Q , Q \}$. We will now show how this works explicitly for the $\mathcal{N}=8$ and $\mathcal{N}=6$ Bagger-Lambert Theories.

\section{$\mathcal{N}=8$ Bagger-Lambert Superalgebra }
The first task is to calculate the supercurrent for the $\mathcal{N}=8$ theory. Using the general expression for the current derived in \eqref{nun} as well as \eqref{act}, \eqref{gyg} and \eqref{sfc} the supercurrent is calculated to be
\begin{align}
i \be J^\mu =  &= \frac{\partial \mathcal{L}}{\partial (\partial_\mu \Lambda )} \delta \Lambda - V^\mu \nn \\
&= - i \be \g^\nu \g^I \g^\mu \psi^a D_\nu X^I_a - \frac{i}{6}\be  \g^{IJK} \g^\mu \psi_d X^I_a X^J_b X^K_c f^{abcd} \label{cur}
\end{align}
where $\Lambda$ is summed over and represents all the fields appearing in the Bagger-Lambert Lagrangian \eqref{act}. From \eqref{cur} we see\footnote{Note that the supercurrent can be derived simply as $-\be J^{\mu} = \bp^a \g^\mu \delta \psi_a$.}
\begin{equation}
J^\mu = - D_\nu X^I_a \g^\nu \g^I \g^\mu \psi^a - \frac{1}{6} X^I_a X^J_b X^K_c f^{abcd} \g^{IJK} \g^\mu \psi_d.
\end{equation}
The validity of this expression can be tested by observing whether the supercharge generates the expected supersymmetry transformations. The supercharge is the integral over the spatial worldvolume coordinates of the timelike component of the supercurrent
\begin{align}
Q &= \int d^2 \sigma J^0 \nn \\
&= - \int d^2 \sigma (D_\nu X^I_a \g^\nu \g^I \g^0 \psi^a + \frac{1}{6} X^I_a X^J_b X^K_c f^{abcd} \g^{IJK} \g^0 \psi_d).
\end{align}
As an example let us generate the scalar field supersymmetry transformation
\begin{align}
\delta X^I &= i\be [Q , X^I] \nn \\
&= i\be [ - \int d^2 \sigma ( \partial_\nu X^J (\sigma ) \g^\nu \g^J \g^0 \psi (\sigma )) , X^I  (\sigma') ] \nn \\
&= - i\be \sigma \g^0 \g^J \g^0  \int d^2 \psi (\sigma) [\partial_0 X^J (\sigma) , X^I (\sigma') ] \nn \\
&= i\be \g^J \int d^2 \sigma \psi (\sigma ) \delta^{IJ} \delta (\sigma - \sigma') \nn \\
&= i\be \g^I \psi.
\end{align}
We see that that this matches the transformation appearing in \eqref{gyg}. Using the fact that the supercharge $Q$ is the generator of supersymmetry transformations it is possible to calculate the anticommutator of supercharges by considering
\begin{equation}
\be_\alpha \{ Q^\alpha , Q^\beta \} = \int d^2 \sigma \be_\alpha \{ Q^\alpha , J^{0 \beta} (\sigma ) \} = \int d^2 \sigma (\delta J^{0 \beta} (\sigma ))
\end{equation}
The supersymmetry variation of the zeroth component of the supercurrent is computed in the appendix. For the case in which fermion fields have been set to zero, the superalgebra takes the form  
\begin{align}
\{ Q, Q \} = &-2 P_\mu \g^\mu \g^0  +  Z_{IJ} \g^{IJ} \g^0  +  Z_{\alpha IJKL} \g^{IJKL} \g^\alpha \g^0  \nn \\
&+  Z_{IJKL} \g^{IJKL}  \label{algebra}
\end{align} 
where the charges are given by
\begin{align}
Z_{IJ} &= - \int d^2 x \Tr (D_\alpha X^I D_\beta X^J \e^{\alpha \beta} - D_0 X^K [X^I, X^J, X^K])  \\
Z_{\alpha IJKL} &= \frac{1}{3} \int d^2 x \Tr (D_\beta X^{I} [X^J, X^K, X^{L}] \e^{\alpha \beta}) \label{bhc} \\
Z_{IJKL} &= \frac{1}{4} \int d^2 x \Tr ([X^M, X^I, X^J][X^M, X^K, X^L]) \label{charges}
\end{align}
with $\alpha = 1,2$ labeling the spatial coordinates of the M2-brane worldvolume. The trace defines the inner product in terms of the 3-algebra generators.  It is possible to use the Bagger-Lambert equations of motion to re-write $Z_{IJ}$ and $Z_{\alpha IJKL}$ as two surface integrals \cite{Passerini:2008qt}. These topological terms correspond to the charges of $(0|M2, M2)$ and $(1| M2, M5)$ intersections respectively. In the next sub-section we will see that the Basu-Harvey
solution excites the one-form central charge \eqref{bhc}, in agreement with the interpretation of this soliton as the quarter-supersymmetric M2-M5 intersection.\footnote{For details of the BPS equation and solution corresponding to the charge $Z_{IJ}$ see \cite{Passerini:2008qt, Jeon:2008zj}.} 

The $Z_{IJKL}$ central charge terms were first considered in \cite{Furuuchi:2008ki} where the distinction was made between so-called trace elements and non-trace elements\footnote{In \cite{Furuuchi:2008ki} trace elements are defined as elements of a linear vector space with an inner product whereas non-trace elements are without a notion of inner-product.}. If one only considers trace elements then $Z_{IJKL}$ vanishes  as a result of the total antisymmetry in $I, J, K, L$ indices and the fundamental identity of the BLG theory. However in \cite{Furuuchi:2008ki} it was pointed out that constant background configurations of $X^I$'s which take values in non-trace elements should give rise to BPS brane charges. Configurations with non-trace elements are familiar in the Matrix theory conjecture for M-theory in the light-cone quantization. The large $N$ limit of supersymmetric Yang-Mills quantum mechanics is proposed as a non-perturbative description of M-theory and M2-brane solutions are described as infinite-sized matrices. For example, application of the Higgs mechanism outlined in the previous chapter would reduce the charge $Z_{IJKL}$ to the form $\textrm{tr} [X^I, X^J] [X^K , X^L]$ where the trace is over the matrices appearing in the commutator and the matrix is taken to be infinite-dimensional. This term is analogous to the D4-brane charge (as well as the charges of the D0-branes within the D4-branes) in the matrix model for M-theory \cite{Banks:1996nn}. In the case of the Bagger-Lambert theory, the action reduces to the D2-brane action rather than the D0-brane action of the matrix model and so this charge should be interpreted as a D6-brane charge.\footnote{A complete classification of BPS states in Bagger-Lambert theory is carried out in \cite{Jeon:2008zj} including a detailed discussion of the charge $Z_{IJKL}$ and its relation to configurations involving the M5-brane.} The M-theory uplift of the D6-brane is the M-KK monopole so perhaps we can think of this charge as the energy bound of the spacetime configuration $(0|M2, KK)$. As we will see in the next chapter, the charge $Z_{IJKL}$ is crucial in obtaining the M5-brane worldvolume superalgebra in the context of the Nambu-Poisson M5-brane theory which is based on an infinite-dimensional representation of the 3-algebra. 

\subsection{Hamiltonian Analysis and BPS equation}
Let us now demonstrate that the central charge $Z_{\alpha IJKL}$ corresponds to the spacetime intersection $(1| M2, M5)$. In order to describe a stack of M2-branes ending on an M5-brane it is necessary to have four non-zero scalar fields parameterising the four spatial worldvolume coordinates of the M5-brane transverse to the self-dual string soliton. Let us call these four scalar fields $X^A$ with $A = 3, 4, 5, 7$. Furthermore we will assume the scalars are functions of only a single spatial coordinate of the M2-brane, $\sigma_2$ which we will label $s$. The BPS condition $\delta \psi = 0$ in this case takes the form
\begin{equation}
\partial_s X^A \g^A \g^2 \e + \frac{1}{6} \e^{ABCD} \g^A [X^B, X^C, X^D] \g^{3456} \e = 0 \label{su1}
\end{equation}  
where we used $\e^{ABCD} \g^D = - \g^{ABC} \g_{3456}$. We can also see this by looking at the Hamiltonian density corresponding to this field configuration which takes the form
\begin{align}
H &= \frac{1}{2} \Tr (\partial_s X^A , \partial_s X^A) + \frac{1}{12} \Tr ([X^A, X^B , X^C] , [X^A, X^B, X^C]) \nn \\
&= \frac{1}{2} \Tr |\partial_s X^A - \frac{1}{6} \e^{ABCD} [X^B, X^C , X^D]|^2 + \frac{1}{6} \e^{ABCD} \Tr ( \partial_s X^A , [X^B , X^C, X^D]) \nn \\
&\ge    \frac{1}{6} \e^{ABCD} \Tr ( \partial_s X^A , [X^B , X^C, X^D]).
\end{align}
Thus when
\begin{equation}
\partial_s X^A = \frac{1}{6} \e^{ABCD} [X^B, X^C , X^D]      \label{bh1}
\end{equation}
the energy bound 
\begin{equation}
E =  \frac{1}{6} \e^{ABCD} \Tr ( \partial_s X^A , [X^B , X^C, X^D]) \label{bund}
\end{equation} 
is saturated. We recognise \eqref{bh1} as the Basu-Harvey equation. When this equation is satisfied we see from \eqref{su1} that the field configuration is half BPS and the preserved supesymmetries satisfy
\begin{equation}
\g^2 \e = \g^{3456} \e.
\end{equation}
We see that the energy bound \eqref{bund} is exactly of the form $Z_{\alpha IJKL}$ appearing in \eqref{bhc}.

\section{$\mathcal{N}=6$ Bagger-Lambert Superalgebra} 
In this section we will use the theory outlined in Chapter 3 to calculate the superalgebra associated with the general $\mathcal{N}=6$ Bagger-Lambert Lagrangian. For a particular choice of 3-algebra we will determine the ABJM superalgebra. Given the invariance of the Lagrangian \eqref{vfl} under the supersymmetry variations \eqref{susy1}, Noether's theorem implies the existence of a conserved supercurrent $J^\mu$. The  supercharge is the spatial integral over the worldvolume coordinates of the zeroth component of the supercurrent. Since we know that the supercharge is the generator of supersymmetry transformations and that the infinitesimal variation of an anticommuting field is given by $\delta \Phi \propto \{ Q, \Phi \} $ we can write 
\begin{equation}
\int d^2 \sigma \delta J^I_{0 \beta} = {\bar{\epsilon}}^\alpha_J \{Q^I_\alpha ,Q^J_\beta\}\label{qerd}
\end{equation}
In order to make use of \eqref{qerd} in the form presented, we will have to re-write the supersymmetry parameters $\epsilon_{AB}$ in terms of a basis of  $4 \times 4$ gamma matrices,
\begin{equation}
\epsilon_{AB} = \varepsilon^I .(\G^I_{AB}), \label{parem}
\end{equation} 
with $I= 1, \ldots 6$. The $\varepsilon^I$ are carrying a suppressed worldsheet spinor index and represent the $\mathcal{N} = 6$ SUSY generators. The gamma matrices are antisymmetric $(\G^I_{AB} = -\G^I_{BA})$ and satisfy the reality condition 
\begin{equation}
\Gt^{IAB} = \frac{1}{2} \varepsilon^{ABCD} \G^I_{CD} = - (\G^I_{AB})^*. \label{up}
\end{equation}
Furthermore they satisfy\footnote{One explicit realisation in terms of Pauli matrices \cite{Terashima:2008sy} is given by $\G^1 = \sigma_2 \otimes 1_2, \G^2 = -i \sigma_2 \otimes \sigma_3 , \G^3 = i \sigma_2 \otimes \sigma_1 , \G^4 = - \sigma_1 \otimes \sigma_2 , \G^5 = \sigma_3 \otimes \sigma_2 , \G^6 = -i 1_2 \otimes \sigma_2.$}
\begin{equation}
\G^I_{AB} \Gt^{JBC} + \G^J_{AB} \Gt^{IBC} = 2 \delta^{IJ} \delta^C_A.
\end{equation}
The $4 \times 4$ matrices $\G^I$ act on a different vector space to the $2 \times 2$ matrices $\gamma^\mu$ which are defined as worldvolume gamma matrices. The supercurrent can be calculated by the usual Noether method outlined in Chapter 2. For the $\mathcal{N}=6$ Bagger-Lambert theory the supercurrent takes the compact form
\begin{equation}
J_\mu =  {\bar{\varepsilon}}^I J^I_\mu = \Tr (\delta \bp_A  \gamma_\mu , \psi^A ) + \Tr (\delta \bp^A \gamma_\mu , \psi_A), \label{supercurrent}
\end{equation}
where $J^{I}_\mu$ is the component supercurrent which appears in \eqref{qerd}. For future reference we use the basis decomposition \eqref{parem} to re-write the fermion supersymmetry variation appearing in \eqref{susy1} as
\begin{align}
\delta \psi_{A} &= -\G^I_{AB} \gamma^\mu D_\mu Z^B \varepsilon^I - N^I_A \varepsilon^I \nn \\
\delta \bp^A &= - \Gt^{IAB} {\bar{\varepsilon}}^I \gamma^\mu D_\mu \bz_{B} - N^{IA} \varepsilon^I \nn \\ 
\delta \psi^A &= \Gt^{IAB} \gamma^\mu D_\mu \bz_{B} \varepsilon^I - N^{IA} \varepsilon^I \\
\delta \bp_{A} &= \G^I_{AB} {\bar{\varepsilon}}^I \gamma^\mu D_\mu Z^B - N^I_A {\bar{\varepsilon}}^I.\nn
\end{align} 
with
\begin{align}
N^I_{A} &= \G^I_{AB} [Z^C , Z^B ; \bz_C ] - \G^I_{CD} [Z^C , Z^D ; \bz_A ] ; \\
N^{IA} &= \Gt^{IAB} [\bz_C , \bz_B ; Z^C] - \Gt^{ICD} [\bz_C , \bz_D ; Z^A].
\end{align}
We have deliberately written these variations in terms of the general 3-bracket introduced in Chapter 3. The benefit of this formalism is that one can easily derive the ABJM superalgebra by choosing a particular representation of the 3-bracket. Since we are only interested in bosonic backgrounds we set the fermions to zero. The result is (see Appendix)
\begin{align}
\delta J^{0,I} =  &-2\delta^{IJ} T^0_\mu \gamma^\mu \varepsilon^J + 2\delta^{IJ} V_1 \gamma^0 \varepsilon^J \nn \\
&+2 \delta^{IJ} ( \Tr (D_i \bz_B , [Z^D , Z^B ; \bz_D] ) -  \Tr (D_i Z^B , [\bz_D , \bz_B ; Z^D] ) \varepsilon^{ij}  \gamma^j \varepsilon^J \nn \\
&- \G^{C[IJ]}_{B}\Tr ( D_i Z^B , D_j \bz_C  )\varepsilon^{ij} \gamma^0 \varepsilon^J \nn \\
&- \G^{C[IJ]}_{B}( \Tr (D_0 \bz_A , [Z^B, Z^A ; \bz_C ] ) + \Tr ( D_0 Z^A, [\bz_C , \bz_A ; Z^B] )) \varepsilon^J \nn \\
&+ \G^{CD(IJ)}_{AB}( \Tr ( D^i \bz^B , [\bz_C , \bz_D ; Z^A] ) - \Tr ( D_i \bz_D ,[Z^A , Z^B ; \bz_C ] ))  \varepsilon^{ij} \gamma^j \varepsilon^J \nn \\
&- \G^{EF(IJ)}_{AB}\Tr ( [Z^C, Z^B ; \bz_C ], [\bz_E , \bz_F ; Z^A ])\varepsilon^J  \nn \\
&- \G^{EF(IJ)}_{AB}\Tr ( [Z^A , Z^B ; \bz_E ],[\bz_C , \bz_F ; Z^C])\varepsilon^J \nn  \\
&+ \G^{EF(IJ)}_{AB}\Tr ( [Z^A, Z^B ; \bz_C ],[\bz_E, \bz_F ; Z^C])\varepsilon^J , \nn
\end{align}
where we have defined
\begin{align}
\G^{CD(IJ)}_{AB} &= \G^I_{AB} \Gt^{JCD} + \Gt^{ICD} \G^J_{AB} ; \\
\Gt^{A[IJ]}_{D} &= \G^I_{DE} \Gt^{JAE} - \Gt^{IAE} \G^J_{DE}.
\end{align}
In order to determine the superalgebra from this expression we need to integrate $\delta J^{0, I}$ over the spatial worldvolume coordinates, and pull off the supersymmetry parameters $\varepsilon^J$, remembering that for Majorana spinors ${\bar{\varepsilon}} = \varepsilon^T C $. We know that $\int d^2 \sigma T^0_{\mu} = P_\mu$ so we see that the first term above will give us the usual momentum term. The other terms will form the charges of the algebra. We can write the superalgebra as
\begin{align}
\{ Q_\alpha^I , Q_\beta^J \} = &-2\delta^{IJ} (P_\mu (\gamma^\mu C)_{\alpha \beta}+ {\mathcal{Z}}_i(\gamma^i C)_{\alpha \beta}  -V_1 (\gamma^0 C)_{\alpha \beta} ) \nn \\
&-  \Gamma^{C[IJ]}_{B} ( {\mathcal{Z}}^B_{C,0} (C)_{\alpha \beta} +  {\mathcal{Z}}^{B}_{C} (\gamma^0 C )_{\alpha \beta} ) \label{minge} \\
&+ \Gamma^{EF(IJ)}_{AB}( {\mathcal{Z}}^{AB}_{EF,i} (\gamma^iC)_{\alpha \beta} \nn + {\mathcal{Z}}^{AB}_{EF} (\gamma^0 C)_{\alpha \beta}) \nn
\end{align}
where $\alpha , \beta$ are spinor indices and $i= x^1, x^2$ are the spatial coordinates of the worldvolume. The charges are given by
\begin{align}
{\mathcal{Z}}_i &= \int d^2 \sigma \Tr (D_i \bz_B , [Z^D , Z^B ; \bz_D] ) -  \Tr (D_i Z^B , [\bz_D , \bz_B ; Z^D] )\varepsilon^{ij} \label{Z1}\\
\nn \\
{\mathcal{Z}}^B_{C} &= \int d^2 \sigma \Tr ( D_i Z^B , D_j \bz_C  )\varepsilon^{ij} \\
\nn \\
{\mathcal{Z}}^B_{C,0} &= \int d^2 \sigma ( \Tr (D_0 \bz_A , [Z^B, Z^A ; \bz_C ] ) + \Tr ( D_0 Z^A, [\bz_C , \bz_A ; Z^B] ) \\
\nn \\
{\mathcal{Z}}^{AB}_{EF,i} &= \int d^2 \sigma  \Tr ( D_i \bz^B , [\bz_E , \bz_F ; Z^A] ) - \Tr ( D_i \bz_F ,[Z^A , Z^B ; \bz_E ] ) \varepsilon^{ij} \\
\nn \\
{\mathcal{Z}}^{CE}_{BA}&= \int d^2 \sigma  \Tr ( [Z^A, Z^B ; \bz_C ],[\bz_E, \bz_F ; Z^C]) - [Z^C, Z^B ; \bz_C ], [\bz_E , \bz_F ; Z^A ]) \nn\\
\quad \quad &- \Tr ( [Z^A , Z^B ; \bz_E ],[\bz_C , \bz_F ; Z^C]) .
\end{align}
\\
In order to interpret these charges we require a particular realisation of the three-bracket. In the next section we derive the ABJM charges by defining the three-bracket as \eqref{bracket}. 

\subsection{ABJM Superalgebra}
In this sub-section we will use the particular form of three-bracket defined in \eqref{bracket} to map the central charge terms of the $\mathcal{N}=6$ Bagger-Lambert theory to the ABJM  theory. This will work in the same way that the Bagger-Lambert Lagrangian is mapped to the ABJM Lagrangian. The structure of the superalgebra presented in \eqref{minge} remains unchanged. Only the central charge terms are affected by the 3-bracket prescription.  Firstly we define $\Tr(X,Y) = \tr(X^\dagger Y)$ and then we write the 3-bracket as $[X,Y;Z]= XZ^\dagger Y - Y Z^\dagger X$. In order to emphasise the change from the Bagger-Lambert to ABJM picture we will relabel our fields as $Z^{A \dagger} \rightarrow X^A$ and $\bz_A \rightarrow X_A$. This matches the conventions of \cite{Bandres:2008ry}. An enjoyable calculation results in the following charges for the ABJM theory
\begin{align}
{\mathcal{Z}}_i &= \int d^2 \sigma \tr D_i(X^B X_B X^D X_D  - X_B X^B X_D X^D ) \\
\nn \\
{\mathcal{Z}}^B_{C} &= \int d^2 \sigma \tr ( D_i X^B  D_j X_C  )\varepsilon^{ij} \\
\nn \\
{\mathcal{Z}}^B_{C,0} &= \int d^2 \sigma  \tr (D_0 X_A (X^B X_C X^A - X^A X_C X^B) - D_0 X^A(X_C X^B X_A - X_A X^B X_C) )\nn\\
\nn \\
{\mathcal{Z}}^{AB}_{EF,i} &= \int d^2 \sigma  \tr  D_i (X^B X_E X^A X_F) \varepsilon^{ij} \\
\nn \\
{\mathcal{Z}}^{AB}_{EF}&= 4\int d^2 \sigma   \tr (X^A X_C X^B X_F X^C X_E - X^B X_E X^A (X_CX^CX_F - X_F X^C X_C))
\end{align}
Let us now consider each of these in turn. First we note that ${\mathcal{Z}}_i$ and ${\mathcal{Z}}^{AB}_{EF,i}$ take the form of topological charges. One might anticipate that these charges represent the energy bound of an M2-M5 configuration; this follows from the fact that both charges are worldvolume one-forms and both charges contain four real scalar fields. As discussed in Chapter 2, this is the expected form of the charge corresponding to $(1 | M2, M5)$ from the M2-brane perspective. In the next sub-section we will identify ${\mathcal{Z}}_i$ as the energy bound of the fuzzy-funnel configuration \cite{Terashima:2008sy, Gomis:2008vc, Hanaki:2008cu} of the ABJM theory representing M2-branes growing into an M5-brane. The BPS solution corresponding to this fuzzy-funnel bound state was first found in \cite{Terashima:2008sy} and subsequently studied in \cite{Gomis:2008vc, Hanaki:2008cu}\footnote{Other BPS states in the ABJM theory have been discussed in \cite{Berman:2009xd, Berman:2009kj, Fujimori:2010ec, Fujimori:2008ga, Arai:2008kv, Kawai:2009rc, Kim:2009ny}.}. The charge ${\mathcal{Z}}^B_{C}$ can be interpreted as the energy bound of a vortex configuration of M2-branes intersecting M2-branes $(0|M2, M2)$ (see for example \cite{Arai:2008kv, Kim:2009ny, Auzzi:2009es, Mohammed:2010eb}). This interpretation can be inferred from the structure of the charges. For example, ${\mathcal{Z}}^B_{C}$ is a worldvolume zero-form which includes dependence on two scalar fields. In the space transverse to the M2-brane worldvolume this charge can be considered a two-form parameterised by two scalars and therefore we identify this charge with the energy bound of the spacetime configuration $(0|M2, M2)$. 

What about the ${\mathcal{Z}}^{AB}_{EF}$ charge? This would appear to be the $\mathcal{N} = 6$ analogue of the charge $Z_{IJKL}$ appearing in the $\mathcal{N} = 8$ algebra. In which case it can be thought of as the charge of an uplifted D6-brane. However it is interesting to note another possible interpretation of this charge due to Terashima \cite{Terashima:2009fy}. We know that the M2-M5 Basu-Harvey configuration can be thought of as the M-theory uplift of the D2-D4 configuration (which is described by the Nahm equation from the D2-brane point of view). However from the D-brane perspective there also exists a bound state of D2-branes and D4-branes. This results from considering D4-branes with constant magnetic field or alternatively infinitely many D2-branes with $[X^1 , X^2]= \textrm{const}$ where $X^i$ are matrix valued scalar fields representing the positions of the D2-branes.  In \cite{Terashima:2009fy} a classical solution was found in the ABJM theory corresponding to a bound state of M5-branes and M2-branes which becomes the bound state of D4-branes and D2-branes in the scaling limit $k \rightarrow \infty$. From the M5-brane perspective this bound state corresponds to the M5-brane with non-zero self-dual three-form flux. In \cite{Terashima:2009fy} it was conjectured that the charge ${\mathcal{Z}}^{AB}_{EF}$ might correspond to the energy bound of this state. 
The fact that the ABJM theory does not possess a manifest $SO(8)$ R-symmetry makes the interpretation of the charges slightly tricky. In \cite{Gustavsson:2009pm, Kwon:2009ar} it was shown that, for Chern-Simons levels $k=1,2$, the $\mathcal{N}=6$ supersymmetry of the ABJM theory is enhanced to $\mathcal{N}=8$ supersymmetry through the use of monopole operators. It should be possible to use similar techniques to investigate the enhancement of the R-symmetry of the charges of the ABJM theory. This would allow for direct comparison with the $\mathcal{N}=8$ Bagger-Lambert superalgebra. Before investigating the BPS fuzzy-funnel configuration in the next sub-section we note that the superalgebra may be re-written in terms of trace, anti-symmetric and symmetric traceless parts
\begin{equation}
\{ Q^I_\alpha , Q^J_\beta \} =  \delta^{IJ} X_{\alpha \beta} + {\tilde{Z}}^{(IJ)}_{\alpha \beta} + {\tilde{Z}}^{[IJ]}_{\alpha \beta} \label{superduper}
\end{equation}
where $X_{\alpha \beta}$ is a singlet, ${\tilde{Z}}^{(IJ)}_{\alpha \beta}$ is symmetric traceless and ${\tilde{Z}}^{[IJ]}_{\alpha \beta}$ antisymmetric in $I,J$ respectively. Explicitly we have
\begin{align}
X_{\alpha \beta} &= -2P_\mu (\gamma^\mu C)_{\alpha \beta} - \frac{4}{3} {\mathcal{Z}}_i (\gamma^i C)_{\alpha \beta}, \nn \\
 {\tilde{Z}}^{(IJ)}_{\alpha \beta} &=  ( \G^{EF(IJ)}_{AB} {\mathcal{Z}}^{AB}_{EF,i}   - \frac{2}{3} \delta^{IJ} {\mathcal{Z}}_i ) (\gamma^i C)_{\alpha \beta} + ( \G^{EF(IJ)}_{AB}{\mathcal{Z}}^{AB}_{EF} + 2 \delta^{IJ} V_1)(\gamma^0 C)_{\alpha\beta} \nn \\
{\tilde{Z}}^{[IJ]}_{\alpha \beta} &= -  \Gamma^{C[IJ]}_{B} ( {\mathcal{Z}}^B_{C,0} C_{\alpha \beta} +  {\mathcal{Z}}^{B}_{C} (\gamma^0 C )_{\alpha \beta} ).
\end{align}
It is interesting to observe what happens when we act with $\delta_{IJ}$ on the superalgebra. In this case $\G^{C[IJ]}_B = 0$ since it is antisymmetric in $I,J$ and so ${\mathcal{Z}}^B_{C}$ and ${\mathcal{Z}}^B_{C,0} $ disappear from the algebra. Similarly ${\tilde{Z}}^{(IJ)}_{\alpha \beta}= 0$ since it is symmetric traceless. This can be confirmed by  using the fact that 
\begin{equation}
\delta_{IJ} \G^{EF(IJ)}_{AB} = \G^I_{AB} \Gt^{IEF} + \Gt^{IEF} \G^I_{AB} = -4\delta^{EF}_{AB}.
\end{equation}
Thus the only term that survives is the trace part $X_{\alpha \beta}$. We can therefore write
\begin{equation}
\delta_{IJ} \{ Q^I_\alpha, Q^J_\beta \} =  -12P_\mu (\gamma^\mu C)_{\alpha \beta}  + 8 \tr \int d^2 \sigma D^i( X_AX^AX_BX^B - X^AX_AX^BX_B) \varepsilon_{ij} (\gamma^j C)_{\alpha \beta}. \label{6}
\end{equation}
We see that the trace of the algebra contains a single charge, namely the one-form central charge ${\mathcal{Z}}_i$. As already mentioned, this charge corresponds to the energy of the BPS Fuzzy-Funnel configuration calculated in \cite{Hanaki:2008cu}. Let us show this explicitly.
\subsection{Hamiltonian Analysis and Fuzzy-funnel BPS equation}
The fuzzy-funnel ABJM BPS equation can be obtained by combining the kinetic and potential terms in the Hamiltonian and rewriting the expression as a modulus squared term plus a topological term. The squared term tells us the BPS equations and the topological term tells us the energy bound of the BPS configuration when the BPS equations are satisfied. In \cite{Hanaki:2008cu} the ABJM potential was written as
\begin{align}
V = &\frac{4 \pi^2}{k^2} \tr (|Z^A Z_A Z^B - Z^B Z_A Z^A - W^A W_A Z^B + Z^B W_A W^A |^2 \nn\\
&+ |W^A W_A W^B - W^B W_A W^A - Z^A Z_A W^B + W^B Z_A Z^A|^2 ) \\
&+ \frac{16\pi^2}{k^2} \tr (|\epsilon_{AC}\epsilon^{BD} W_B Z^C W_D|^2  + |\epsilon^{AC} \epsilon_{BD}Z^B W_C Z^D|^2).\nn
\end{align}
where $Z^A$ and $W^A$ are the upper and lower two components respectively of the 4 component complex scalar $X^A$. The first two lines correspond to D-term potential pieces whereas the last line corresponds to F-term potential pieces (from the superspace perspective \cite{Benna:2008zy}). In \cite{Hanaki:2008cu} the potential and kinetic terms were combined in two different ways, depending on whether the F-term or D-term potential is used in conjunction with the kinetic term. This leads to two sets of BPS equations. For the case in which $W^A=0$ the scalar part of the Hamiltonian only contains D-term contributions and takes the form
\begin{align}
H = &\int dx^1 ds \tr (|\partial_s Z^A + \frac{2\pi}{k}(Z^B Z_B Z^A - Z^A Z_B Z^B)|^2) \nn \\
&+\frac{\pi}{k} \tr \partial_s (Z_A Z^A Z_B Z^B - Z^A Z_A Z^B Z_B),
\end{align}
where $x^2 = s$. As usual, the first line gives the  BPS equation
\begin{equation}
\partial_s Z^A + \frac{2\pi}{k}(Z^B Z_B Z^A - Z^A Z_B Z^B) = 0, \label{ff1}
\end{equation}
and the second line gives the energy of the system when the BPS equation is satisfied
\begin{equation}
E = \frac{\pi}{k} \tr  \int ds dx^1  \partial_s (Z_A Z^A Z_B Z^B - Z^A Z_A Z^B Z_B). \label{bd2}
\end{equation} 
The general procedure for finding a solution to \eqref{ff1} is to consider an ansatz in which the complex scalar fields separate into an s-dependent and s-independent part
\begin{equation}
Z^A = f(s) G^A , \quad f(s) = \sqrt{\frac{k}{4 \pi s}} , \label{ansatz}
\end{equation}
where, according to \eqref{ff1}, the $G^A$ satisfy
\begin{equation}
G^A = G^B G^\dagger_B G^A - G^A G^\dagger_B G^B.
\end{equation}
In \cite{Gomis:2008vc, Hanaki:2008cu} a solution to this equation was presented and interpreted as describing a fuzzy $S^3 / Z_k$.\footnote{Note that in \cite{Nastase:2009zu, Nastase:2009ny} the M2-M5 fuzzy funnel solution of ABJM was shown to describe a fuzzy $S^2$ as opposed to the conjectured fuzzy $S^3$. The solution is consistent with a spacetime picture in which an M5-brane wraps an $S^3/Z_k$, with the three-sphere given by the Hopf fibration of an $S^1/Z_k$ over $S^2$, and identified $S^1/Z_k$ with the M-theory circle. In the $k \rightarrow \infty$ limit one has a (double) dimensional reduction and a D4-brane wrapping an $S^2$ in Type IIA string theory, as opposed to an M5-brane wrapping the $S^3/Z_k$ in M-theory.} The energy bound corresponding to this fuzzy-funnel configuration \eqref{bd2} exactly matches the central charge term appearing in \eqref{6} (when $W^A = 0$). Thus we see that the physical information corresponding to the energy bound of the fuzzy funnel configuration appears in the trace expression of the algebra \eqref{superduper}. One might wonder about the charge ${\mathcal{Z}}^{AB}_{EF,i}$ appearing in \eqref{superduper}. Interestingly, this charge corresponds to the energy bound of the F-term configuration considered in \cite{Hanaki:2008cu}. In other words, minimising the Hamiltonian using the F-term potential results in an energy bound of the form ${\mathcal{Z}}^{AB}_{EF,i}$. It would be informative to solve the correpsonding BPS equation and provide a space-time interpretation of this charge.

In this chapter we investigated the worldvolume superalgebra of the $\mathcal{N}=6$ and $\mathcal{N}=8$ Bagger-Lambert theories. As expected from the M2-brane perspective we found charges corresponding to $(1|M2, M5)$ and $(0| M2, M2)$ intersections. It should also be possible to see the $(1|M2, M5)$ configuration from the M5-brane worldvolume superalgebra (as well as the $(3| M5, M5)$ intersection which is related to $(0| M2, M2)$ through spacetime duality). In the next chapter we will investigate the superalgebra of the Nambu-Poisson M5-brane which has been proposed as a model of an M5-brane in a background of three-form flux. 

\chapter{Nambu-Poisson M5-brane superalgebra}
In this chapter we calculate the worldvolume superalgebra of the Nambu-Poisson M5-brane theory as presented in \cite{Low:2009de}. We begin with a review of the M5-brane model proposed in \cite{Ho:2008nn, Ho:2008ve} which is believed to describe an M5-brane in a three-form flux background. We then investigate the superalgebra associated with the theory. In particular we derive the central charges corresponding to M5-brane solitons in 3-form backgrounds. We show that double dimensional reduction of the superalgebra results in the Poisson bracket terms of a non-commutative D4-brane superalgebra. We provide interpretations of the D4-brane charges in terms of spacetime intersections. 

\section{Nambu-Poisson M5-brane Theory}
In this section we provide a brief overview of the Nambu-Poisson M5-brane model \cite{Ho:2008nn, Ho:2008ve}.\footnote{For alternative reviews see for example \cite{Ho:2009zt, Ho:2009mi, Pasti:2009xc}. For more recent developments see \cite{Chen:2010br, Chen:2010jg}.} For an n-dimensional manifold $\mathcal{N}$, a tri-linear map $\{ . , . , . \}$ that maps three functions on $\mathcal{N}$ to a single function on $\mathcal{N}$ is called a Nambu-Poisson bracket if in addition to the fundamental identity \eqref{vidi} and complete antisymmetry it satisfies the Leibniz rule
\begin{equation}
\{fg , a, b \} = f \{ g , a, b\} + g \{ f , a, b\}.
\end{equation}
One of the most important properties of the Nambu-Poisson bracket is the decomposition theorem\footnote{See for example \cite{Marmo:1997tb}.}  which states that by a suitable choice of local coordinates on $\mathcal{N}$, the Nambu-Poisson bracket can be reduced to Jacobian form
\begin{equation}
\{ f  , g , h \} = \e^{\mud \nud \rhod} \frac{\partial f}{\partial y^{\mud}} \frac{\partial g}{\partial y^{\nud}} \frac{\partial h}{\partial y^{\rhod}}
\end{equation}
where $y^{\mud}$ represents 3 coordinates out of $n$-coordinates on the manifold. The basic idea of \cite{Ho:2008nn} was to consider the 3-manifold $\mathcal{N}$ on which Nambu-Poisson bracket is defined as an internal manifold from the perspective of the M2-brane worldvolume $\mathcal{M}$. The generators of Bagger-Lambert theory are taken to be an infinite dimensional basis of functions on $\mathcal{N}$ such that
\begin{equation}
\{ \chi^a , \chi^b , \chi^c \} = \e^{\mud \nud \rhod} \partial_{\mud} \chi^a \partial_{\nud} \chi^b \partial_{\rhod} \chi^c = f^{abc}_{\ \ \ \ d} \chi^d. \label{a}
\end{equation}
One then expands the 3 dimensional fields appearing in the Bagger-Lambert theory in terms of the basis generators $\{ \chi^a (y) \}$. This results in 6-dimensional fields defined on $\mathcal{M} \times \mathcal{N}$. For example the scalar field takes the form $X^I (x, y) = \sum_a X^I_a (x)\chi^a (y)$.
The inner product may be defined as integration over the manifold $\mathcal{N}$ 
\begin{equation}
\langle f, g \rangle = \frac{1}{g^2} \int_{\mathcal{N}} d^3 y f(y) g(y).
\end{equation}
The inner product between basis elements defines the metric
\begin{equation}
h^{ab} = \langle \chi^a , \chi^b \rangle.
\end{equation}
Of particular importance is the Bagger-Lambert gauge field which may be expressed as a bi-local field defined by
\begin{equation}
A_{\mu} (x, y ,y') = A_\mu^{ab} (x) \chi^a (y) \chi^b (y'). \label{b}
\end{equation}
Taylor expanding the left hand side of this expression around $\Delta y^{\mud} = y^{' \mud} - y^{\mud}$ results in
\begin{equation}
A_{\mu} (x, y, y') = a_\mu (x,y) + b_{\lambda \mud } (x,y) \Delta y^{\mud} + \frac{1}{2} c_{\lambda \mud \nud} (x,y) \Delta y^{\mud} \Delta y^{\nud} + \ldots
\end{equation}
where 
\begin{equation}
b_{\lambda \mud} (x, y) = \frac{\partial}{\partial y^{'\mud}} A_{\lambda} (x, y ,y') |_{y'=y} \label{c}
\end{equation}
It turns out that this is the only component in the Taylor expansion that ever appears in the Bagger-Lambert action. This follows from the fact that $A_{\lambda ab}$ always appears in the action in the form $f^{bcd}_{\ \ \ a}A_{\lambda bc}$ and using \eqref{a}, \eqref{b} and \eqref{c}
\begin{align}
f^{bcd}_{\ \ \ a} A_{\lambda bc} &= \e^{\mud \nud \rhod} \langle \partial_{\mud} b_{\lambda \nud} \partial_{\rhod} \chi^d , \chi_a \rangle.
\end{align}
So for example the covariant derivative appearing in the Bagger-Lambert theory \eqref{cov4} can be written as
\begin{align}
\D_\lambda X^I (x,y) &\equiv [\partial_\lambda X^{Ia} (x) - g f^{bcd}_{\ \ \ a} A_{\lambda bc} X^I_d (x)] \chi^a (y) \nn \\
&= \partial_\lambda X^I (x,y) - g \e^{\mud \nud \rhod} \partial_{\mud} b_{\lambda \nud} (x,y)\partial_{\rhod} X^I (x,y) \nn \\
&= \partial_\lambda X^I - g \{ b_{\lambda \nud} , y^{\nud} , X^I \}
\end{align}
We see that the only term from the Taylor expansion that appears in the covariant derivative is the two-form gauge field $b_{\lambda \nud}$. In \cite{Ho:2008ve} this two-form gauge field is interpreted as the gauge potential associated with volume preserving diffeomorphisms of $\mathcal{N}$. Why volume preserving diffeomorphisms? Re-writing the Bagger-Lambert gauge transformation in terms of Nambu-Poisson bracket one finds that it takes the form (for scalars and fermions)
\begin{equation}
\delta_\Lambda \Phi (x,y) = g (\delta_\Lambda y^{\mud}) \partial_{\mud} \Phi (x,y).
\end{equation}
This may be considered the infinitesimal transformation of the field under the reparameterisation 
\begin{equation}
y^{\mud '} = y^{\mud} - g \delta y^{\mud}
\end{equation}
subject to the condition $\partial_{\mud} (\delta y^{\mud}) = 0$ which implies it is a volume preserving diffeomorphism since
\begin{equation}
\textrm{det} |\frac{\partial y'}{\partial y}| = 1 + \frac{\partial}{\partial y^{\mud}} (\delta y^{\mud}).
\end{equation}
Thus we have the situation in which the gauge symmetry of the original Bagger-Lambert theory is interpreted as a volume preserving diffeomorphism of the internal manifold $\mathcal{N}$. It is possible to define another type of covariant derivative $\D_{\mud}$ which acts on $\mathcal{N}$ by observing that, for a scalar field $\Phi$, the object $\{X^{\nud}, X^{\rhod}, \Phi \}$ transforms covariantly under volume preserving diffeomorphisms. Thus we can define $\D_{\mud}$ as
\begin{align}
\D_{\mud} \Phi &= \frac{g^2}{2} \e_{\mud \nud \rhod} \{ X^{\nud} , X^{\rhod}, \Phi \} \nn \\
&= \partial_{\mud} \Phi + g (\partial_{\lambdad} b^{\lambdad} \partial_{\mud} \Phi- \partial_{\mud} b^{\lambdad} \partial_{\lambdad} \Phi) + \frac{g^2}{2} \e_{\mud \nud \rhod} \{ b^{\nud} , b^{\rhod}, \Phi \}
\end{align}
where the field $b^{\mud}$ is defined through the relation
\begin{equation}
X^{\mud} = \frac{1}{g}y^{\mud} + b^{\mud}, \quad \quad b_{\mud \nud} = \e_{\mud \nud \rhod} b^{\rhod}.
\end{equation}
This may be interpreted as a weakened form of the static gauge condition relating to the fact that $y^{\mud}$ space does not possess full diffeomorphism invariance but only volume preserving diffeomorphism invariance. This implies that one cannot completely fix the $X^{\mud}$ fields, resulting in residual degrees of freedom parameterised by $b^{\mud}$. The quantity $X^{\mud}$ defined above can be shown to transform as a scalar under volume preserving diffeomorphisms. This fact allows one to construct gauge field strengths which transform as scalars under gauge variations. We now have two types of covariant derivative, $\D_{\mu}$ and $\D_{\mud}$ which together constitute a set of derivatives acting on the 6-dimensional space $\mathcal{N} \times \mathcal{M}$. Just as with ordinary gauge theory the field strength arises in the commutator of covariant derivatives
\begin{align}
[\D_{\mud} , \D_{\nud}] \Phi &= g^2 \e_{\nud \mud \sigmad} \{ \mH_{\dot{1} \dot{2} \dot{3}} , X^{\sigmad} , \Phi \}, \nn \\
[\D_\lambda , \D_{\lambdad}] \Phi &= g^2 \{ \mH_{\lambda \nud \lambdad} , X^{\nud} , \Phi \}.
\end{align}
where
\begin{align}
\mH_{\lambda \mud \nud} &= \e_{\mud \nud \lambdad} \D_{\lambda} X^{\lambdad} \\ \nn 
&= H_{\lambda \mud \nud} - g \e^{\rhod \sigmad \lambdad} \partial_{\rhod} b_{\lambda \sigmad} \partial_{\rhod} b_{\mud \nud} 
\end{align}
and
\begin{align}
\mH_{\dot{1} \dot{2} \dot{3}} &= g^2 \{ X^{\dot{1}}, X^{\dot{2}}, X^{\dot{3}} \} - \frac{1}{g} = \frac{1}{g} (V-1) \nn \\
&= H_{\dot{1} \dot{2} \dot{3}} + \frac{g}{2} (\partial_{\mud} b^{\mud} \partial_{\nud} b^{\nud} - \partial_{\mud} b^{\nud} \partial_{\nud} b^{\mud} ) + g^{2} \{b^{\dot{1}} , b^{\dot{2}} , b^{\dot{3}} \}
\end{align}
where V is the induced volume defined by
\begin{equation}
V = g^3 \{X^{\dot{1}}, X^{\dot{2}}, X^{\dot{3}} \}.
\end{equation}
In \cite{Ho:2008ve}, these were the only components of the field strength ${\mathcal{H}}_{\bar{\mu} \bar{\nu} \bar{\rho}}$ ($\bar{\mu} = 0, \ldots 5$) appearing in the action derived from the BLG theory. In the absence of scalar and fermion matter fields, the non-linear chiral field action of \cite{Ho:2008ve} took the form
\begin{equation}
S = - \int d^3x d^3y \left( \frac{1}{4} \mH_{\mu \nud \rhod} \mH^{\mu \nud \rhod} + \frac{1}{12} \mH_{\mud \nud \rhod} \mH^{\mud \nud \rhod} + \frac{1}{2} \e^{\mu \nu \rho} B_{\mu}^{\ \mud} \partial_{\nu} b_{\rho \mud} + g \textrm{ det} B_{\mu}^{\ \mud}\right)
\end{equation} 
where $B_{\mu}^{\ \mud} = \e^{\nud \rhod \mud} \partial_{\nud} b_{\mu \rhod}$. This action is invariant under volume preserving diffeomorphisms but does not possess gauge covariance due to the last two terms (which originate from the BLG Chern-Simons term). In \cite{Ho:2008ve} it was assumed that the field strength components ${\mathcal{H}}_{\mu \nu \rho}$ and ${\mathcal{H}}_{\mu \nu \rhod}$ are dual to $\mH_{\mud \nud \rhod}$ and $\mH_{\mu \nud \rhod}$ respectively. In other words
\begin{equation}
\mH_{\mu \nud \rhod} = \frac{1}{2} \e_{\mu \nu \rho} \e_{\nud \rhod \mud} \mH^{\nu \rho \mud}, \quad \mH_{\mud \nud \rhod} = -\frac{1}{6} \e_{\mud \nud \rhod} \e^{\mu \nu \rho} \mH_{\mu \nu \rho}. \label{birt}
\end{equation}
This was confirmed in \cite{Pasti:2009xc} where it was shown that solving the field equations associated with $b_{\mu \nud}$ and $b_{\mud \nud}$ is tantamount to imposing the Hodge self-duality condition on the non-linear field strength ${\mathcal{H}}_{\bar{\mu} \bar{\nu} \bar{\rho}}$. This allowed the authors to re-write the gauge-field Lagrangian in a gauge-covariant form as
\begin{equation}
S = - \int d^3x d^3 y \left(  \frac{1}{8} \mH_{\mu \nud \rhod} \mH^{\mu \nud \rhod} + \frac{1}{12} \mH_{\mud \nud \rhod} \mH^{\mud \nud \rhod} - \frac{1}{144} \e^{\mu \nu \rho} \e^{\mud \nud \rhod} \mH_{\mu \nu \rho} \mH_{\mud \nud \rhod} - \frac{1}{12g} \e^{\mu \nu \rho} \mH_{\mu \nu \rho} \right).
\end{equation}
The last term in this expression can be interpreted as a coupling of the M5-brane to the constant background $C_3$ field which has non-zero components $C_{\mud \nud \rhod} = \frac{1}{g} \e_{\mud \nud \rhod}$. Following \cite{Pasti:2009xc}, it is possible to re-write this as
\begin{equation}
\int d^3 x d^3 y \frac{1}{12g} \e^{\mu \nu \rho} \mH_{\mu \nu \rho} = \frac{1}{2} \int \mH_3 \wedge C_3. \label{couple}
\end{equation}
This action possesses full volume preserving diffeomorphism invariance. However the Lorentz symmetry is broken by the presence of the three-form field. This concludes the brief overview of the Nambu-Poisson M5-brane model.

\section{Nambu-Poisson M5-brane superalgebra}
Our plan is to take the central charges of the Bagger-Lambert M2-brane superalgebra and re-express the 3-dimensional fields in terms of 6-dimensional fields using the conventions outlined in the previous section. We will see that in doing so, the M2-brane central charges will recombine to form M5-brane central charges corresponding to solitons of the worldvolume theory. We will also look at the Bogomoly'ni completion of the Hamiltonian and derive the BPS equations corresponding to the self-dual string soliton and the 3-brane vortex. To begin with we will calculate the Nambu-Poisson BLG supercharge. This will highlight the methodology and introduce useful notation and conventions.
\subsection{Nambu-Poisson BLG Supercharge}
We begin by deriving an expression for the supercharge of the Nambu-Poisson BLG theory. The basic idea is to take the original BLG M2-brane supercharge and expand the fields in terms of the basis $\{\chi^a \}$. As we noted above, the spatial integral of the zeroth component of the supercurrent represents the supercharge, and the worldvolume supercurrent takes the simple form 
\begin{equation}
-\be J^{\mu} = \bp^a \g^\mu \delta \psi_a \label{alpha}
\end{equation}
where $\delta \psi_a$ refers to the supersymmetric variation of the fermion field. 
Evaluating \eqref{alpha} using the original BLG supersymmetry transformation one finds
\begin{equation}
\be J^\mu = - \be \Tr (D_\nu X^I, \g^\nu \g^I \g^\mu \psi )  -  \frac{1}{6} \be  \g^{IJK} \g^{\mu} \Tr ( \{ X^I, X^J, X^K \}, \psi ).
\end{equation}
Here the trace defines an inner product. We wish to re-express this in terms of the 6-dimensional covariant derivatives and field strengths introduced in the previous section. This can be achieved by splitting the scalar field index $I \rightarrow (\mud , i)$ and making the replacements
$[ * , * , * ] \rightarrow g^2 \{ * , * , * \} $ and $\textrm{Tr} \rightarrow \la \rangle $. Using this prescription along with the definitions presented in the previous section we can write the supercharge as
\begin{align}
\be Q  = &- \be\g^\nu \g^i \g^0 \la \D_\nu X^i , \psi \rangle - \be\g^{\nud} \g^i \g^0 \g_{\dot{1} \dot{2} \dot{3}}\la \D_{\nud} X^i , \psi \rangle \nn \\
&- \frac{1}{2} \be \g^\nu \g^{\nud \lambdad} \g^0 \g_{\dot{1} \dot{2} \dot{3}}\la \mH_{\nu \nud \lambdad} , \psi \rangle - \be \g_{\dot{1} \dot{2} \dot{3}} \g^0 \la  (\mH_{\dot{1} \dot{2} \dot{3}} + \frac{1}{g}) , \psi \rangle \nn \\
&- \frac{g^2}{6} \be \g^{ijk} \g^{0} \la \{ X^i, X^j , X^k \}, \psi \rangle - \frac{g^2}{2}\be \g^{\mud} \g^{ij} \g^{0} \la \{X^{\mud}, X^i, X^j \}, \psi \rangle \label{ace}
\end{align}
where we have suppressed the integral over the worldvolume coordinates of the M2-brane worldvolume. In deriving this expression we made use of the fact that
\begin{align}
\g_{\mud \nud \rhod} = \g_{\dot{1} \dot{2} \dot{3}} \e_{\mud \nud \rhod}  \quad  \textrm{and} \quad (\g_{\dot{1} \dot{2} \dot{3}})^2 = -1
\end{align}
from which it follows that
\begin{equation}
\e_{\mud \nud \rhod} \g^{\mud \nud} = 2 \g_{\rhod} \g_{\dot{1} \dot{2} \dot{3}}.
\end{equation}
The presence of $\g_{\dot{1} \dot{2} \dot{3}}$ in the second and third terms of \eqref{ace} means that only the SO(1,2)$\times$SO(3) subgroup of the full 6 dimensional Lorentz symmetry is manifest. A similar difficulty was encountered for the fermion kinetic terms in \cite{Ho:2008ve}. There it was shown that it is possible to perform a unitary transformation of the spinor variables 
\begin{equation}
\be = \be' U, \quad \psi = U \psi' \label{qw}
\end{equation}
where U is the matrix
\begin{equation}
U = \frac{1}{\sqrt{2}} (1 - \g_{\dot{1} \dot{2} \dot{3}}).
\end{equation}
Performing this transformation on the supercharge, and using the fact that $[\g^{\mud} , \g_{\dot{1} \dot{2} \dot{3}}] = \{ \g^i , \g_{\dot{1} \dot{2} \dot{3}} \} = \{ \g^\mu , \g_{\dot{1} \dot{2} \dot{3}}\} = 0$ results in
\begin{align}
\be' Q' = &- \be'\g^\nu \g^i \g^0 \la \D_\nu X^i , \psi' \rangle - \be' \g^{\nud} \g^i \g^0 \la \D_{\nud} X^i , \psi' \rangle \nn \\
&- \frac{1}{2} \be' \g^\nu \g^{\nud \lambdad} \g^0 \la \mH_{\nu \nud \lambdad} , \psi' \rangle - \be' \g_{\dot{1} \dot{2} \dot{3}} \g^0 \la  (\mH_{\dot{1} \dot{2} \dot{3}} + \frac{1}{g}) , \psi' \rangle \nn \\
&+ \frac{g^2}{6} \be' \g^{ijk} \g^{0} \g_{\dot{1} \dot{2} \dot{3}}\la \{ X^i, X^j , X^k \}, \psi' \rangle - \frac{g^2}{2}\be' \g^{\mud} \g^{ij} \g^{0} \la \{X^{\mud}, X^i, X^j \}, \psi' \rangle.  \label{ace2}
\end{align}
It is possible to look at the weak coupling limit in which $g \rightarrow 0$. In this limit $\mH \rightarrow H$ and $\D \rightarrow \partial$ and the supercharge becomes\footnote{Note that in deriving the weak coupling limit we have used the additional fermionic shift symmetry introduced in \cite{Ho:2008ve} to eliminate the $1/g$ term from the fermion supersymmetry transformation.}
\begin{equation}
\be'Q' = -\be'  \g^{\bar{\nu}} \g^i \g^0 \la \partial_{\bar{\nu}} X^i , \psi' \rangle - \frac{1}{12} \be' \g^{\bar{\mu} \bar{\nu} \bar{\lambda}} \g^0 \la H_{\bar{\mu} \bar{\nu} \bar{\lambda}} , \psi'\rangle \label{lol}
\end{equation}
with $\bar{\mu} = (\mu , \mud)$. This expression agrees with the supercharge associated with an abelian $\mathcal{N} = (2,0)$ tensor multiplet \cite{Romans:1986er}. In order to calculate the Nambu-Poisson M5-brane superalgbera one could in principle calculate the anticommutator of the supercharge \eqref{ace2}. In this chapter we will adopt a different approach. We will make use of the M2-brane superalgebra \eqref{algebra} calculated in the previous chapter and re-express the charges \eqref{charges} in terms of M5-brane fields (using the method outlined above). To this we now turn.

\subsection{Re-writing of M2-brane central charges}
We begin by by splitting the scalar field index $I \rightarrow (\mud , i)$ and making the replacements
$[ * , * , * ] \rightarrow g^2 \{ * , * , * \} $ and $\textrm{Tr} \rightarrow \la \rangle$. In what follows, for the sake of clarity, we will suppress the angle bracket $\langle \rangle$ which denotes the inner product. Making repeated use of the conventions outlined in the previous section and the Appendix results in the following expressions for the charges \eqref{charges}
\begin{align}
-D_\alpha X^I D_\beta X^J \e^{\alpha \beta} \g^{IJ} \g^0 = &- \D_\alpha X^i \D_\beta X^j \g^{\alpha \beta} \g^{ij}  + \frac{1}{2} \mH_{\alpha \rhod \lambdad} \mH_{\beta \sigmad \omegad} \g^{\rhod} \g^{\lambdad \sigmad \omegad} \g^{\alpha \beta} \nn \\
 &+ \D_\beta X^i \mH_{\alpha \nud \lambdad} \g^{\beta \alpha \nud \lambdad} \g^i \g_{\dot{1} \dot{2} \dot{3}}. \label{beast1} \\
 \nn \\
D_0 X^I \{ X^I, X^J , X^K \} \g^{JK} \g^0 = &- 2 \D_0 X^i \D_{\lambdad} X^i \g^{\lambdad}\g^0 \g_{\dot{1} \dot{2} \dot{3}} + 2 \D_{\omegad} X^i \mH_{0 \nud \omegad}  \g^{\nud} \g^i \g^0 \nn \\
&+ \frac{g^2}{2} \mH_{0 \sigmad \lambdad} \g^0 \left( \{ X^{\mud} , X^{\nud} , X^{\rhod} \} \e^{\mud \sigmad \lambdad} \g^{\nud \rhod} +   \{ X^{\mud}, X^j , X^k \} \g^{\mud \sigmad \lambdad} \g^{jk} \g_{\dot{1} \dot{2} \dot{3}} \right)\nn \\
&+ g^2D_0 X^i \{ X^i,  X^{\nud} , X^j \} \g^{\nud} \g^j \g^0 + g^2 \D_0 X^i \{ X^i , X^j , X^k \} \g^{jk} \g^0. \label{beast2} \\
\nn \\
Z_{\alpha IJKL} \g^{IJKL} \g^{\alpha 0} = &- \D_{\sigmad} X^i \mH_{\beta \lambdad \omegad} \g^{\sigmad \beta \lambdad \omegad} \g^i + 2 \D_\beta X^i (\mH_{\dot{1} \dot{2} \dot{3}} + \frac{1}{g}) \g^\beta \g^i \g_{\dot{1} \dot{2} \dot{3}}  \nn \\
&+ 2 \D_\beta X^i \D_{\lambdad} X^j \g^{\beta \lambdad} \g^{ij} \g_{\dot{1} \dot{2} \dot{3}} - g^2 \mH_{\beta \rhod \nud} \{ X^{\nud} , X^i , X^j \} \g^{\beta \rhod} \g^{ij} \g_{\dot{1} \dot{2} \dot{3}} \nn \\
&- g^2 \D_\beta X^i \{ X^{\mud} , X^j , X^k \} \g^{\beta \mud} \g^{ijk} + \frac{g^2}{3} \D_\beta X^i \{ X^j, X^k ,X^l \} \g^\beta \g^{ijkl} \nn \\
&+ \frac{g^2}{6} \mH_{\beta \rhod \lambdad} \{ X^i , X^j , X^k \} \g^{\beta \rhod \lambdad} \g^{ijk} \g_{\dot{1} \dot{2} \dot{3}}. \label{beast3} \\
\nn \\
Z_{IJKL} \g^{IJKL} = &- \D_{\lambdad} X^i D_{\nud} X^j \g^{\lambdad \nud} \g^{ij} + g^2 \D_{\rhod} X^j \{ X^{\mud} , X^k , X^l\} \g^{\rhod \mud} \g^{jkl} \g_{\dot{1} \dot{2} \dot{3}} \nn \\
&+ g^2 \D_{\lambdad} X^i \{X^i , X^j, X^k  \} \g^{\lambdad} \g^{jk} \g_{\dot{1} \dot{2} \dot{3}} + 2g^2 \D_{\mud} X^i \{ X^{\mud}, X^i , X^j \} \g^j \g_{\dot{1} \dot{2} \dot{3}} \nn \\
&+ g^4 \{ X^{\mud} , X^i , X^j \} \left( \frac{1}{2}\{X^{\mud} , X^{\nud} , X^{\rhod} \}  \g^{\nud \rhod} \g^{ij} -  \{ X^i , X^k , X^{\nud} \} \g^{\mud \nud} \g^{jk} \right) \nn \\
& + g^4  \{X^{\mud} , X^i , X^j \} \left( \frac{1}{4}\{ X^{\mud} , X^k , X^l \} \g^{ijkl} -  \{ X^i , X^k , X^l \} \g^{\mud} \g^{jkl} \right) \nn \\
&+ \frac{g^4}{4} \{ X^i , X^j , X^k \} \{ X^i , X^l, X^m \} \g^{jklm}.
\label{beast4}
\end{align}
Although these terms look complicated we notice that many of them share the same structure despite originating from different M2-brane charges. Importantly, we notice that $Z_{IJKL}$ plays a crucial role in obtaining the M5-brane charges. For example we see that the first term in \eqref{beast1} will combine with the first term in \eqref{beast4} to give the 3-brane vortex charge. The hope is that all of these terms will combine to form central charges of the M5-brane worldvolume theory. However as it stands, many terms sharing the same structure in \eqref{beast1}-\eqref{beast4} have the wrong relative sign to combine. It turns out that this problem can be resolved by multiplying the central charges from the left and right by the unitary matrix U. This is a reasonable thing to do since, as mentioned earlier, the method for calculating the M2-brane superalgebra involves calculating $\be \delta J^0 = \be \{ Q , Q \} \e$. We have  seen that when calculating spinor quantities in terms of Nambu-Poisson brackets it is necessary to perform unitary transformations on the spinor fields to bring them to the correct form (for example the fermion kinetic terms, fermion supersymmetry transformation and supercharge).  Therefore one might expect that the correct expression for the M5-brane superalgbera would result from evaluating $\be' \{ Q' , Q' \} \e'$. This is equivalent to multiplying the central charges in \eqref{beast1}-\eqref{beast4} from the left and right by the unitary matrix $U$. We will see in the next section that by performing this unitary transformation, all terms of similar structure combine into M5-brane charges.

\subsection{Nambu-Poisson BLG Superalgebra}
In this section we will combine the terms appearing in the previous section into M5-brane central charges and write the Bagger-Lambert Nambu-Poisson superalgebra. At lowest order in the coupling $g$ we will see that the superalgebra contains the charges \eqref{xx21} and \eqref{hhh21} expected from the worldvolume superalgebra of the M5-brane. In addition we will find higher order coupling central charges, some of which involve the background 3-form $C$-field. Since the resulting expression is long and complicated we will split the superalgebra into three parts depending on the order of the coupling of the central charge. Having performed the unitary transformation described in the previous section one finds
\begin{equation}
\{ Q, Q \}_{\textrm{central}} \propto  Z (g^0)\g_{\dot{1} \dot{2} \dot{3}} + Z(g^2) \g_{\dot{1} \dot{2} \dot{3}} + Z(g^4) \g_{\dot{1} \dot{2} \dot{3}}
\end{equation}
We will take each term separately and try to provide an interpretation of the central charges. To begin with we look at $Z(g^0)$. Remarkably one finds the compact expression 
\begin{align}
Z(g^0) = &+ \D_a X^i \D_b X^j \g^{ab} \g^{ij} + 2 \D_{0} X^i D_{\lambdad} X^i \g^0 \g^{\lambdad} + \frac{1}{3}\D_a X^i \mH_{bcd} \g^{abcd} \g^i \nn \\
 &-\frac{2}{g} \D_{\beta} X^i \g^{\beta} \g^i \g_{\dot{1} \dot{2} \dot{3}} + \mH_{0 \mud \nud} (\mH^{\mud \nud \rhod} + \frac{1}{g} \e^{\mud \nud \rhod}) \g^0 \g^{\rhod} -\frac{1}{2} \mH_{\alpha \rhod \lambdad} \mH_{\beta \sigmad \omegad} \g^{\rhod} \g^{\lambdad \sigmad \omegad} \g^{\alpha \beta} \label{lowest}
\end{align}
with $a = (\alpha , \mud)$ representing the spatial coordinates on the worldvolume of the M5-brane. Note that in obtaining this result we made use of $\mH^{\alpha \beta \lambdad} = - \frac{1}{2} \e^{\alpha \beta} \e^{\nud \rhod \lambdad} \mH_{0 \nud \rhod}$ as well as 
\begin{align}
\frac{1}{3}\D_a X^i \mH_{bcd} \g^{abcd} = &+ \D_\alpha X^i \mH_{\beta \mud \nud} \g^{\alpha \beta \mud \nud}+  \D_{\mud} X^i \mH_{\beta \nud \lambdad} \g^{\mud \beta \nud \lambdad} \nn \\
&+  \D_{\mud} X^i \mH_{\beta \gamma \nud} \g^{\mud \beta \gamma \nud} + \frac{1}{3}\D_\alpha X^i \mH_{\nud \sigmad \lambdad} \g^{\alpha \nud \sigmad \lambdad}.
\end{align} 
We see that there are three types of term in $Z(g^0)$: Charges of the form $\D X \D X$, those of the form $\D X \mH$ and finally those of the form $\mH \mH$. We see that the first term $\D_a X^i \D_b X^j$ corresponds to the charge of the 3-brane vortex living on the M5-brane worldvolume. When only two scalar fields are active (call them $X$ and $Y$) we can identify it as the charge $Z$ appearing in \eqref{xx21}. The $\D \mH$ term corresponds to the self-dual string charge. If we consider the situation in which only one scalar field is active (call it $Y$) and assume that this scalar is a function of only four of the spatial worldvolume coordinates of the M5 brane, namely $\hat{\mu} = (2, \dot{1}, \dot{2}, \dot{3})$, we see that the $\D \mH$ term becomes $\e^{\hat{\mu} \hat{\nu} \hat{\rho} \hat{\sigma}} \D_{\hat{\mu}} X  \mH_{\hat{\nu} \hat{\rho} \hat{\sigma}}$. This exactly corresponds to the energy bound $Y$ appearing in \eqref{hhh21}. The  $\D \mH$ term with coefficient $2/g$ appearing in \eqref{lowest} can be thought of as a contribution from the background 3-form gauge field $C$. Making the identification $C_{\dot{1} \dot{2} \dot{3}} = \frac{1}{6} \e^{\mud \nud \rhod} C_{\mud \nud \rhod} \propto 1/g$ we have  
\begin{equation}
-\frac{2}{g} \D_{\beta} X^i \g^{\beta} \g^i \g_{\dot{1} \dot{2} \dot{3}} = \frac{1}{3} \D_\beta X^i C_{\mud \nud \rhod} \g^{\beta \mud \nud \rhod} \g^i.
\end{equation}
We can think of this term as representing a C-field modification of the self-dual string charge in the directions $\dot{1} \dot{2} \dot{3}$. We can provide an interpretation of the $\mH \mH$ term by thinking about its double dimensional reduction. This will be carried out explicitly in the next section but for now we note that compactifying along the $\dot{3}$ direction of the M5-brane reduces the term $\frac{1}{2}\mH_{\alpha \rhod \lambdad} \mH^{\alpha 0 \lambdad} + \mH^{0 \mud \nud} \mH_{\mud \nud \rhod}$ appearing in \eqref{lowest} to the charge of a D4-brane instanton of the form $ F\wedge {F}$. From the D4-brane perspective this can be thought of as the charge of a D0-brane within the worldvolume of the D4-brane. Thus from the M5-brane perspective it would appear that this charge describes an M-wave intersecting an M5-brane. Note that the factor of $1/g$ appearing in the $\mH \mH$ term in \eqref{lowest} can be thought of as a C-field modification of the MW-M5 charge along the $\dot{1} \dot{2} \dot{3}$ direction. We now turn to the $Z(g^2)$ charges,
\begin{align}
Z(g^2) = &+ g^2 \D_a X^i \{ X^{\mud} , X^j , X^k \} \g^{a \mud} \g^{ijk} - g^2 D_{\bar{a}} X^i \{ X^i , X^j, X^k \} \g^{\bar{a}} \g^{jk} \g_{\dot{1} \dot{2} \dot{3}} \nn \\
&- \frac{g^2}{3} \D_\beta X^i \{ X^j , X^k , X^l \} \g^\beta \g^{ijkl} \g_{\dot{1} \dot{2} \dot{3}} + 2g^2 \D_{\mud} X^i \{ X^{\mud} , X^i , X^j \} \g^j \nn \\
&+ g^2 \D_0 X^i \{ X^{\nud} , X^i , X^j \} \g^0 \g^{\nud} \g^j + \frac{g^2}{2} \mH_{0 \rhod \lambdad} \{X^{\mud} , X^j , X^k \} \g^{\mud \rhod \lambdad} \g^{jk} \g^0 \nn \\
&- g^2 \mH_{\beta \rhod \nud} \{ X^{\nud} , X^i , X^j \} \g^{\beta \rhod} \g^{ij} - \frac{g^2}{6} \mH_{\beta \rhod \lambdad} \{ X^i , X^j , X^k \}\g^{\beta \rhod \lambdad} \g^{ijk} \nn \\
&-2g^2 (\mH_{\dot{1} \dot{2} \dot{3}} + \frac{1}{g}) \{ X^{\mud}, X^i , X^j \} \g^{\mud} \g^{ij}. \label{Basu1}
\end{align}
In the second term the label $\bar{a} = (0 , \mud)$. A few comments are in order. As we will see in the next section, double dimensional reduction of the first term gives rise to a charge $g\tD_a X^i \{ X^j, X^k \}$. 
This term bares a structural similarity to the charge corresponding to a Nahm equation configuration in which multiple D2-branes intersect a D4-brane (in the case where Poisson-brackets have been replaced with matrix commutators). Thus one might expect to find a Nambu-Poisson analogue of the Basu-Harvey energy bound in the C-field modified M5-brane superalgebra. Indeed the charge $\frac{g^2}{3} \D_\beta X^i \{ X^j , X^k , X^l \}$ appearing in \eqref{Basu1} is reminiscent of the Basu-Harvey charge expressed in terms of Nambu-Poisson bracket. We will comment further on the interpretation of these charges later in the chapter when we consider the double-dimensional reduction of the algebra. 
For a discussion of the geometry of the M5-brane in the presence of a constant C-field as well as the derivation of the C-field modified Basu-Harvey equation as a boundary condition of the multiple M2-brane theory see for example \cite{Bergshoeff:2000jn, Bergshoeff:2000ai, Chu:2009iv}. 
Finally we have the $Z(g^4)$ charges which take the form
\begin{align}
Z(g^4) = &+ g^4 \{ X^i , X^j , X^{\mud}\} \{ X^i , X^k , X^{\nud} \} \g^{\mud \nud} \g^{jk} + g^4 \{X^{\mud} , X^i , X^j \} \{ X^i , X^k , X^l \} \g^{\mud} \g^{jkl} \g_{\dot{1} \dot{2} \dot{3}}  \nn \\
&- \frac{g^4}{4} \{ X^{\mud} , X^i , X^j \} \{ X^{\mud} , X^k , X^l\} \g^{ijkl} - \frac{g^4}{4} \{X^i , X^j , X^k \} \{ X^i , X^l , X^m \} \g^{jklm}. 
\end{align}
We will see that only the first and third terms survive the dimensional reduction and give rise to a charge analogous to the D4-brane charge found in Matrix theory (expressed in terms of Poisson brackets). The M5-brane analogue of this interpretation requires further investigation. Most importantly in this section we have seen that the BLG superalgebra based on Nambu-Poisson bracket contains the expected central charges corresponding to an M5-brane. Namely the 3-brane charge and self-dual string charge, suitably modified by the presence of the background gauge field. In the next section we will confirm the existence of these charges as energy bounds from the Hamiltonian perspective.

\subsection{Hamiltonian analysis and BPS equations}
We would like to derive the energy bounds corresponding to M5-M2 and M5-M5 intersections by looking at the Bogomoly'ni completion of the Nambu-Poisson BLG Hamiltonian. For the original Bagger-Lambert theory the energy-momentum tensor with fermions set to zero takes the form
\begin{equation}
T_{\mu \nu} = D_\mu X^I D_\nu X^I - \eta_{\mu \nu} (\frac{1}{2} D_\rho X^I D^\rho X^I + V).
\end{equation}
The Chern-Simons term does not contribute due to the fact that it is topological in nature and does not depend on the worldvolume metric. For static configurations the energy density takes the form
\begin{equation}
E = \frac{1}{2} D_\alpha X^I D_\alpha X^I + \frac{1}{12} [X^I , X^J , X^K]^2.
\end{equation}
Re-writing this expression in terms of M5-brane fields one finds
\begin{align}
E = &+\frac{1}{2}(\D_{\alpha} X^i)^2 + \frac{1}{2} (\D_{\mud} X^i)^2 + \frac{1}{4} \mH^2_{\alpha \mud \nud} + \frac{1}{2}(\mH_{\dot{1} \dot{2} \dot{3}} + C_{\dot{1} \dot{2} \dot{3}})^2 \nn \\
&+ \frac{g^4}{4} \{ X^{\mud} , X^i , X^j \}^2 + \frac{g^4}{12} \{ X^i , X^j , X^k \}^2 . \label{energy}
\end{align}
In order to find the energy bound corresponding to the self-dual string soliton we consider the situation in which there is only one active scalar field which we call $X$. Furthermore we assume that this scalar is a function of only four of the spatial worldvolume coordinates of the M5 brane, namely $\hat{\mu} = (2, \dot{1}, \dot{2}, \dot{3})$. In what follows we will assume $C_{\dot{1} \dot{2} \dot{3}} = 0$. In this case the energy density takes the following form
\begin{equation}
E = \frac{1}{2}(\D_{\hat{\mu}} X)^2 + \frac{1}{12} \mH^2_{\hat{\mu} \hat{\nu} \hat{\rho}} 
\end{equation}
We can re-write this as 
\begin{equation}
E = \frac{1}{2}|\D_{\hat{\mu}} X \pm \frac{1}{6}\e_{\hat{\mu}}^{\ \hat{\nu} \hat{\rho} \hat{\sigma}} \mH_{\hat{\nu} \hat{\rho} \hat{\sigma}}|^2 \mp \frac{1}{6}\e^{\hat{\mu} \hat{\nu} \hat{\rho} \hat{\sigma}} \D_{\hat{\mu}} X  \mH_{\hat{\nu} \hat{\rho} \hat{\sigma}} 
\end{equation}
We see that the energy density is minimised when the BPS equation
\begin{equation}
\D_{\hat{\mu}} X \pm \frac{1}{6}\e_{\hat{\mu}}^{\ \hat{\nu} \hat{\rho} \hat{\sigma}} \mH_{\hat{\nu} \hat{\rho} \hat{\sigma}} = 0 \label{cunt}
\end{equation}
is satisfied. In this case the energy is bounded by the central charge
\begin{equation}
Z = \mp \frac{1}{6}\e^{\hat{\mu} \hat{\nu} \hat{\rho} \hat{\sigma}} \D_{\hat{\mu}} X  \mH_{\hat{\nu} \hat{\rho} \hat{\sigma}}. \label{cunt2}
\end{equation}
The BPS equation \eqref{cunt} matches the result found recently in \cite{Furuuchi:2009zx}, where the author used supersymmetry arguments to derive the BPS equation corresponding to the self dual string soliton. Furthermore the energy bound \eqref{cunt2} matches the central charge found in the superalgebra of the previous section. The effect of turning on the $C_{\dot{1} \dot{2} \dot{3}}$ field is to modify the BPS equation and energy bound  by shifting the $\mH_{\dot{1} \dot{2} \dot{3}}$ component of the field strength.  In \cite{Ho:2008ve} a generalisation of the Seiberg-Witten map was shown to relate the theory of M5-branes based on Nambu-Poisson bracket with the theory of an M5-brane in constant 3-form background. In \cite{Furuuchi:2009zx},  solutions to the BPS equation \eqref{cunt} were found up to first order in the coupling $g$. The generalised Seiberg-Witten map  was then used to match the solution with the known results derived in \cite{Michishita:2000hu}, \cite{Youm:2000kr}. We have shown that the energy bound corresponding to this solution can be seen from the Hamiltonian and superalgebra perspective. Next we will consider the 3-brane soliton corresponding to an M5-M5 intersection. We will let the 3-brane lie in the $(0, 1, 2, \dot{1})$ plane with transverse directions in the $\dot{a} = (\dot{2}, \dot{3})$ directions. We allow two scalars to be non-zero and label them $X$ and $Y$.  These scalars are only functions of the transverse directions $(\dot{2}, \dot{3})$. In this case the energy density \eqref{energy} becomes
\begin{align}
E &= \frac{1}{2}(\D_{\dot{a}} X)^2  + \frac{1}{2}(\D_{\dot{a}} Y)^2  \nn \\
&= \frac{1}{2}|\D_{\dot{a}} X \pm \e_{\dot{a} \dot{b}}\D_{\dot{b}} Y |^2 \mp \e_{\dot{a} \dot{b}} \D_{\dot{a}} X \D_{\dot{b}} Y
\end{align}
Thus we see that the energy is bounded by 
\begin{equation}
Z = \e_{\dot{a} \dot{b}} \D_{\dot{a}} X \D_{\dot{b}} Y.
\end{equation}
This matches the central charge of the 3-brane vortex found in the superalgebra of the previous section. This concludes our discussion of the M5-brane central charges. In the next section we will investigate the double dimensional reduction of the M5-brane Nambu-Poisson superalgebra.
\\
\section{Double Dimensional Reduction of Superalgebra}
In this section we perform a double-dimensional reduction of the Nambu-Poisson M5-brane superalgebra. We then attempt to provide an interpretation of the corresponding charges in terms of spacetime intersections involving the D4-brane. Importantly we will see that the algebra we derive consists of only the `Poisson-bracket' terms of the full non-commutative D4-brane superalgebra. 
\subsection{Dimensional Reduction Conventions}
In order to perform the double-dimensional reduction we choose the compactification direction to be $X^{\dot{3}}$. Following the conventions of \cite{Ho:2008ve} we define the gauge potential as
\begin{equation}
{\hat{a}}_\mu = b_{\mu \dot{3}}, \quad \quad {\hat{a}}_{\dot{\alpha}} = b_{\dot{\alpha} \dot{3}}. 
\end{equation}
As a result the covariant derivatives become
\begin{equation}
D_\mu X^{\dot{\alpha}} = - \e^{\dot{\alpha} \dot{\beta}} {\hat{F}}_{\mu \dot{\beta}}, \quad D_\mu X^{\dot{3}} = - {\tilde{a}}_\mu, \quad D_\mu X^i = {\hat{D}}_\mu X^i,
\end{equation}
where we define
\begin{align}
{\hat{F}}_{ab} &= \partial_a {\hat{a}}_b - \partial_b {\hat{a}}_a +g \{{\hat{a}}_a , {\hat{a}}_b \}, \nn \\
{\tilde{a}}_\mu &= \e^{\dot{\alpha} \dot{\beta}} \partial_{\dot{\alpha}} b_{\mu \dot{\beta}}, \nn \\
{\hat{D}}_\mu \Phi &= \partial_\mu \Phi + g \{{\hat{a}}_\mu , \Phi \}.
\end{align}
The Poisson bracket $\{ *, * \}$ is defined as the reduction of the Nambu-Poisson bracket
\begin{equation}
\{f,g \} = \{y^{\dot{3}}, f, g \}.
\end{equation}
The relevant Nambu-Poisson Brackets are
\begin{align}
\{X^{\dot{1}}, X^{\dot{2}}, X^{\dot{3}} \} &= \frac{1}{g^2} {\hat{F}}_{\dot{1} \dot{2}} + \frac{1}{g^3} \label{please} \\
\{ X^{\dot{3}}, X^{\dot{\alpha}} , X^i \} &= \frac{1}{g^2} \e^{\dot{\alpha} \dot{\beta}} {\hat{D}}_{\dot{\beta}} X^i  \\
\{X^{\dot{3}}, X^i , X^j \} &= \frac{1}{g} \{ X^i , X^j \}
\end{align}
To perform the dimensional reduction it will prove easiest if we start with the M2-brane central charges and re-write them in terms of the D4-brane variables directly\footnote{This is equivalent to performing the reduction on the M5-brane superalgebra directly.}. 

\subsection{D4-brane Superalgebra}
We begin by using the conventions of the previous section to re-write the central charges of the M2-brane theory. One finds
\begin{align}
-D_\alpha X^I D_\beta X^J \e^{\alpha \beta} \g^{IJ} \g^0 \rightarrow &+ \tD_\alpha X^i \tD_\beta X^j \g^{\alpha \beta} \g^{ij} + 2 \tF_{\betad \alpha} \tD_\beta X^i \g^{\betad \alphad \beta} \g^i \g^{\dot{3}} \nn \\
&+ \tF_{\alpha \gammad} \tF_{\beta \deltad} \g^{\alpha \gammad \beta \deltad} + 2 \tF_{0 \beta} \tF_{\beta \gammad} \g^{0 \gammad} \nn \\ 
&+ 2 \tD_\beta X^i \tF_{0 \beta } \g^{\dot{3} i } \g^0. \nn \\
\nn \\
D_0 X^K \{ X^K, X^I, X^J \} \g^{IJ} \g^0 \rightarrow &+ \frac{1}{2} \tF_{\alpha \beta} \tF_{\alphad \betad} \g^{\alpha \beta \alphad \betad} + \frac{1}{g} \tF_{\alpha \beta} \g^{\dot{1} \dot{2}} \g^{\alpha \beta} \nn \\
&+ \tF_{\alpha \beta} \tD_{\gammad} X^i \g^{\alpha \beta \gammad} \g^i \g^{\dot{3}} + \frac{g}{2} \tF^{\alpha \beta} \{X^i, X^j \} \g^{\alpha \beta} \g^{ij} \nn \\
&+ \tF_{0 \betad} \tF_{\alphad \betad} \g^{0 \betad \alphad \gammad} + 2\tF_{0 \betad} \tF_{\betad \gammad} \g^{0 \gammad} \nn \\
&+ \frac{2}{g} \tF_{0 \betad} \g^{0 \betad} \g^{\dot{1} \dot{2}} + 2 \tD_{\betad} X^i \tF_{0 \betad } \g^{\dot{3} i}\g^0  \nn \\ 
&+ 2g \tD_0 X^i \{ X^i , X^j \} \g^0 \g^j \g^{\dot{3}} - 2\tD_0 X^i \tD_{\betad} X^i \g^{0 \betad} . \nn \\
\end{align}
\begin{align}
\frac{1}{3} \tD_\beta X^I \{ X^J, X^K, X^L \} \g^{IJKL} \e^{\alpha \beta} \g^{\alpha 0} \rightarrow &+ \tF_{\alphad \betad} \tD_{\gamma} X^i \g^{\alphad \betad \gamma} \g^i \g^{\dot{3}} + \frac{2}{g} \tD_{\beta} X^i \g^i \g^{\dot{1} \dot{2} \dot{3}} \g^\beta \nn \\
&+ g \tF_{\alpha \betad} \{ X^i, X^j \} \g^{\alpha \betad} \g^{ij} + g \tD_\beta X^i \{ X^j , X^k \} \g^{\dot{3}} \g^{\beta} \g^{ijk} \nn \\
&+ 2\tF_{\alpha \betad} \tD_{\gammad} X^i \g^{\alpha \betad \gammad} \g^i \g^{\dot{3}} + 2 \tD_\alpha X^i \tD_{\betad} X^j \g^{\alpha \betad}. \nn \\
\nn \\
\frac{1}{4} \{X^M, X^I, X^J \} \{ X^M, X^K, X^L \} \g^{IJKL} \rightarrow &+ \frac{g}{2}\tF_{\alphad \betad} \{ X^i , X^j \} \g^{\alphad \betad} \g^{ij} +  \{ X^i , X^j \} \g^{\dot{1} \dot{2}} \g^{ij} \nn \\
&+ g \tD_{\alphad} X^i \{ X^j , X^k \} \g^{\dot{3}} \g^{\betad} \g^{ijk} + \tD_{\alphad} X^i \tD_{\betad} X^j \g^{\alphad \betad} \g^{ij} \nn \\
&+ \frac{g^2}{4} \{ X^i , X^j \} \{ X^k , X^l \} \g^{ijkl}.
\end{align}
Looking at these terms we notice that many of them share the same structure despite originating from different M2-brane charges. Combining these terms it is possible to write the D4-brane superalgbera as
\begin{align}
\{Q , Q \}_{\textrm{central}} = &+ \tD_a X^i \tD_b X^j \g^{ab} \g^{ij}  + \tD_a X^i \tF_{bc}  \g^{abc} \g^i \g^{\dot{3}} - 2 \tD_a X^i \tF_{0a} \g^0 \g^i \g^{\dot{3}}\nn \\
&+ \frac{1}{4}\tF_{ab}\tF_{cd} \g^{abcd} + \frac{g}{2} \tF_{ab} \{ X^i, X^j \} \g^{ab} \g^{ij} + g \tD_a X^i \{ X^j, X^k \}  \g^{a} \g^{ijk} \g^{\dot{3}} \nn \\
&+\frac{g^2}{4} \{ X^i , X^j \} \{ X^k , X^l \} \g^{ijkl} + \frac{1}{g} \tF_{\alpha \beta} \g^{\dot{1} \dot{2}} \g^{\alpha \beta} +\frac{2}{g} \tD_{\beta} X^i \g^i \g^{\dot{1} \dot{2} \dot{3}} \g^\beta \nn \\
&+ \{ X^i , X^j \} \g^{\dot{1} \dot{2}} \g^{ij}+ 2g \tD_0 X^i \{ X^i , X^j \} \g^0 \g^j \g^{\dot{3}} - 2\tD_0 X^i \tD_{\betad} X^i \g^{0 \betad}  \label{eight}
\end{align}
where the spatial coordinates of the brane are labeled by $a = (\alpha, \alphad)$ .
\subsection{Interpretation of charges}
In \cite{Ho:2008ve} (see also \cite{Ho:2008ei}) it was argued that the Nambu-Poisson BLG theory is a theory of an M5-brane in a strong C-field background. This was motivated by the fact that, upon double dimensional reduction, the theory shares a structural similarity to a non-commutative D4-brane theory. 
It is important however to emphasise a few points relating to this interpretation. The double dimensional reduction of the Nambu-Poisson M5-brane only captures the Poisson bracket structure of the non-commutative D4-brane action, but misses all the higher order terms in the $\theta$-expansion of the Moyal bracket. Recall that the non-commutative D4-brane theory is described using the Moyal $*$ product (see for example \cite{Seiberg:1999vs}). It is possible to expand the Moyal bracket in $\theta$. One finds at lowest order in $\theta$ the Poisson bracket structure,
\begin{equation}
[f, g]_{\textrm{Moyal}} = f * g - g * f = \theta^{ij} \partial_i f \partial_j g + \mathcal{O} (\theta^3).
\end{equation}
where
\begin{equation}
f(x) * g(x) = e^{\frac{i}{2} \theta^{ij} \frac{\partial}{\partial \xi^i} \frac{\partial }{\partial \chi^j}} f(x + \xi ) g(x + \chi )|_{\xi = \chi =0}.
\end{equation}
The fact that the double-dimensional reduction of the Nambu-Poisson theory only captures the lowest order term in the Moyal-bracket means that one is not able to identify this theory as the full non-commutative D4-brane theory. Rather it may be viewed as a Poisson-bracket truncation of the full non-commutative field theory.\footnote{In a recent paper \cite{Chen:2010ny} the question was asked whether it is possible to deform the Nambu-Poisson M5-brane theory such that its double dimensional reduction gives rise to the full non-commutative Yang-Mills theory to all orders in the non-commutativity parameter $\theta$. They found that there is no way to deform the Nambu-Poisson gauge symmetry such that the full non-commutative gauge symmetry was recovered upon double-dimensional reduction.} Given these considerations, plus the fact that higher order terms in $g$ are considered in this paper, it is important to emphasise that the results derived for the central charges in \eqref{eight} are only valid for the Poisson-bracket D4-brane theory. 

We know that the D4-brane results from the double dimensional reduction of the M5-brane and therefore one would expect to see remnants of the M5-brane solitons in the worldvolume superalgebra of the D4-brane. Furthermore, the M5-brane couplings to the background 3-form should appear as couplings to the NS-NS 2-form B-field from the D4-brane perspective. As outlined in \cite{Howe:1997ue}, the self dual string soliton from the M5-brane worldvolume theory can be reduced to a 0-brane or 1-brane solution on the D4-brane worldvolume depending on whether the string is wrapped or un-wrapped around the compact dimension. In the abelian case the BPS equation corresponding to the 0-brane BPS soliton on the D4-brane takes the form $F_{0a} \propto \partial_a X$ where $X$ is the only active scalar field parameterising a single direction transverse to the brane. From this one can read off the energy bound when the BPS equation is satisfied which is proportional to $ \partial_a X F_{0a}$. We see the covariant generalisation of this term appearing in the superlagebra derived above in the form $D_a X^i F_{0a}$. This bound can naturally be interpreted as the endpoint of a fundamental string on the D4-brane. Alternatively, the 1-brane soliton resulting from the un-wrapped self-dual string has BPS equation $F_{bc} \propto \e_{abc} \partial^a X$ with corresponding energy bound proportional to $ \e_{abc} \partial_a  X F_{bc}$. This term can naturally be compared with $D_a X^i F_{bc}  \g^{abc}$ appearing in \eqref{eight}. One can think of this as the bound associated with a D2-brane intersecting a D4-brane which appears as a 1-brane soliton from the D4-brane perspective. It is worth noting that the D4-brane worldvolume superalgebra allows scalar central charges in the $\bf{1} + \bf{5}$ representation of the $\textrm{Spin}(5)$ R-symmetry group which can be interpreted as the transverse rotation group. As noted above, the scalars in the $\bf{5}$ representation can be interpreted as the endpoints of fundamental strings on the D4-brane. The $\textrm{Spin} (5)$ singlet scalar is a magnetic charge which from the spacetime perspective corresponds to a D0-brane intersecting a D4-brane. Because it is a singlet its charge cannot depend on any transverse scalars. Following arguments similar to those presented in \cite{Gauntlett:1997ss}, one can show that the total energy $E$ of this configuration, relative to the worldvolume vacuum, is subject to the bound $E \ge |Z|$ where Z is the topological charge
\begin{equation}
Z = \frac{1}{4} \int_{\textrm{D4}} \textrm{tr} F \tilde{F}.
\end{equation}
This corresponds to the central charge appearing in the superalgebra
\begin{align}
\frac{1}{4}\tF_{ab}\tF_{cd} \g^{abcd} &\propto \frac{1}{4}\tF_{ab}\tF_{cd} \e^{abcd} \nn \\
&= \frac{1}{4} \tF {\tilde{\tF}}.
\end{align}
The central charge in \eqref{eight} involving two covariant derivatives, $\tD_a X^i \tD_b X^j$, derives from the 3-brane vortex on the M5-brane worldvolume  corresponding to an M5-M5 intersection. The double dimensional reduction of the 3-brane gives rise to a 2-brane on the D4-brane worldvolume. This may be interpreted as the intersection of two D4-branes over a 2-brane. 
It is interesting to notice that the superalgebra contains a central charge term proportional to $\tD_a X^i \{ X^j, X^k \}$. This charge would appear to represent the Poisson-bracket analogue of the energy bound associated with the Nahm equation which describes the intersection of multiple D2-branes with a D4-brane. However it is well known that the Nahm equation is the BPS equation associated with the non-abelian worldvolume theory of D-branes. It therefore seems queer that any information regarding the energy bound of such a configuration should appear in the worldvolume theory of a single D4-brane. It is not immediately clear how to interpret this charge (or the M-theory uplift appearing in the Nambu-Poisson M5-brane algebra). Furthermore, the central charge proportional to $\{ X^i, X^j \} \{ X^k, X^l\}$ seems analogous to the D4-brane charge 
appearing in the matrix model for M-theory found in \cite{Banks:1996nn} (with matrix commutator replaced by Poisson-bracket). 

In this section we have seen that the D4-brane algebra \eqref{eight} contains a wealth of information regarding spacetime configurations of the D4-brane, including information about intersections involving multiple D0-branes and multiple D2-branes  from the perspective of the D4-brane. We have also seen Poisson-bracket analogues of charges that one would expect to find in the non-abelian worldvolume superalgebra of the D0-brane and D2-brane.

\subsection{Comments on $1/g$ terms in D4-brane theory}
It is worth making a few comments regarding the interpretation of the $1/g$ terms appearing in the superalgebra \eqref{eight}. In \cite{Ho:2008ve}, the coupling of the dimensionally reduced theory was identified with the non-commutativity parameter of the D4-brane theory, $g = \theta$. Based on this identification one can think of the $1/g$ terms as representing $1/ \theta$ modifications of certain central charge terms. More precisely, any term in the algebra \eqref{eight} deriving from the dimensional reduction rule \eqref{please} will contain a term of the form $(F_{\dot{1} \dot{2}} + 1/ \theta)$ (once we have identified $g = \theta$). 
Ultimately we can think of this shift in the $(\dot{1} \dot{2})$ components of the field strength as being related to the presence of a background B-field. After all, the non-commutative theory is meant to be an effective description of the D4-brane in B-field background. In order to move from the non-commutative description (without background flux) to the `commutative' description involving the background B-field it is necessary to use the Seiberg-Witten map \cite{Seiberg:1999vs}. 
In principle this can be done for the central charge terms involving $1/\theta$ appearing in \eqref{eight}. It is known for example that from the perspective of the 1-brane soliton on the D4-brane worldvolume, the presence of a background $B$-field  causes a tilting of the 1-brane as it extends from the D4-brane. This can be shown to result from a shift in the field strength by the B-field. This shift leads to a change in the form of the BPS equation and therefore the corresponding BPS energy bound (central charge). 

We end this chapter by noting that there is a slightly puzzling feature of the D4-brane algebra appearing in \eqref{eight}. This is related to the fact that, for the case studied in this chapter, it would seem the correct non-commutative description of the D4-brane theory is the matrix model description. This might be inferred from the fact that the Nambu-Poisson M5-brane theory is constructed from the M2-brane action (The analogous situation in type IIA string theory would be constructing the D4-brane from the D2-brane action and we know the matrix theory description naturally arises in this setting). In this case, following \cite{Seiberg:2000zk}, one expects to find the combination $(F - 1/ \theta)$ appearing in the action, not $(F + 1/ \theta)$. This discrepancy may be related to the fact that the reduction of the Nambu-Poisson algebra only gives the Poisson terms of the non-commutative D4-brane theory, whereas the combination $(F- 1 / \theta)$ appears in the full non-commutative theory. We hope to return to this issue in a future publication. We would like to thank the referee of \cite{Low:2009de} for bringing this point to our attention. 

\chapter{Higher-Order Bagger-Lambert theory}
The Bagger-Lambert Lagrangian and supersymmetry transformations presented in Chapter 2 can be thought of as representing the leading order terms in an $l_p$ expansion of a non-linear M2-brane theory. This is analogous to the fact that super Yang-Mills theory represents the leading order terms of the non-abelian Born-Infeld action, which is believed to describe the dynamics of coincident D-branes.\footnote{Note that the symmetrised trace prescription of the non-abelian Born-Infeld action \cite{Tseytlin:1997csa}  breaks down at sixth order and higher in the worldvolume field strength \cite{Hashimoto:1997gm}.}  Ultimately one would like to determine the full theory, of which the leading order terms are those of the Bagger-Lambert Lagrangian. Toward this end it is constructive to consider the next order in $l_p$ corrections to the theory. At the level of the Lagrangian this analysis has been performed in the literature \cite{Alishahiha:2008rs, Ezhuthachan:2009sr} using two complimentary methods.\footnote{For other discussions on non-linear corrections to Bagger-Lambert theory see \cite{Iengo:2008cq, Kluson:2008nw, Sasaki:2009ij, Garousi:2008xn}.} 

The first method involves a duality transformation due to de-Witt, Nicholai and Samtleben (dNS). This duality is based on the fact that in (2+1) dimensions, a gauge field is dual to a scalar, and it is therefore possible to replace the gauge-field with a scalar field such that the theory possesses a manifest SO(8), rather than SO(7) symmetry. In \cite{Ezhuthachan:2008ch}, it was shown that applying this procedure to the D2-brane Lagrangian, it is possible to re-write the theory as a Lorentzian Bagger-Lambert theory. This technique was then applied to the ${\alpha'}^2$ terms of the D2-brane Lagrangian in order to determine the $l_p^3$ corrections to the Lorentzian Bagger-Lambert theory\footnote{Lorentzian Bagger-Lambert theories are considered in \cite{Gomis:2008uv, Ho:2008ei, Benvenuti:2008bt, Bandres:2008kj, Gomis:2008be}. See also \cite{Honma:2008un, Honma:2008ef, Honma:2008jd, Antonyan:2008jf}.}. Remarkably, all higher order Lagrangian terms were expressible in terms of basic building blocks involving covariant derivatives, $D_\mu X^I$ and three-brackets $[X^I, X^J, X^K]$. This led the authors of \cite{Alishahiha:2008rs} to conjecture that the higher derivative Lagrangian they had derived would also apply to the ${\mathcal{A}}_4$ Bagger-Lambert Theory. 

This conjecture was confirmed in \cite{Ezhuthachan:2009sr} where the novel Higgs mechanism  was used to determine the ${\mathcal{A}}_4$ theory Lagrangian at order $l_p^3$. This involved using dimensional analysis to write down all possible $l_p^3$ corrections to the Bagger-Lambert Lagrangian with arbitrary coefficients. The coefficients were fixed by applying the novel Higgs mechanism to the higher order terms and matching them to the ${\alpha'}^2$ terms of the D2-brane theory. It was shown that the structure of the higher order terms in both the ${\mathcal{A}}_4$ and Lorentzian theories take the same form. 

Given that the $l_p^3$ corrections to the Bagger-Lambert theory have been calculated, one might ask whether these terms are maximally supersymmetric, and if so, to determine the structure of the higher order supersymmetry transformations. In this chapter we begin the task of calculating the $l_p^3$ corrections to the supersymmetry transformations of the Bagger-Lambert theory.
The hope is that closure of the higher order supersymmetry transformations would uniquely determine the higher order corrections to the Bagger-Lambert equations of motion which can then be `integrated' to determine the higher order Lagrangian, which by definition, would be supersymmetric. One could in principle write down all possible $l_p^3$ corrections to the supersymmetry transformations and then try and fix the coefficients by demanding the closure of the supersymmetry algebra. However the plethora of possible terms at order $l_p^3$ would make the closure of the algebra a mammoth task. To try and simplify the problem we will use the non-abelian D2-brane theory as a guide. We know that dNS duality transformation allows us to map the non-abelian D2-brane Lagrangian into the Lorentzian Bagger-Lambert Lagrangian. Furtherore we know that the structure of this Lagrangian is the same as the structure of the ${\mathcal{A}}_4$ theory Lagrangian. It is natural to ask whether this methodology can tell us anything about how the higher order D2-brane supersymmetry transformations are related to the $l_p^3$ corrections to the Bagger-Lambert supersymmetry transformations. 

In the first part of this chapter we will review the dNS duality transformation \cite{Nicolai:2003bp, deWit:2003ja, deWit:2004yr} by considering how the Lorentzian Bagger-Lambert Lagrangian can be derived from the D2-brane theory. We will then attempt to apply the duality transformation at the level of supersymmetry transformations. To simplify the task we will begin by only considering the `abelian' Bagger-Lambert theory. We will see that the duality transformation works for the fermion variation but fails to work for the scalar variation. Therefore in order to calculate the scalar variation we have to use a different approach. This involves using dimensional analysis to write the most general scalar variation with arbitrary coefficients. Invariance of the higher order Lagrangian is then used to fix the values of the coefficients. 
In the final part of this chapter we begin the task of calculating the full `non-abelian' Bagger-Lambert supersymmetry transformations at $\mathcal{O} (l_p^3)$. We are able to uniquely determine the higher order fermion variation but unable to uniquely determine the scalar variation. As a result, this chapter represents work in progress. The content of this chapter is based on the original work presented in \cite{Low:2010ie}.
\section{Non-abelian duality in 2+1 dimensions}
We begin by reviewing a prescription for dualising non-abelian gauge fields in (2+1) dimensions due to de Wit, Nicolai and Samtleben (dNS)\cite{Nicolai:2003bp, deWit:2003ja, deWit:2004yr}. We will follow the presentation of \cite{Ezhuthachan:2008ch}. According to the dNS prescription the Yang-Mills gauge field $A_\mu$ gets replaced by two non-dynamical gauge fields $A_\mu$ and $B_\mu$ with a $B \wedge F$ type kinetic term, plus an extra scalar which ends up carrying the dynamical degrees of freedom of the original Yang-Mills gauge field. The duality transformation is enforced by making the replacement
\begin{equation}
\Tr \left(-\frac{1}{4 g^2_{YM}} F_{\mu \nu} F^{\mu \nu} \right) \rightarrow \Tr \left(\frac{1}{2} \epsilon^{\mu \nu \lambda}B_\mu F_{\nu \lambda} - \frac{1}{2}(D_\mu \phi-g_{YM}B_\mu)^2 \right). \label{kunt}
\end{equation}
We wish to consider the effect of this transformation on the multiple D2-brane theory. The low energy Lagrangian for this theory is obtained by reducing ten-dimensional $U(N)$ super Yang-Mills theory to (2+1) dimensions. In this case, making the replacement \eqref{kunt} in the D2-brane Lagrangian results in the dNS transformed Lagrangian\footnote{This action exhibits an abelian gauge symmetry  allowing one to pick a gauge in which either $D^\mu B_\mu = 0$ or $\phi = 0$. In the latter case $B_\mu$ becomes an auxiliary field that can be integrated out thereby showing the equivalence of the LHS and RHS of \eqref{kunt}. For explicit details see \cite{Ezhuthachan:2008ch}. } 
\begin{align}
\mathcal{L}= &\Tr ( \frac{1}{2} \epsilon^{\mu \nu \lambda}B_\mu F_{\nu \lambda} - \frac{1}{2}(D_\mu \phi-g_{YM}B_\mu )^2- \frac{1}{2} D_\mu X^i D^\mu X^i \nn \\
&- \frac{g^2_{YM}}{4}[X^i, X^j][X^i, X^j]  + \frac{i}{2} \bp \g^\mu D_\mu \psi + \frac{i}{2} g_{YM} \bp \g_i [X^i, \psi ] ).
\end{align}
The gauge invariant kinetic terms for the eight scalars can be shown to possess an SO(8) invariance by renaming $\phi \rightarrow X^8$ and writing
\begin{align}
{\tilde{D}}^\mu X^i &= D_\mu X^i = \partial_\mu X^i - [A_\mu, X^i], \quad i=1,2, \ldots, 7 \\
 {\tilde{D}}^\mu X^8 &= D_\mu X^8 - g_{YM}B_\mu = \partial_\mu X^8 - [A_\mu, X^8] - g_{YM}B_\mu.
\end{align}
Defining the constant 8-vector
\begin{equation}
g^I_{YM} = (0, \ldots, 0, g_{YM}), \quad I=1,2, \ldots, 8,
\end{equation}
allows one to define the covariant derivative
\begin{equation}
{\tilde{D}}^\mu X^I = D_\mu X^I - g^I_{YM} B_\mu.
\end{equation}
It is then possible write the super Yang-Mills action in a form that is $SO(8)$ invariant under transformations that rotate both the fields $X^I$ and the coupling constant vector $g^I_{YM}$:
\begin{align}
\mathcal{L} = \Tr \left( \frac{1}{2} \e^{\mu \nu \lambda} B_\mu F_{\nu \lambda} - \frac{1}{2} ({\tilde{D}}_\mu X^I)^2 + \frac{i}{2} \bp \g^\mu D_\mu \psi + \frac{i}{2} g^I_{YM} \bp \g_{IJ} [X^J , \psi ] -\frac{1}{12} ({X^{IJK}})^2 \right)
\end{align}
where the three-bracket $X^{IJK}$ is defined as
\begin{equation}
X^{IJK} = g^I_{YM} [X^J, X^K] + g^J_{YM} [X^K, X^I] + g_{YM}^K [X^I , X^J].
\end{equation}
This theory is only formally SO(8) invariant, as the transformations must act on the coupling constants as well as the fields. However, one can replace the vector of coupling constants $g^I_{YM}$ by a new (gauge singlet) scalar $X^I_+$ provided that the new scalar field has an equation of motion that renders it constant.  Constancy of $X^I_+$ is imposed by adding a new term to the Lagrangian involving a set of abelian gauge fields and scalars $C_\mu^I$ and $X_-^I$: 
\begin{equation}
{\mathcal{L}}_C = (C^\mu_I - \partial^\mu X^I_-) \partial_\mu X^I_+.
\end{equation}
As explained in \cite{Ezhuthachan:2008ch, Gomis:2008be}, this term has the effect of constraining the vector $X_+^I$ to be an arbitrary constant which can be identified with $g^I_{YM}$. In this way one recovers the gauge-fixed Lorentzian models of \cite{Bandres:2008kj, Gomis:2008be}. One might wonder whether this non-abelian duality works when higher order (in $\alpha'$) corrections are included in the D2-brane theory. In particular, does the 3-algebra structure survive $\alpha'$ corrections? In \cite{Alishahiha:2008rs, Ezhuthachan:2009sr} it was shown that at $\mathcal{O} ({\alpha'}^2)$ the duality does work and all terms in the resulting $l_p^3$ corrected M2-brane theory are expressible in terms of ${\tilde{D}}_\mu X^I$ and $X^{IJK}$ building blocks. Another question one might ask is whether this duality works at the level of supersymmetry transformations and if so, would it be possible to derive the $\mathcal{O} (l_p^3)$ corrections to the Bagger-Lambert supersymmetry transformations? The first step towards answering this question is to consider \textit{how} abelian duality in (2+1) dimensions can be implemented at the level of supersymmetry transformations. To this we now turn.

\section{Abelian Duality and Supersymmetry}
Our ultimate objective is to determine higher order supersymmetry transformations in Bagger-Lambert theory by using the dNS procedure outlined in the previous section. As a warm-up exercise we will consider dualising abelian gauge-fields to scalars in 2+1 dimensions and see how this works at the level of supersymmetry transformations for a single D2-brane. Let us begin by considering abelian duality at the level of the Lagrangian.
\subsection{Abelian Duality}
Consider the following 2+1 dimensional action involving a Lagrange multiplier field $X$
\begin{equation}
S = - \int d^3 \sigma (\frac{1}{4} F_{\mu \nu} F^{\mu \nu} + \frac{1}{2} \e_{\mu \nu \lambda} F^{\mu \nu} \partial^\lambda X )\label{lagrange}
\end{equation}
We see that the gauge field equation of motion takes the form
\begin{equation}
F_{\mu \nu} = - \e_{\mu \nu \lambda} \partial^\lambda X \label{gf}
\end{equation}
whereas the $X$ equation of motion takes the form of the Bianchi identity
\begin{equation}
\e_{\mu \nu \lambda} \partial^\mu F^{\nu \lambda} = 0.
\end{equation}
If we substitute the gauge field equation of motion into \eqref{lagrange} then we find a kinetic term for $X$
\begin{equation}
\int d^3 \sigma (\frac{1}{4} F_{\mu \nu} F^{\mu \nu} + \frac{1}{2} \e_{\mu \nu \lambda} F^{\mu \nu} \partial^\lambda X ) \rightarrow  \int d^3 \sigma \frac{1}{2}\partial_\mu X \partial^\mu X.
\end{equation} 
Alternatively, use of the Bianchi identity in \eqref{lagrange} results in 
\begin{equation}
\int d^3 \sigma (\frac{1}{4} F_{\mu \nu} F^{\mu \nu} + \frac{1}{2} \e_{\mu \nu \lambda} F^{\mu \nu} \partial^\lambda X ) \rightarrow  \int d^3 \frac{1}{4} F_{\mu \nu} F^{\mu \nu}. 
\end{equation} 
with $F = dA$. So how does this relate to the D2-brane theory? The leading order Lagrangian for a single D2-brane can be obtained by dimensional reduction of super Yang-Mills theory in ten dimensions. The bosonic D2-brane action can be expressed as
\begin{equation}
S = \int d^3\sigma (-\frac{1}{4}F_{\mu \nu} F^{\mu \nu} - \frac{1}{2} \partial_\mu X^i \partial^\mu X^i). \label{vjy}
\end{equation}
Abelian duality is implemented by making the replacement 
\begin{equation}
-\frac{1}{4} F_{\mu \nu} F^{\mu \nu} \rightarrow - (\frac{1}{4} F_{\mu \nu} F^{\mu \nu} + \frac{1}{2} \e_{\mu \nu \lambda} \partial^\mu X F^{\nu \lambda})
\end{equation}
in the action \eqref{vjy}. Use of the gauge field equation of motion \eqref{gf} then results in
\begin{align}
S &= \int d^3\sigma (-\frac{1}{4}F_{\mu \nu} F^{\mu \nu} - \frac{1}{2} \e_{\mu \nu \lambda} F^{\mu \nu} \partial^\lambda X - \frac{1}{2} \partial_\mu X^i \partial^\mu X^i) \nn \\
&= \int d^3 \sigma ( -\frac{1}{2} \partial_\mu X \partial^\mu X - \frac{1}{2} \partial_\mu X^i \partial^\mu X^i) \nn \\
&= \int d^3 \sigma (- \frac{1}{2} \partial_\mu X^I \partial^\mu X^I)
\end{align}
where in obtaining the last line we identified $X = X^8$ as the eighth scalar field. We see that the scalar kinetic term now has the desired $SO(8)$ invariant form. Note that it is possible to implement abelian duality in (2+1) dimensions at the level of the full DBI action \cite{Bergshoeff:1996tu}. In this way one is able to derive a non-linear Lagrangian for a membrane in the static gauge with the expected SO(8) symmetry. Now that we have seen how abelian duality works at the level of the Lagrangian let us consider applying this to the supersymmetry transformations of a single D2-brane.
\subsection{supersymmetry transformations}
The D2-brane supersymmetry transformation can be obtained by dimensionally reducing the supersymmetry transformations of ten dimensional super Yang-Mills. The spinors appearing in the 10-dimensional theory are Majorana-Weyl and satisfy
\begin{equation}
\g^{(10)} \chi = \chi
\end{equation}
where $\g^{(10)}$ is the ten dimensional chirality matrix. Since we are interested in uplifting the D2-brane theory to M-theory it is desirable to look for an embedding of $SO(1,9)$ into $SO(1,10)$ in which $\g^{(10)}$ becomes the eleventh gamma matrix. We denote the gamma matrices of $SO(1,10)$ as $\g^M (M = 0, \ldots , 9, 10 )$. In eleven dimensions the spinors will be Majorana. However we know that the presence of the M2-brane breaks the Lorentz symmetry as $SO(1,10) \rightarrow SO(1,2) \times SO(8)$ and therefore we can have a Weyl spinor of $SO(8)$. Let us denote the chirality matrix of $SO(8)$ by $\g$ where
\begin{equation}
\g = \g^{3 \ldots 9(10)}
\end{equation}
The M2-brane breaks half the supersymmetry of the vacuum. We choose conventions consistent with chapter 3 in which 
\begin{equation}
\g \e = \e, \qquad \g \psi = - \psi.
\end{equation}
Under dimensional reduction the (9+1) dimensional gauge field will split into a (2+1)-dimensional gauge field $A_\mu$ and a scalar field $X^i$ transforming under $SO(7)$. As usual with dimensional reduction, the fields are independent of the circle directions such that one can set $\partial_i = 0$. In what follows, for reasons that will become clear shortly, we will label $\mu = 0,1,2$ and $i = 1, \ldots 7$ with the ten-dimensional chirality matrix relabeled as $\g^{(10)} = \g^8$. Dimensional reduction of the ten-dimensional super Yang-Mills transformations
\begin{align}
\delta A^M &= i\be \g^M \psi \nn \\
\delta \psi &= \frac{1}{2} \g^{MN} F_{MN} \e
\end{align}
results in the following D2-brane transformations 
\begin{align}
\delta X^i &= i \be \g^i \psi  \\
\delta A_\mu &= i \be \g_\mu \g^8 \psi  \label{one} \\
\delta \psi &= \frac{1}{2} \g^{\mu \nu} F_{\mu \nu} \g^8\e + \g^\mu \g^i \partial_\mu X^i \e \label{too}
\end{align}
We now consider the effect of applying abelian duality at the level of supersymmetry transformations. This can be achieved by using \eqref{gf} to write
\begin{equation}
\partial_\mu X^8 = \frac{1}{2} \e_{\mu \nu \lambda} F^{\nu \lambda} \label{pohg}
\end{equation}
where we have relabeled the scalar appearing in \eqref{gf} as $X^8$ (this will provide the `eighth' scalar which will combine with the other seven to give an SO(8) invariant supersymmetry transformation). Performing the duality transformation on the fermion variation involves substituting \eqref{pohg} into \eqref{too} 
\begin{align}
\delta \psi &= \frac{1}{2} \e_{\mu \nu \lambda} \g^{\mu \nu} \partial^\lambda X^8 \g^8 \e + \g^\mu \g^i \partial_\mu X^i \e \nn \\
&= \g^\mu \g^8 \partial_\mu X^8 \e + \g^\mu \g^i \partial_\mu X^i \e \nn \\
&= \g^\mu \g^I \partial_\mu X^I \e.
\end{align}
We see that this now takes the desired SO(8) form.  In order to determine the SO(8) transformation of the scalar field $\delta X^I$ we need to consider \eqref{one} re-written as
\begin{equation}
\delta F_{\mu \nu} = - 2i \be \g_{[ \mu} \g^8 \partial_{\nu ]} \psi.
\end{equation}
Substituting \eqref{pohg} into the left-hand side of this transformation allows us to write
\begin{align}
 \partial^\lambda \delta X^8 &= - i \be \e^{\mu \nu \lambda} \g_{[\mu} \g^8 \partial_{\nu ]} \psi \nn \\
&= i \be \g^{\nu \lambda } \g^8 \partial_\nu \psi \nn \\
&= i \be (\eta^{\nu \lambda} - \g^\lambda \g^{\nu}) \g^8 \partial_\nu \psi \nn \\
&= i \be \g^8 \partial^\lambda \psi \label{bt}
\end{align}
where we have made use of the lowest order fermion equation of motion $\g^\mu \partial_\mu \psi = 0$. This relation implies that $\delta X^8 = i \be \g^8 \psi$ which can be combined with $\delta X^i = i \be \g^i \psi$ to give
\begin{equation}
\delta X^I = \be \g^I \psi.
\end{equation}
In summary we see that at lowest order it is possible to re-write the D2-brane supersymmetry transformations in an SO(8) invariant form. For the fermion variation this simply involved substituting \eqref{pohg} into the D2-brane expression. For the scalar field variation it was necessary to `dualise' the gauge field variation $\delta F_{\mu \nu}$ to form the eighth scalar. In the next section we extend our analysis to higher order abelian supersymmetry transformations. 

\section{Higher Order Abelian Supersymmetry }
In this section we will determine the $l_p^3$ corrections to the abelian M2-brane supersymmetry transformations (excluding bi-linear and tri-linear fermion terms). We begin by using dimensional arguments to determine the structure of the supersymmetry transformations. We will then apply the duality transformation outlined in the previous section to the $\mathcal{O}({\alpha'}^2)$ D2-brane supersymmetry transformations. We will see that this procedure uniquely determines the fermion variation but fails to work for the scalar variation. This will motivate us to try a different approach.   
\subsection{Dimensional Analysis}
Dimensional analysis tells us that the mass dimensions of the fields appearing in the Bagger-Lambert theory are
$ [X] = \frac{1}{2} , [\psi ] = [ A_\mu ] = 1$. The supersymmetry parameter $\e$ carries mass dimension $[\e] = -\frac{1}{2}$. We expect the first non-trivial corrections to the supersymmetry transformations to appear at $\mathcal{O} (l_p^3)$. Therefore we see that the $\mathcal{O} (l_p^3)$ terms in $\delta \psi$ must be mass dimension $4$. In a similar manner the correction terms in $\delta X$ must be mass dimension $3 \frac{1}{2}$. For the sake of simplicity we will neglect bi-linear fermion terms in the scalar variation and tri-linear fermion terms in the fermion variation. In terms of the basic building blocks of scalar fields and derivatives, the only possible types of term appearing in the fermion variation at $\mathcal{O} (l_p^3)$ are those involving three derivatives and three scalar fields. If we assume that derivatives must always act on scalars (with at most one derivative) then a little thought reveals that the higher-order fermion variation takes the form
\begin{align}
\delta \psi = &+ a_1 l_p^3\g_\mu \g^I \partial_\nu X^J \partial^\nu X^J \partial^\mu X^I \e + a_2l_p^3 \g_\mu \g^I \partial^\mu X^J \partial_\nu X^J \partial^\nu X^I \e \nn \\
&+ a_3l_p^3\e^{\mu \nu \rho} \g^{IJK} \partial_\mu X^I \partial_\nu X^J \partial_\rho X^K \e. \label{fermi}
\end{align}
The motivation for assuming that scalars are always acted on by derivatives is based on the form of the ${\alpha'}^2$ D2-brane supersymmetry transformations (derived in the next section) which have no free scalar terms. Let us now consider the scalar transformation. Based on dimensional analysis and the reasons already outlined, the only types of term appearing in the scalar variation are those involving two derivatives, two scalar fields and a fermion. Considering all independent index contractions one arrives at the following expression
\begin{align}
\delta X^I =  &+ b_1 l_p^3\be \g^I \psi \partial_\mu X^J \partial^\mu X^J + b_2 l_p^3 \be \g^J \psi \partial_\mu X^I \partial^\mu X^J  \nn \\
&+ b_3 l_p^3\be \g^J \g^{\mu \nu} \psi \partial_\mu X^I \partial_\nu X^J + b_4 l_p^3\be \g^{\mu \nu} \g^{IJK} \psi \partial_\mu X^J \partial_\nu X^K. \label{bose}
\end{align}
Our task in the remainder of this section is to fix the coefficients appearing in \eqref{fermi} and \eqref{bose}. There are a number of ways that this can be achieved. In the next section we will attempt to fix these coefficients through abelian duality. We will see that this only works for the fermion variation. In order to determine the scalar variation we will have to use a different approach. This will involve checking that the higher order abelian Lagrangian derived in \cite{Ezhuthachan:2009sr} is invariant under \eqref{fermi} and \eqref{bose}. Not only will this allow us to determine the scalar variation coefficients but it will also provide a test for the fermion terms derived using the duality approach. 

\subsection{Higher Order Abelian Supersymmetry via Dualisation}
In this section we will attempt to derive the higher order abelian supersymmetry transformations using abelian duality. Our starting point will be the $\mathcal{O} (\alpha'^2)$ supersymmetry transformations of the ten-dimensional super Yang-Mills theory. These were first discovered by Bergshoeff and collaborators in \cite{Bergshoeff:1986jm}\footnote{The form of these transformations was later confirmed in \cite{Bergshoeff:2001dc, Cederwall:2001bt, Cederwall:2001td}.}
\begin{align}
\delta \psi &= \alpha'^2 (\lambda_1 \g^{MN} F_{PQ} F^{PQ} F_{MN} \e + \lambda_2 \g^{MN} F_{MP} F^{PQ} F_{QN} \e + \lambda_3 \g^{MNPQR}F_{MN}F_{PQ}F_{RS} \e). \nn \\
\delta A_M &= \alpha'^2 ( \alpha_1 \be \g_M  F_{NP} F^{NP} \psi + \alpha_2 \be \g_N  F_{MP}F^{PN} \psi + \alpha_3\be \g^{NPQ} F_{MN}F_{PQ} \psi \nn \\
&\quad + \alpha_4\be \g_{MNPQR}  F^{NP}F^{QR} \psi ).\label{ymtran}
\end{align}
Note that in \cite{Bergshoeff:1986jm} the fermion variation also included tri-linear fermion terms and the gauge field variation included bi-linear fermion terms which we have not included for the sake of simplicity. We have purposely left the coefficients unspecified. The hope is that these coefficients will be fixed by the requirement that the (2+1) dimensional transformations collect into $SO(8)$ invariant terms under the duality transformation. Next we wish to reduce these expressions to (2+1) dimensions. We will first focus on the fermion. Performing the dimensional reduction one finds
\begin{align}
\lambda_1 \g^{MN} F_{PQ} F^{PQ} F_{MN} \e \rightarrow &+ \lambda_1 \g^{\mu \nu} F_{\mu \nu} F^{\rho \sigma} F_{\rho \sigma} \e + 2\lambda_1 \g^{\mu \nu} F_{\mu \nu} \partial^\rho X^i \partial_\rho X^i \e \nn \\
&+ 2 \lambda_1 \g^\mu \g^i \partial_{\mu} X^i F_{\rho \sigma} F^{\rho \sigma} \e + 4 \lambda_1 \g^{\mu} \g^i \partial_\mu X^i \partial^\nu X^j \partial_\nu X^j \e.  \nn \\
\end{align}
\begin{align}
\lambda_2 \g^{MN} F_{MP} F^{PQ} F_{QN} \e \rightarrow &+ \lambda_2 \g^{\mu \nu} F_{\mu \rho} F^{\rho \sigma} F_{\sigma \nu} \e - 2\lambda_2 \g^{\mu \nu} F_{\mu \rho} \partial^\rho X^i \partial_\nu X^i \e  \nn \\
&+ 2\lambda_2 \g^\mu \g^i F_{\mu \rho} F^{\rho \sigma} \partial_\sigma X^i \e \nn \\
&- 2 \lambda_2\g^\mu \g^i \partial_\mu X^j \partial^\rho X^j \partial_\rho X^i \e  - \lambda_2 \g^{ij} \partial_\rho X^i F^{\rho \sigma} \partial_\sigma X^j \e. \nn 
\end{align}
\begin{align}
\lambda_3 \g^{MNPQRS} F_{MN} F_{PQ} F_{RS} \e \rightarrow &- 8 \lambda_3 \g^{\mu \nu \rho} \g^{ijk} \partial_\mu X^i \partial_\nu X^j \partial_\rho X^k \e. 
\end{align}
Next we dualise the gauge field using \eqref{pohg}. After a small amount of algebra one finds
\begin{align}
\lambda_1\g^{MN} F_{PQ} F^{PQ} F_{MN} \e \rightarrow &+4\lambda_1 \g^\mu \partial_\mu X^8 \partial^\nu X^8 \partial_\nu X^8\e - 4\lambda_1 \g^\mu \partial_\mu X^8 \partial^\nu X^i \partial_\nu X^i\e  \nn \\
&- 4 \lambda_1\g^\mu \g^i \partial_\mu X^i \partial^\nu X^8 \partial_\nu X^8 \e + 4 \lambda_1\g^{\mu} \g^i \partial_\mu X^i \partial^\nu X^j \partial_\nu X^j \e. \label{dad}
\end{align}
\begin{align}
\lambda_2 \g^{MN} F_{MP} F^{PQ} F_{QN} \e \rightarrow &-2 \lambda_2 \g^\mu \partial_\nu X^8 \partial^\nu X^8 \partial_\mu X^8\e - 2 \lambda_2 \g^\mu \partial^\nu X^8 \partial_\mu X^i \partial^\nu X^i\e \nn \\
&+ 2 \lambda_2 \g^\mu \partial_\mu X^8 \partial^\nu X^i \partial_\nu X^i \e + 2 \lambda_2 \g^\mu \g^i \partial_\nu X^8 \partial^\nu X^8 \partial_\mu X^i \e \nn \\
&-2 \lambda_2 \g^\mu \g^i \partial^\nu X^8 \partial_\mu X^8 \partial^\nu X^i \e - 2 \lambda_2 \g^\mu \g^i \partial_\mu X^j \partial^\rho X^j \partial_\rho X^i \e \nn \\
&- \lambda_2 \g^{ij} \e^{\rho \sigma \lambda} \partial_\rho X^i \partial_\lambda X^8 \partial_\sigma X^j \e.
\end{align}
\begin{align}
\lambda_3 \g^{MNPQRS} F_{MN} F_{PQ} F_{RS} \e &\rightarrow - 8 \lambda_3 \g^{\mu \nu \rho} \g^{ijk} \partial_\mu X^i \partial_\nu X^j \partial_\rho X^k \e. \label{mum}
\end{align}
We would like to re-write these transformed expressions in terms of $SO(8)$ objects. The only possible $SO(8)$ objects involving three derivatives are those contained in \eqref{fermi}. The hope is that the fermion supersymmetry transformation should be expressible as a particular combination of these basic objects. More specifically, by noting that
\begin{align}
\g_\mu \g^I \partial_\nu X^J \partial^\nu X^J \partial^\mu X^I \e &\rightarrow  \g_\mu \g^i \partial_\nu X^j \partial^\nu X^j D^\mu X^i \e + \g_\mu \g^i \partial_\nu X^8 \partial^\nu X^8 \partial^\mu X^i \e \nn \\
& \quad + \g_\mu \g^8 \partial_\nu X^j \partial^\nu X^j \partial^\mu X^8 \e + \g_\mu \g^8 \partial_\nu X^8 \partial^\nu X^8 \partial^\mu X^8 \e \nn \\
\nn \\
\g_\mu \g^I \partial^\mu X^J \partial_\nu X^J \partial^\nu X^I &\rightarrow \g_\mu \g^i \partial^\mu X^j \partial_\nu X^j \partial^\nu X^i \e + \g_\mu \g^i \partial^\mu X^8 \partial_\nu X^8 \partial^\nu X^i \e \nn \\
& \quad + \g_\mu \g^8 \partial^\mu X^j \partial_\nu X^j \partial^\nu X^8 \e + \g_\mu \g^8 \partial^\mu X^8 \partial_\nu X^8 \partial^\nu X^8  \e 
\end{align}
we can write the terms in \eqref{dad}-\eqref{mum} as
\begin{align}
\delta \psi_{\textrm{abelian}} = &+4 \lambda_1 \g_\mu \g^I \partial_\nu X^J \partial^\nu X^J \partial^\mu X^I\e - 2 \lambda_2 \g_\mu \g^I  \partial^\mu X^J \partial_\nu X^J \partial^\nu X^I \e\nn \\
&+ (2 \lambda_2 - 8\lambda_1 ) (\g_\mu  \partial_\nu X^j \partial^\nu X^j \partial^\mu X^8 \e + \g_\mu \g^j  \partial_\nu X^8 \partial^\nu X^8 \partial^\mu X^j\e) \nn \\
&- 8 \lambda_3 \g^{\mu \nu \rho} \g^{ijk} \partial_\mu X^i \partial_\nu X^j \partial_\rho X^k \e - \lambda_2 \g^{ij} \e^{\rho \sigma \lambda} \partial_\rho X^i \partial_\lambda X^8 \partial_\sigma X^j \e. \label{magic}
\end{align}
The last line in \eqref{magic} can be expressed in $SO(8)$ form by noting   
\begin{equation}
\e^{\mu \nu \rho} \g^{IJK}  \partial_\mu X^I \partial_\nu X^J \partial_\rho X^K  \e \rightarrow \e^{\mu \nu \rho} \g^{ijk}   \partial_\mu X^i \partial_\nu X^j \partial_\rho X^k \e + 3 \e^{\mu \nu \rho} \g^{ij} \g^8 \partial_\mu X^i \partial_\nu X^j \partial_\rho X^8 \e.  \nn
\end{equation}
Provided with this information we see that it's possible to write \eqref{magic} in SO(8) form provided the coefficients are related as
\begin{equation}
\lambda_2 = 4 \lambda_1 ; \quad \lambda_2 = - 24 \lambda_3.
\end{equation}
The final result for the abelian fermion variation is
\begin{align}
\delta \psi = &+4 \lambda_1 l_p^3 \g_\mu \g^I \partial_\nu X^J \partial^\nu X^J \partial^\mu X^I\e - 8 \lambda_1 l_p^3\g_\mu \g^I  \partial^\mu X^J \partial_\nu X^J \partial^\nu X^I \e\nn \\
&-\frac{4}{3}\lambda_1 l_p^3\e^{\mu \nu \rho} \g^{IJK}  \partial_\mu X^I \partial_\nu X^J \partial_\rho X^K \e \label{magic2}.
\end{align}
A few comments are in order. Firstly we see that the structure of these terms exactly matches the structure of the terms appearing in \eqref{fermi} with the coefficients fixed as $a_1 = 4\lambda_1$, $a_2 = - 8\lambda_1$ and $a_3 = - 4/3 \lambda_1$. It remains to determine $\lambda_1$. Most remarkably we see that the requirement of SO(8) invariance has placed a constraint on the coefficients of the ten-dimensional supersymmetry transformations! Furthermore the ratios of the coefficients exactly matches the literature \cite{Collinucci:2002gd, Bergshoeff:1986jm, Cederwall:2001td, Collinucci:2002ac}. Thus it would appear that abelian duality does indeed work at the level of the fermion supersymmetry transformation. So what about the scalar variation? One would expect the higher order scalar supersymmetry transformation to work in a similar way to the lower order transformation. In other words, one expects the (2+1) dimensional gauge field transformation to contribute (after dualisation) to the `eighth' component of the scalar transformation $\delta X^I$. In order to see how this works we will need to determine $\delta F_{\mu \nu}$ in (2+1) dimensions. This can be constructed from our knowledge of $\delta A_\mu$. Therefore the first thing we need to do is dimensionally reduce the ten dimensional gauge field transformation $\delta A_M$ appearing in \eqref{ymtran}. Performing the reduction results in a scalar field supersymmetry transformation
\begin{align}
\delta X^i = &+\alpha_1 \be \g^i F_{\mu \nu} F^{\mu \nu} \psi + 2 \alpha_1 \be \g^i \partial_\mu X^j \partial^\mu X^j \nn \\
&+ \alpha_2 \be \g_\mu \partial_\rho X^i F^{\mu \rho} \psi - \alpha_2 \be \g^j \partial_\rho X^i \partial^\rho X^j \psi \nn \\
&- \alpha_3 \be \g^{\mu \nu \rho} \partial_\mu X^i F_{\nu \rho} \psi - \alpha_3 \be \g^{\mu \nu} \g^j \partial_\mu X^i \partial_\nu X^j \psi \nn \\
&- 4 \alpha_4 \be \g^{ijk} \g^{\mu \nu} \partial_\mu X^j \partial_\nu X^k \psi - 4 \alpha_4 \be \g^{ij} \g^{\mu \nu \rho} F_{\mu \nu} \partial_\rho X^j
\end{align}
and a gauge field supersymmetry transformation 
\begin{align}
\delta A_\mu = &+ \alpha_1 \be \g_\mu F_{\nu \rho} F^{\nu \rho} \psi + 2 \alpha_1 \be \g_\mu \partial_\rho X^i \partial^\rho X^i \psi \nn \\
&+ \alpha_2 \be \g_\rho F_{\mu \nu} F^{\nu \rho} \psi + \alpha_2 \be \g^i F_{\mu \nu} \partial^\nu X^i \psi \nn \\
&- \alpha_2 \be \g_\rho \partial^\rho X^i \partial_\mu X^i \psi + \alpha_3 \be \g^{\nu \rho \sigma} F_{\mu \nu} F_{\rho \sigma} \psi \nn \\
&+ 2 \alpha_3 \be \g^{\nu \rho} \g^i F_{\mu \nu} \partial_\rho X^i + \alpha_3 \be \g^{\rho \sigma} \g^i \partial_\mu X^i F_{\rho \sigma} \nn \\
&- 2 \alpha_3 \be \g^{ij} \g^\rho \partial_\mu X^i \partial_\rho X^j - 4 \alpha_4 \be \g_{\mu \nu \rho} \g^{ij} \partial^\nu X^i \partial^\rho X^j.
\end{align}
Performing the dualisation of the gauge field results in the scalar transformation 
\begin{align}
\delta X^i = &- 2 \alpha_1 \be \g^i \partial_\mu X^8 \partial^\mu X^8 \psi + 2 \alpha_1 \be \g^i \partial_\mu X^j \partial^\mu X^j \psi \nn \\
&+ \alpha_2 \be \g^{\mu \nu} \partial_\mu X^i \partial_\nu X^8 \psi - \alpha_2 \be \g^j \partial_\mu X^i \partial^\mu X^j \psi \nn \\
&- 2 \alpha_3 \be \partial_\mu X^i \partial^\mu X^8 \psi - 2 \alpha_3 \be \g^{\mu \nu} \g^j \partial_\mu X^i \partial_\nu X^j \psi \nn \\
&-8 \alpha_4 \be \g^{ij} \partial_\mu X^j \partial^\mu X^8 \psi - 4 \alpha_4 \be \g^{ijk} \g^{\mu \nu} \partial_\mu X^j
 \partial_\nu X^k \psi. \label{failure1}
\end{align}
Similarly for the gauge field one finds
\begin{align}
\delta A_\mu = &+ (\alpha_2 - 2 \alpha_1) \be \g_\mu \psi \partial_\nu X^8 \partial^\nu X^8  - \alpha_2 \be \g_\nu \psi \partial^\nu X^8 \partial_\mu X^8 \nn \\
&+ 2 \alpha_1 \be \g_\mu \psi \partial_\nu X^i \partial^\nu X^i - \alpha_2 \be \g_\nu \psi \partial_\mu X^i \partial^\nu X^i   \nn \\
&- 2\alpha_3 \be \g^{ij} \g^\sigma \psi \partial_\mu X^i \partial_\sigma X^j - 2 \alpha_3 \be \g_\mu \g^j \psi \partial^\nu X^8 \partial_\nu X^j \nn \\
 &+ \alpha_2 \e_{\mu \nu \lambda} \be \g^j \psi \partial^\lambda X^8 \partial^\nu X^j+ 4 \alpha_4 \e_{\mu \nu \rho} \be \g^{ij} \psi \partial^\nu X^i \partial^\rho X^j. \label{failure2}
\end{align}
The hope is that, just as for the fermion, these terms will combine into SO(8) invariant objects and in doing so fix the ratios of the coefficients. However we immediately encounter a problem which did not exist for the fermion variation. To see this let us focus on the first two terms appearing in \eqref{failure1}. These are the only two terms in \eqref{failure1} with the correct index structure to form the SO(8) term $\be \g^I \psi \partial_\mu X^J \partial^\mu X^J$ appearing in \eqref{bose}. However there is a relative minus sign appearing in these two terms meaning they are unable to combine. The problem can be traced back to the ten-dimensional gauge-field transformation term $\be \g_M  F_{NP} F^{NP} \psi$. 
Why is this happening? After all, we know that abelian duality works for the D2-brane Lagrangian and we know that the D2-brane Lagrangian derives from the ten-dimensional Yang-Mills term $-\frac{1}{4}F_{MN}F^{MN}$. Indeed, upon dimensional reduction of the ten-dimensional Yang-Mills Lagrangian and application of \eqref{pohg} one is left with a term $\frac{1}{2}\partial_\mu X^8 \partial^\mu X^8 -\frac{1}{2} \partial_\mu X^i \partial^\mu X^i$ which will not combine to form an SO(8) invariant scalar kinetic term. The way this problem is solved at Lagrangian level is by adding a Lagrange multiplier term $\frac{1}{2} \e_{\mu \nu \lambda} \partial^\mu X^8 F^{\nu \lambda}$ which under dualisation (according to \eqref{pohg}) combines with $\frac{1}{2}\partial_\mu X^8 \partial^\mu X^8$ in such a way as to change the sign of this term thereby allowing it to combine with $-\frac{1}{2} \partial_\mu X^i \partial^\mu X^i$ to form the desired SO(8) invariant scalar kinetic term. This suggests that the problem may in fact be the prescription \eqref{pohg}. In order to implement the duality on $F^2$ terms it may be necessary to make the replacement $\frac{1}{4}F^2 \rightarrow \frac{1}{4}F^2 + \frac{1}{2} \e_{\mu \nu \lambda} \partial^\mu X^8 F^{\nu \lambda}$. However, if this is true then it's unclear why making the replacement \eqref{pohg} works for the fermion variation. Perhaps the reason we had no problem with the fermion is related to the fact that the ten dimensional Yang-Mills fermion variation contains terms of order $F^3$ whereas the gauge-field variation contains terms of order $F^2$. The result being that the dualised fermion variation contains terms with the \textit{same} structure that derive from \textit{different} ten-dimensional terms. This allows for the coefficients to be related in such a way that unwanted terms are eliminated. This is not true for the scalars. Furthermore, by observing how the terms in \eqref{bose} break-up into SO(7) objects
\begin{align}
\be \g^I \psi \partial_\mu X^J \partial^\mu X^J &\ra \be \g^i \psi \partial_\mu X^8 \partial^\mu X^8 + \be \g^i \psi \partial_\mu X^j \partial^\mu X^j \nn \\
\be \g^{\mu \nu}\g^{IJK} \psi \partial_\mu X^J \partial_\nu X^K &\ra  2\be \g^{\mu \nu} \g^{ij} \psi \partial_\mu X^j \partial_\nu X^8 + \be \g^{\mu \nu} \g^{ijk} \psi \partial_\mu X^j \partial_\nu X^k \nn \\
\be \g^J \psi \partial_\mu X^I \partial^\mu X^J &\ra \be \g^8 \psi \partial_\mu X^i \partial^\mu X^8 + \be \g^j \psi \partial_\mu X^i \partial^\mu X^j \nn \\
\be \g^J \g^{\mu \nu} \psi \partial_\mu X^I \partial_\nu X^J &\ra \be \g^8 \g^{\mu \nu} \psi \partial_\mu X^i \partial_\nu X^8 + \be \g^j \g^{\mu \nu} \psi \partial_\mu X^i \partial_\nu X^j \nn
\end{align}
we see that there are terms appearing in \eqref{failure1} which do not appear in \eqref{bose}. Therefore, until we know how to modify the abelian duality transformation such that the scalar fields combine into SO(8) objects, we will have to follow a different path to determine $\delta X^I$. In the next section we will use our knowledge of the higher order abelian Lagrangian to determine both $\delta \psi$ and $\delta X^I$ by requiring invariance of the action. 
\subsection{Invariance of Higher Order Lagrangian}
The $l_p^3$ corrected abelian M2-brane Lagrangian takes the form \cite{Alishahiha:2008rs, Ezhuthachan:2009sr}
\begin{align}
S_{BLG} = \int d^3 x &- \frac{1}{2} \partial_\mu X^I \partial^\mu X^I + \frac{i}{2} \bp \g^\mu \partial_\mu \psi \nn \\
&+ \frac{1}{4}l_p^3 (\partial^\mu X^I \partial_\mu X^J \partial^\nu X^J \partial_\nu X^I - \frac{1}{2} \partial^\mu X^I \partial_\mu X^I \partial^\nu X^J \partial_\nu X^J) \nn \\
&+ \frac{i}{4}l_p^3 ( \bp \g^\mu \g^{IJ} \partial_\nu \psi \partial_\mu X^I \partial^\nu X^J - \bp \g^\mu \partial_\nu \psi \partial_\mu X^I \partial^\nu X^I) \nn \\
&-\frac{1}{16}  l_p^3 \bp \g^\mu \partial^\nu \psi \bp \g_\nu \partial_\mu \psi
\end{align}
The supersymmetry transformations at lowest order are
\begin{align}
\delta X^I &= i \be \g^I \psi. \nn \\
\delta \psi &= \partial_\mu X^I \g^\mu \g^I \e.
\end{align}
At higher order we will consider the transformations \eqref{fermi} and \eqref{bose} (neglecting bi-linear and tri-linear fermion terms). To recap, for the fermion we have
\begin{align}
\delta \psi = &+ a_1 \g_\mu \g^I \partial_\nu X^J \partial^\nu X^J \partial^\mu X^I \e + a_2 \g_\mu \g^I \partial^\mu X^J \partial_\nu X^J \partial^\nu X^I \e \nn \\
&+ a_3\g^{\mu \nu \rho} \g^{IJK} \partial_\mu X^I \partial_\nu X^J \partial_\rho X^K \e 
\end{align}
and for the scalar 
\begin{align}
\delta X^I =  &+ b_1 \be \g^I \psi \partial_\mu X^J \partial^\mu X^J + b_2 \be \g^J \psi \partial_\mu X^I \partial^\mu X^J \nn \\
&+ b_3 \be \g^J \g^{\mu \nu} \psi \partial_\mu X^I \partial_\nu X^J + b_4 \be \g^{\mu \nu} \g^{IJK} \psi \partial_\mu X^J \partial_\nu X^K.
\end{align}
In the variation of the action there will be terms coming from the higher order supersymmetry variation of the lower order Lagrangian and there will be terms coming from the lower order supersymmetry variation of the higher order Lagrangian. These terms should cancel against each other up to a surface term. Demanding invariance of the action will put constraints on the coefficients. Not only will we determine $\delta X^I$ but also $\delta \psi$ allowing for comparison with the result derived in the previous section using abelian duality. Let us begin by considering the higher order supersymmetry variation of the lower order Lagrangian. This results in
\begin{align}
- \partial_\mu X^I \partial^\mu (\delta X^I) = &-i b_1 \be \g^I \partial_\mu \psi \partial^\mu X^I \partial^\nu X^J \partial_\nu X^J  - i b_1 \be \g^I \psi \partial_\mu (\partial_\nu X^J \partial^\nu X^J) \partial^\mu X^I \nn \\
&- i b_2 \be \g^I \partial_\mu \psi \partial^\mu X^J \partial^\nu X^J \partial_\nu X^I - i b_2 \be \g^I \psi \partial_\mu (\partial^\nu X^J \partial_\nu X^I) \partial^\mu X^J \nn \\
&- i b_3 \be \g^I \g^{\mu \nu} \partial_\rho \psi \partial_\mu X^J \partial_\nu X^I \partial^\rho X^J - i b_3 \be \g^I \g^{\mu \nu} \psi \partial_\rho (\partial_\mu X^J \partial_\nu X^I) \partial^\rho X^J \nn \\
&- i b_4 \be \g^{\rho \nu} \g^{IJK} \partial_\mu \psi \partial^\mu X^I \partial_\rho X^J \partial_\nu X^K  - i b_4 \be \g^{\rho \nu } \g^{IJK} \psi \partial_\mu (\partial_\rho X^J \partial_\nu X^K) \partial^\mu X^I 
\end{align}
\begin{align}
\delta (\frac{i}{2} \bp \g^\mu \partial_\mu \psi) = &+ \frac{i}{2} a_1 \be \g^I \partial_\mu \psi \partial^\mu  X^I \partial_\nu X^J \partial^\nu X^J - \frac{i}{2} a_1 \be \g^I \psi \partial_\mu (\partial_\nu X^J \partial^\nu X^J \partial^\mu X^I)  \nn \\
&+ \frac{i}{2} a_2 \be \g^I \partial_\mu \psi \partial^\mu X^J \partial_\nu X^J \partial^\nu X^I - \frac{i}{2} a_2 \be \g^I \psi \partial_\mu (\partial^\mu X^J \partial_\nu X^J \partial^\nu X^I )  \nn \\
&- \frac{i}{2} a_1 \be  \g^I \g^{\mu \nu} \partial_\mu \psi \partial_\nu X^I \partial^\rho X^J \partial_\rho X^J + \frac{i}{2} a_1 \be  \g^I \g^{\mu \nu} \psi \partial_\mu (\partial_\rho X^J \partial^\rho X^J \partial_\nu X^I)  \nn \\
&- \frac{i}{2} a_2 \be  \g^I \g^{\mu \nu} \partial_\mu \psi \partial_\nu X^J \partial^\rho X^J \partial_\rho X^I + \frac{i}{2} a_2 \be \g^I \g^{\mu \nu} \psi \partial_\mu (\partial_\nu X^J \partial^\rho X^J \partial_\rho X^I)  \nn \\
&- \frac{3i}{2} a_3 \be \g^{\rho \nu } \g^{IJK} \partial^\mu \psi \partial_\mu X^I \partial_\nu X^J \partial_\rho X^K  + \frac{3i}{2} a_3 \be \g^{\rho \nu } \g^{IJK} \psi \partial_\mu (\partial^\mu X^I \partial_\nu X^J \partial_\rho X^K) 
\end{align}
Let us now look at the lower order supersymmetry variation of the higher-order Lagrangian terms. We have
\begin{align}
\delta {\mathcal{L}}_{higher} = &+i \be \g^I \partial_\mu \psi \partial^\mu X^J \partial_\nu X^J \partial^\nu X^I - \frac{i}{2} \be \g^I \partial^\mu \psi \partial_\mu X^I \partial^\nu X^J \partial_\nu X^J \nn \\
&+ \frac{i}{4}  \be \g^I \partial_\mu \psi \partial^\mu X^I \partial_\nu X^J \partial^\nu X^J - \frac{i}{2} (2 \pi )^2 \be \g^I \partial_\mu \psi \partial^\mu X^J \partial_\nu X^J \partial^\nu X^I \nn \\
&- \frac{i}{4}  \be \g^{\mu \nu} \g^I \psi \partial_\mu X^J \partial^\rho X^I (\partial_\nu \partial_\rho X^J ) + \frac{i}{4}  \be \g^{\mu \nu} \g^I \psi  \partial_\mu X^I \partial^\rho X^J (\partial_\nu \partial_\rho X^J) \nn \\
&- \frac{i}{4}  \be \g^{\mu \nu} \g^I \psi \partial_\mu X^J \partial^\rho X^J (\partial_\nu \partial_\rho X^I) + \frac{i}{4} \be \g^I \psi \partial^\mu X^J \partial^\rho X^J (\partial_\mu \partial_\rho X^I ) \nn \\
&- \frac{i}{4} \be \g^{\rho \nu} \g^{IJK} \psi \partial_\mu X^I \partial_\nu X^J (\partial^\mu \partial_\rho X^K) + \frac{i}{4} \be \g^{\rho \nu} \g^{IJK} \partial_\mu \psi \partial^\mu X^I \partial^\rho X^J \partial_\nu X^K \nn \\
&+ \psi^3  \textrm{terms}
\end{align}
How will these terms cancel against each other? Firstly we observe that there are three `types' of term appearing in the above, depending on the gamma matrix structure. It is clear that terms involving the same gamma matrix structure should cancel against each other (up to total derivatives). We begin by focusing on $\g^{\nu \rho} \g^{IJK}$ terms.
Collecting these terms together we can write them as
\begin{align}
&(\frac{3i}{2} a_3 - i b_4 + \frac{i}{4} ) \be \g^{\rho \nu} \g^{IJK} \partial_\mu \psi \partial^\mu X^I \partial_\rho X^J \partial_\nu X^K - \frac{3i}{2} a_3 \be \g^{\rho \nu} \g^{IJK} \psi \partial_\mu (\partial^\mu X^I \partial_\nu X^K \partial_\rho X^J) \nn \\
&+ 2i b_4 \be \g^{\rho \nu} \g^{IJK} \partial_\mu X^I \partial_\nu X^J (\partial^\mu \partial_\rho X^K) \psi - \frac{i}{4}  \be \g^{\rho \nu} \g^{IJK} \psi \partial_\mu X^I \partial_\nu X^J (\partial^\mu \partial_\rho X^K) 
\end{align}
We notice that the first line can be expressed as a total derivative and the second line vanishes provided that
\begin{align}
b_4 &= \frac{1}{8} , \qquad a_3 = -\frac{1}{24} 
\end{align}
Now let us look at the $\g^I$ terms. We can write these as
\begin{align}
&+i(\frac{1}{2} a_1 - b_1 - \frac{1}{4}) \be \g^I \partial^\mu \psi \partial_\mu X^I \partial^\nu X^J \partial_\nu X^J - \frac{i}{2} a_1 \be \g^I \psi \partial_\mu (\partial_\nu X^J \partial^\nu X^J \partial^\mu X^I) \nn \\
&+i (\frac{1}{2} a_2 - b_2 + \frac{1}{2}) \be \g^I \partial_\mu \psi \partial^\mu X^J \partial_\nu X^J \partial^\nu X^I - \frac{i}{2} a_2 \be \g^I \psi \partial_\mu (\partial^\mu X^J \partial_\nu X^J \partial^\nu X^I ) \nn \\
&+ \frac{i}{4} \be \g^I \psi \partial^\mu X^J \partial^\nu X^J (\partial_\mu \partial_\nu X^I) - i b_2 \be \g^I \psi \partial^\nu X^J \partial^\mu X^J (\partial_\mu \partial_\nu X^I) \nn \\
&-i b_2 \be \g^I \psi (\partial_\mu \partial_\nu X^J) \partial^\nu X^I \partial^\mu X^J - 2 i b_1 \be \g^I (\partial_\mu \partial_\nu X^J) \partial^\nu X^J \partial^\mu X^I \label{bzo}
\end{align}
The first two lines can be expressed as total derivatives provided that
\begin{align}
a_1 &= b_1 + \frac{1}{4}  \nn \\
a_2 &= b_2 - \frac{1}{2} 
\end{align}
We see that the last two lines in \eqref{bzo} are equal to zero provided
\begin{align}
b_1 = -\frac{1}{8}, \qquad b_2 = \frac{1}{4} 
\end{align}
Putting this information together we conclude 
\begin{equation}
a_2 = - 2 a_1. \label{boya}
\end{equation}
This agrees with the result derived from the duality transformation method. Now we just need to check that the remaining terms cancel against each other. Focusing on the $\g^J \g^{\rho \mu}$ terms we see they can be written as
\begin{align}
&- b_3 l_p^3 \be \g^J \g^{\rho \mu} \partial_\nu \psi \partial_\rho X^I \partial_\mu X^J \partial^\nu X^I + \frac{i}{2} a_1 l_p^3 \be \g^J \g^{\rho \mu} \partial_\mu \psi \partial_\nu X^I \partial^\nu X^I \partial_\rho X^J \nn \\
&+ \frac{i}{2} l_p^3 a_2 \be \g^J \g^{\rho \mu} \partial_\mu \psi \partial_\rho X^I \partial^\nu X^I \partial_\nu X^J + (\frac{i}{4} + b_3 + \frac{i}{2} a_2 ) l_p^3 \be \g^J \g^{\rho \mu} \psi (\partial_\nu \partial_\rho X^J) \partial_\mu X^I \partial^\nu X^I  \nn \\
&+ (i a_1 - \frac{i}{4} - b_3 ) l_p^3 \be \g^J \g^{\rho \mu} \psi (\partial_\nu \partial_\rho X^I ) \partial^\nu X^I \partial_\mu X^J + (\frac{i}{4} + \frac{i}{2} a_2 ) l_p^3 \be \g^J \g^{\rho \mu} \psi (\partial_\nu \partial_\rho X^I) \partial_\mu X^I \partial^\nu X^J 
\end{align}
If we set $b_3 = 0$ then it is possible to write the remaining terms as total derivatives provided that
\begin{align}
a_1 = \frac{1}{8}, \qquad a_2 = - \frac{1}{4}
\end{align}
which is consistent with \eqref{boya}. We are now in a position to write down expressions for the $\mathcal{O} (l_p^3)$ corrections to the abelian supersymmetry transformations (excluding bi-linear and tri-linear fermion terms)
\begin{align}
\delta X^I &= i \be \g^I \psi - \frac{i}{8} l_p^3 \be \g^I \psi \partial_\mu X^J \partial^\mu X^J + \frac{i}{4} l_p^3 \be \g^J \psi \partial_\mu X^I \partial^\mu X^J + \frac{i}{8} l_p^3 \be \g^{\mu \nu} \g^{IJK} \psi \partial_\mu X^J \partial_\nu X^K. \nn \\
\delta \psi &= \partial_\mu X^I \g^\mu \g^I \e + \frac{1}{8}l_p^3 \g^\mu \g^I \partial_\mu X^I \partial^\nu X^J \partial_\nu X^J\e - \frac{1}{4}l_p^3  \g^\mu \g^I \partial_\mu X^J \partial^\nu X^J \partial_\nu X^I \e \nn \\
&\quad - \frac{1}{24} l_p^3 \g^{\mu \nu \rho} \g^{IJK} \partial_\mu X^I \partial_\nu X^J \partial_\rho X^K \e.
\end{align}
Looking at the fermion variation we see that it is possible to fix the undetermined overall coefficient in \eqref{magic2} as $\lambda_1 = \frac{1}{32}$. In the next section we will consider extending this analysis to the non-abelian Bagger-Lambert M2-brane theory.
\section{Non-Abelian Extension}
In this section we begin an investigation into the non-abelian supersymmetry transformation of the Bagger-Lambert theory at $\mathcal{O} (l_p^3)$. We will see that using the non-abelian dNS duality transformation outlined at the beginning of the chapter it is possible to uniquely determine the higher order fermion variation. We begin by using dimensional analysis to determine the types of terms that can appear in the M2-brane supersymmetry transformations.
\subsection{Dimensional Analysis}
We can write the most general variation of the fermion field as
\begin{equation}
\delta \psi_a = D_\mu X^I_a \g^\mu \g^I \e + \frac{1}{6} X^{IJK} \g^{IJK} \e + l^3_p D^m X^n \psi^{2l} \e.
\end{equation}
Dimensional analysis then tells us that
\begin{equation}
2m + n + 4l = 9
\end{equation}
This gives rise to potentially nine types of term. However we can restrict our attention by making use of our knowledge of the D2-brane supersymmetry transformations. In other words we will only consider terms that match the D2-brane corrections upon application of the novel Higgs mechanism. This leaves us with
\begin{enumerate}
\item $X^9$ .
\item $(DX)X^6$.
\item $(DX) (DX) X^3$.
\item  $(DX)(DX)(DX)$.
\end{enumerate}
working out all independent index contractions one finds an expression of the form
\begin{align}
\delta \psi = l_p^3 [ &a_1 \g_\mu \g^I  D_\nu X^J D^\nu X^J D^\mu X^I + a_2 \g_\mu \g^I  D^\mu X^J D_\nu X^J D^\nu X^I + a_3 \e^{\mu \nu \rho} \g^{IJK}  D_\mu X^I D_\nu X^J D_\rho X^K \nn \\
&+ a_4 \g^{\mu \nu} \g^I  D_\mu X^J D_\nu X^K X^{JKI}  + a_5 \g^{IJK}   D_\mu X^L D^\mu X^J X^{ILK} + a_6 \g^{IJK}  D_\mu X^L D^\mu X^L X^{IJK} \nn \\
&+ a_7 \g^{\mu \nu} \g^{IJKLM}  D_\mu X^I D_\nu X^J X^{KLM} + a_8 \g_\mu \g^J  D^\mu X^K X^{KLM} X^{LJM} \nn \\
&+ a_9 \g_\mu \g^{IJKLM}  D^\mu X^M X^{IJN} X^{KLN} + a_{10} \g_\mu \g^J D_\mu X^J X^{KLM} X^{KLM}  \nn \\
&+ a_{11} \g_\mu \g^{IJKLM} D^\mu X^N X^{IJM} X^{KLN} + a_{12} \g^{IJKLMNP}  X^{IJQ} X^{KLQ} X^{MNP}  \nn \\
&+ a_{13}\g^{IJM}  X^{IKN} X^{KLN} X^{LJM} +  a_{14} \g^{IJM} X^{KLN} X^{KLN} X^{IJM} ]\e. \label{hf}
\end{align}
Similarly we can write the most general scalar field variation as
\begin{equation}
\delta X^{I}_a = i\be \g^I \psi_a + l^3_p \be D^m X^n \psi^{2l+1}
\end{equation}
with
\begin{align}
2m + n + 4l = 6.
\end{align}
This leads to the following possible terms
\begin{enumerate}
\item $\psi X^6$.
\item $\psi (DX)X^3$.
\item $\psi (DX)(DX)$.
\end{enumerate}
After a little thought about possible index contractions one arrives at the following expression
\begin{align}
\delta X^I = il_p^3 [&b_1 \be \g^I \psi D_\mu X^J D^\mu X^J + b_2 \be \g^J \psi D_\mu X^I D^\mu X^J  + b_3 \be \g^J \g^{\mu \nu} \psi D_\mu X^I D_\nu X^J   \nn \\
&+ b_4 \be \g^{\mu \nu}\g^{IJK} \psi D_\mu X^J D_\nu X^K + b_5 \be \g_\mu \g^{JKL} \psi D^\mu X^I X^{JKL} + b_6 \be \g_\mu \g^{IJK} \psi D^\mu X^L X^{JKL} \nn \\
&+ b_7 \be \g_\mu \g^J \psi D_\mu X^K X^{IJK} + b_8 \be \g^\mu \g^{IJKLM} \psi D_\mu X^J X^{KLM}  + b_9 \be \g_\mu \g^{JKL} \psi D^\mu X^K X^{IJL} \nn \\ 
&+ b_{10} \be \g^J \psi X^{JKL} X^{IKL}  + b_{11} \be \g^{JKL} \psi X^{KLN} X^{NIJ}  + b_{12} \be \g^{IJKLM} \psi X^{JKN} X^{LMN}]. \label{hj}
\end{align}
Now that we know the types of terms that will appear in the supersymmetry transformations we will use the non-abelian dNS duality transformation outlined at the beginning of the chapter to try and determine their exact form. Our starting point is the non-abelian D2-brane supersymmetry transformations.
\subsection{D2-brane supersymmetry transformations}
We begin by deriving the non-abelian D2-brane $\alpha'^2$ supersymmetry transformations. Our starting point will be the $\alpha'^2$ ten-dimensional $U(N)$ super Yang-Mills transformations \eqref{ymtran}. Because we are now considering the full non-abelian theory we will have to keep all terms in the dimensional reduction, including commutator terms. Upon dimensional reduction to (2+1) dimensions one finds the following expressions
\begin{equation}
\delta X^i = \sum_{j=1}^6  \delta X^i_{(j)} \label{uno}
\end{equation}
with
\begin{align}
\delta X^i_{(1)} &= \frac{1}{g^2_{YM}}\alpha_1 \be \g^i \psi F_{\mu \nu} F^{\mu \nu} \nn \\
\delta X^i_{(2)} &= -\frac{1}{g_{YM}} (\alpha_2 \be \g_\mu \psi D_\rho X^i F^{\rho \mu} - \alpha_3 \e_{\mu \rho \sigma}\be  \psi D^\mu X^i F^{\rho \sigma} - 4 \alpha_4 \e_{\mu \nu \rho} \be \g^{ij}  \psi F^{\mu \nu} D^\rho X^j )\nn \\
\delta X^i_{(3)} &= - \alpha_3 \be \g^{\rho \sigma} \g^j X^{ij} F_{\rho \sigma} - 2 \alpha_4 \be \g^{ijk} \g_{\mu \nu} \psi X^{jk} F^{\mu \nu} \nn \\
\delta X^i_{(4)} &= 2 \alpha_1 \be \g^i \psi D_\mu X^j D^\mu X^j - \alpha_2 \be \g^j \psi D_\rho X^i D^\rho X^j - \alpha_3 \be \g^{\mu \rho} \g^j \psi D_\mu X^i D_\rho X^j \nn \\
&- \alpha_3 \be \g^{\mu \sigma} \g^j \psi D_\mu X^i D_\sigma X^j - 4 \alpha_4 \be \g^{ijk} \g_{\mu \nu} \psi D_\mu X^j D_\nu X^k \nn \\
\delta X^i_{(5)} &= g_{YM} (\alpha_2 \be \g_\mu \psi X^{ij} D_\mu X^j + \alpha_3 \be \g^\mu \g^{jk} \psi D_\mu X^i X^{jk} + \alpha_3 \be \g^\rho \g^{jk} \psi  X^{ij} D_\rho X^k \nn \\
&+ \alpha_3 \be \g^\sigma \g^{jk} \psi X^{ij} D_\sigma X^k + 4 \alpha_4 \be \g_\mu \g^{ijkl} \psi D^\mu X^j X^{kl}) \nn \\
\delta X^i_{(6)} &= g^2_{YM} (\alpha_1 \be \g^i \psi X^{jk} X^{jk} + \alpha_2 \be \g^k \psi X^{ij} X^{jk} + \alpha_3 \be \g^{jkl} \psi X^{ij} X^{kl} + \alpha_4 \be \g^{ijklm} \psi X^{jk} X^{lm} )\nn
\end{align}
\\
\\
\begin{equation}
\delta A_\mu = \sum_{i=1}^6  \delta A_{\mu (i)}  \label{duz}
\end{equation}
with
\begin{align}
\delta A_{\mu (1)} &=  \frac{1}{g_{YM}}(\alpha_1 \be \g_\mu \psi F_{\rho \sigma} F^{\rho \sigma} + \alpha_2 \be \g_\rho \psi F_{\mu \nu} F^{\nu \rho} - \alpha_3 \e^{\nu \rho \sigma} \be \psi F_{\mu \nu} F_{\rho \sigma}) \nn \\
\delta A_{\mu (2)} &= \alpha_2 \be \g^j \psi F_{\mu \nu} D^\nu X^j + \alpha_3 \be \g^{\nu \sigma} \g^j \psi F_{\mu \nu} D_\sigma X^j + \alpha_3 \be \g^{\rho \sigma} \g^j \psi D_\mu X^j F_{\rho \sigma} + \alpha_3 \be \g^{\nu \rho} \g^k \psi F_{\mu \nu} D_\rho X^k \nn \\
\delta A_{\mu (3)} &= - g_{YM} (\alpha_3 \be \g^\nu \g^{jk} \psi F_{\mu \nu} X^{jk}
 - \alpha_4 \e_{\mu \nu \rho} \be \g^{ij} \psi F^{\nu \rho} X^{ij} - \alpha_4 \e_{\mu \nu \rho} \be  \g^{ij} \psi X^{ij} F^{\nu \rho} )\nn \\
 \delta A_{\mu (4)} &= g_{YM} (2 \alpha_1 \be \g_\mu \psi D_\nu X^i D^\nu X^i - \alpha_2 \be \g_\nu \psi D_\mu X^i D^\nu X^i - \alpha_3 \be \g^{ij} \g^\sigma \psi D_\mu X^i D_\sigma X^j  \nn \\
 &- \alpha_3 \be \g^{ij} \g^\rho \psi D_\mu X^i D_\rho X^j + 4 \alpha_4 \e_{\mu \nu \rho} \be \g^{ij} \psi D^\nu X^i D^\rho X^j ) \nn 
\end{align}
\begin{align} 
\delta A_{\mu (5)} &= - g^2_{YM} (\alpha_2 \be \g^j \psi D_\mu X^i X^{ij} + \alpha_3 \be \g^{ijk} \psi D_\mu X^i X^{jk} \nn \\
&+ 2 \alpha_4 \be \g_{\mu \nu} \g^{jkl} \psi D^\nu X^j X^{kl} + 2 \alpha_4 \be \g_{\mu \nu} \g^{jkl} \psi X^{jk} D^\nu X^l ) \nn \\
\delta A_{\mu (6)} &= g^3_{YM} (\alpha_1 \be \g_\mu \psi X^{ij} X^{ij} + \alpha_4 \be \g_\mu \g^{ijkl} \psi X^{ij} X^{kl}) \nn
\end{align}

\begin{equation}
\delta \psi = \sum_{i=1}^{10}  \delta \psi_i \e \label{trez}
\end{equation}
with
\begin{align}
\delta \psi_1 &= \frac{1}{g^3_{YM}}(\lambda_1 \g^{\mu \nu} F_{\rho \sigma} F^{\rho \sigma} F_{\mu \nu} + \lambda_2 \g^{\mu \nu} F_{\mu \rho} F^{\rho \sigma} F_{\sigma \nu})  \nn \\
\delta \psi_2 &= \frac{1}{g^2_{YM}} (2 \lambda_1 \g^\mu \g^j F_{\rho \sigma} F^{\rho \sigma} D_\mu X_j + \lambda_2 \g^\mu \g^j F_{\mu \nu} F^{\nu \rho} D_\rho X_j +  \lambda_2 \g^\nu \g^j D_\rho X_j F^{\rho \sigma} F_{\sigma \nu} ) \nn \\
\delta \psi_3 &= - \frac{1}{g_{YM}}(\lambda_1 \g^{ij} F_{\rho \sigma} F^{\rho \sigma} X^{ij}) \nn \\
\delta \psi_4 &= \frac{1}{g_{YM}} ( \g^{\mu \nu}(2 \lambda_1  D_\rho X^j D^\rho X^j F_{\mu \nu} - \lambda_2  D_\mu X^j D^\sigma X^j F_{\sigma \nu}- \lambda_2 F_{\mu \rho} D^\rho X^k D_\nu X^k) \nn \\
&\quad - \lambda_2 \g^{ij} D_\rho X^i F^{\rho \sigma} D_\sigma X_j )\nn \\
\delta \psi_5 &= -\lambda_2 \g^\mu \g^j F_{\mu \nu} D^\nu X^k X^{kj} - \lambda_2 \g^\nu \g^i X^{ij} D^\rho X^j F_{\rho \nu} - 12 \lambda_3 \e^{\mu \nu \rho} \g^{ijk} F_{\mu \nu} D_\rho X^i X^{jk} \nn \\
\delta \psi_6 &= g_{YM} (\lambda_1 \g^{\mu \nu} X^{ij} X^{ij} F_{\mu \nu} + 3 \lambda_3 \g^{\mu \nu} \g^{ijkl} F_{\mu \nu} X^{ij} X^{kl} ) \nn \\
\delta \psi_7 &= 4 \lambda_1 \g^\mu \g^j D_\nu X^k D^\nu X^k D_\mu X^j - \lambda_2 \g^\mu \g^j D_\mu X^k D^\rho X^k D_\rho X^j \nn \\ &- \lambda_2 \g^\nu \g^i D_\rho X^i D^\rho X^j D_\nu X^j - 8 \lambda_3 \e^{\mu \nu \rho} \g^{ijk} D_\mu X^i D_\nu X^j D_\rho X^k \nn \\
\delta \psi_8 &= g_{YM} (-2 \lambda_1 \g^{ij} D_\mu X^k D^\mu X^k X^{ij} + \lambda_2 \g^{\mu \nu} D_\mu X^j X^{jk} D_\nu X^k + \lambda_2 \g^{ij} D_\rho X^i D^\rho X^k X^{kj} \nn \\
&+ \lambda_2 \g^{ij} X_{ik} D^\rho X^k D_\rho X_j + 12 \lambda_3 \g^{\mu \nu} \g^{ijkl} D_\mu X^i D^\nu X^j X^{kl} )\nn \\
\delta \psi_9 &= g^2_{YM} \g^\mu (2 \lambda_1  \g^j X^{kl} X^{kl} D_\mu X^j + \lambda_2 \g^i X^{ij} X^{jk} D_\mu X^k + \lambda_2  \g^j D_\mu X^k X^{kl} X^{lj} \nn \\
&\quad +  6 \lambda_3  \g^{ijklm} D_\mu X^i X^{jk} X^{lm}) \nn \\
\delta \psi_{10} &= - g^3_{YM} (\lambda_1 \g^{ij} X^{kl} X^{kl} X^{ij} + \lambda_2 \g^{ij} X^{ik} X^{kl} X^{lj} + \lambda_3 \g^{ijklmn} X^{ij} X^{kl} X^{mn}). \nn
\end{align}
Now that we have the non-abelian D2-brane supersymmetry transformations we can attempt to dualise the Yang-Mills gauge field to a scalar. Our method will follow the presentation of \cite{Alishahiha:2008rs} where the $l_p^3$ corrections to the Lorentzian Bagger-Lambert theory were derived using the dNS duality prescription. Let us briefly review the procedure for implementing dNS duality in the higher order Lagrangian.
\subsection{Higher order dNS duality}
In \cite{Alishahiha:2008rs},  higher order corrections to the Lorentzian Bagger-Lambert theory were derived by making use of the dNS duality transformation. As outlined at the beginning of this chapter, implementing the dNS duality involves rewriting the D2-brane Lagrangian in terms of the new fields $B_\mu$ and $X^8$. To see how this works at higher order we will derive the $\mathcal{O} (l_p^3)$ bosonic terms of the Bagger-Lambert theory. Our starting point will be the $(\alpha')^2$ corrections to the non-abelian D2-brane theory. These terms derive from the $F^4$ corrections of ten dimensional super Yang-Mills theory \cite{Tseytlin:1997csa, Cederwall:2001td, Bergshoeff:2001dc}
\begin{align}
\mathcal{L} &= -\frac{1}{4} F^2 + \frac{1}{8} \Str (F^4 - \frac{1}{4} (F^2)^2) \nn \\
&=-\frac{1}{4}F_{MN}F^{MN} + \frac{1}{12} \Tr [F_{MN} F_{RS} F^{MR} F^{NS} + \frac{1}{2} F_{MN} F^{NR} F_{RS} F^{SM} \nn \\
&\quad \quad - \frac{1}{4} F_{MN} F^{MN} F_{RS} F^{RS} - \frac{1}{8} F_{MN} F_{RS} F^{MN} F^{RS}] .\label{10d}
\end{align} 
The next step is to reduce this expression to (2+1) dimensions. We then re-write the (2+1) dimensional field strength $F_{\mu \nu}$ in terms of the dual field strength ${\tilde{F}}_\mu = \e_{\mu \nu \lambda} F^{\nu \lambda}$. In order to implement the dNS duality we replace the dual field strength ${\tilde{F}}_\mu$ by an independent matrix-valued one-form field $B_\mu$. The resulting Lagrangian looks like
\begin{align}
\mathcal{L} = &\Tr [ {\tilde{F}}_\mu B^\mu - \frac{g^2_{YM}}{2} B_\mu B^\mu + \frac{g^4_{YM}}{4} (B_\mu B^\mu B_\nu B^\nu + \frac{1}{2} B_\mu B_\nu B^\mu B^\nu ) \nn \\
&+ \frac{g^2_{YM}}{12} (2 B^\mu B_\nu D^\nu X^i D_\mu X^i - 2B^\mu B_\mu D_\nu X^i D^\nu X^i + 2 B^\mu B^\nu D_\mu X^i D_\nu X^i \nn \\
&+ B^\mu D^\nu X^i B_\nu D_\mu X^i - B^\mu D^\nu X^i B_\mu D_\nu X^i + B^\mu D_\mu X^i B^\nu D_\nu X^i) \nn \\
&+ \frac{g^4_{YM}}{12} (B^\mu B_\mu X^{ij} X_{ij} + \frac{1}{2} B^\mu X^{ij} B_\mu X^{ij}) \nn \\
&+ \frac{g^2_{YM}}{6} \e_{\mu \nu \lambda} (B^\lambda D^\mu X^i D^\nu X^j + D^\nu X^j B^\lambda D^\mu X^i + D^\mu X^i D^\nu X^j B^\lambda ) X^{ij} ]. \label{poor}
\end{align} 
We see that $\tilde{F}$ only appears in the Chern-Simons term ${\tilde{F}}_\mu B^\mu$. To show that this expression is equivalent to the $(\alpha' )^2$ D2-brane Lagrangian one simply integrates out the field $B_\mu$ order by order using its equation of motion. In order to rewrite the Lagrangian in an SO(8) invariant form we introduce the field $X^8$ and replace $B_\mu$ everywhere it occurs by $- 1/g_{YM} (D_\mu X^8 - g_{YM} B_\mu)$. Recall that there is now a shift symmetry allowing one to set $X^8 = 0$ in which case we arrive back at \eqref{poor}. Performing this substitution and collecting the resulting terms into the SO(8) invariant building blocks $\D_\mu X^I$ and $X^{IJK}$ results in the compact expression
\begin{align}
\mathcal{L} = &+\frac{1}{2} \e_{\mu \nu \lambda} B^\mu F^{\nu \lambda} - \frac{1}{2} \D_\mu X^I \D^\mu X^I \nn \\
 &+ \frac{1}{8} l_p^3 \Str [ 2 \D^\mu X^I \D_\mu X^J \D^\nu X^J \D_\nu X^I - \D^\mu X^I \D_\mu X^I \D^\nu X^J \D_\nu X^J  \nn \\
 &- \frac{4}{3} \e^{\mu \nu \lambda} X^{IJK} \D_\mu X^I \D_\nu X^J \D_\lambda X^K \nn \\
 &+ 2 X^{IJK} X^{IJL} \D^\mu X^K \D_\mu X^L - \frac{1}{3} X^{IJK} X^{IJK} \D^\mu X^L \D_\mu X^L \nn \\
 &+ \frac{1}{3} X^{IJM} X^{KLM} X^{IKN} X^{JLN} - \frac{1}{24} X^{IJK} X^{IJK} X^{LMN} X^{LMN} ].
\end{align}
In \cite{Alishahiha:2008rs} it was shown that the same approach can be used to derive the $\mathcal{O} (l_p^3)$ fermion terms. We see that it is possible to implement dNS duality at higher order by applying the following prescription
\begin{enumerate} 
\item Dimensionally reduce 10 dimensional expression to (2+1) dimensions.
\item Write all field strengths in terms of their duals: \quad $F_{\mu \nu} = - \e_{\mu \nu \lambda} {\tilde{F}}^\lambda$.
\item Replace ${\tilde{F}}_\mu$ with the field $B_\mu$.
\item Replace $B_\mu$  with 
$-{g_{YM}} \D_\mu X^8$.
\item Rewrite all expressions in terms of $\D_\mu X^I$ and $X^{IJK}$ building blocks.
\end{enumerate}
In the next section we will test whether this prescription works at the level of supersymmetry transformations. We have already performed the first task on the list. Next we must re-write the D2-brane supersymmetry transformations \eqref{uno}, \eqref{duz} and \eqref{trez} in terms of $\D_\mu X^8$. 

\subsection{dNS transformed supersymmetry}
\begin{align}
\delta X^i = &- \overbrace{2 \alpha_1 \be \g^i \psi \D^\mu X^8 \D_\mu X^8 + \alpha_2 \be \g^{\mu \nu} \psi \D_\mu X^i \D_\nu X^8 -2 \alpha_3 \be \psi \D^\mu X^i \D_\mu X^8}^{\textrm{Two Derivative}} \nn \\
&+2 \alpha_1 \be \g^i \psi \D_\mu X^j \D^\mu X^j - \alpha_2 \be \g^j \psi \D_\mu X^i \D^\mu X^j - 2\alpha_3 \be \g^{\mu \nu} \g^j \psi \D_\mu X^i \D_\nu X^j \nn \\
& - 8 \alpha_4 \be \g^{ij} \psi \D_\mu X^8 \D^\mu X^j - 4 \alpha_4 \be \g^{ijk} \g_{\mu \nu} \psi \D_\mu X^j \D_\nu X^k  \nn \\
&+ \overbrace{ g_{YM}( 2 \alpha_3 \be \g_\mu \g^j X^{ij} \D^\mu X^8 + 4 \alpha_4 \be \g^{ijk} \g^\mu \psi X^{jk} \D_\mu X^8 +\alpha_2 \be \g_\mu \psi X^{ij} \D_\mu X^j }^{\textrm{One Derivative}} \nn \\
&+ \alpha_3 \be \g^\mu \g^{jk} \psi \D_\mu X^i X^{jk} + \alpha_3 \be \g^\rho \g^{jk} \psi  X^{ij} \D_\rho X^k \nn \\
&+ \alpha_3 \be \g^\sigma \g^{jk} \psi X^{ij} \D_\sigma X^k + 4 \alpha_4 \be \g_\mu \g^{ijkl} \psi \D^\mu X^j X^{kl} )\nn \\
&+ \overbrace{ g^2_{YM}( \alpha_1 \be \g^i \psi X^{jk} X^{jk} + \alpha_2 \be \g^k \psi X^{ij} X^{jk} + \alpha_3 \be \g^{jkl} \psi X^{ij} X^{kl} + \alpha_4 \be \g^{ijklm} \psi X^{jk} X^{lm}}^{\textrm{Zero Derivative}}) \label{sc1}
\end{align}

\begin{align}
\delta A_\mu = &+ \overbrace{g_{YM} (- 2 \alpha_1 \be \g_\mu \psi \D_\nu X^8 \D^\nu X^8  + \alpha_2 \be \g_\mu \psi \D^\nu X^8 \D_\nu X^8 - \alpha_2 \be \g_\nu \psi \D^\nu X^8 \D_\mu X^8 }^{\textrm{Two Derivative}} \nn \\
&+ 2 \alpha_3 \e_{\mu \nu \lambda} \be \psi \D^\lambda X^8 \D^\nu X^8   + \alpha_2 \e_{\mu \nu \lambda} \be \g^j \psi \D^\lambda X^8 \D^\nu X^j - 2 \alpha_3 \be \g_\lambda \g^j \psi D_\mu X^j \D^\lambda X^8  \nn \\
&- \alpha_3 \be \g_\mu \g^j \psi \D^\nu X^8 \D_\nu X^j + \alpha_3 \be \g_\lambda \g^j \psi \D^\lambda X^8 \D_\mu X^j - \alpha_3 \be \g_\mu \g^k \psi \D^\rho X^8 \D_\rho X^k  \nn \\
&+ \alpha_3 \be \g_\lambda \g^k \psi \D^\lambda X^8 \D_\mu X^k   + 2 \alpha_1 \be \g_\mu \psi \D_\nu X^i \D^\nu X^i - \alpha_2 \be \g_\nu \psi \D_\mu X^i \D^\nu X^i   \nn \\
 &- \alpha_3 \be \g^{ij} \g^\sigma \psi \D_\mu X^i \D_\sigma X^j   - \alpha_3 \be \g^{ij} \g^\rho \psi \D_\mu X^i \D_\rho X^j +4 \alpha_4 \e_{\mu \nu \rho} \be \g^{ij} \psi \D^\nu X^i \D^\rho X^j )\nn \\
&+ \overbrace{ g^2_{YM} (\alpha_3 \be \g_{\mu \nu} \g^{jk} \psi \D^\nu X^8 X^{jk} - 2 \alpha_4 \be \g^{ij} \psi \D_\mu X^8 X^{ij} - 2 \alpha_4 \be \g_{\mu \nu} \g^{jkl} \psi D^\nu X^j X^{kl}  }^{\textrm{One Derivative}} \nn \\
&- \alpha_2 \be \g^j \psi \D_\mu X^i X^{ij} - \alpha_3 \be \g^{ijk} \psi \D_\mu X^i X^{jk} - 2 \alpha_4 \be \g^{ij} \psi X^{ij} \D_\mu X^8 - 2 \alpha_4 \be \g_{\mu \nu} \g^{jkl} \psi X^{jk} \D^\nu X^l )\nn \\
&+ \overbrace{ g^3_{YM}(\alpha_1 \be \g_\mu \psi X^{ij} X^{ij} + \alpha_4 \be \g_\mu \g^{ijkl} \psi X^{ij} X^{kl} }^{\textrm{Zero Derivative}}) \label{pfm}
\end{align}

\begin{align}
\delta \psi = &+ \overbrace{4 \lambda_1 \g_\mu \e\D^\nu X^8 \D_\nu X^8 \D^\mu X^8 - \lambda_2 \g_\mu \e\D^\nu X^8 \D_\nu X^8 \D^\mu X^8 -\lambda_2 \g_\mu \e \D^\mu X^8 \D_\nu X^8 \D^\nu X^8}^{\textrm{Three Derivative}} \nn \\
&- 4 \lambda_1 \g^\mu \g^j \e \D_\nu X^8 \D^\nu X^8 \D_\mu X^j + \lambda_2 \g^\mu \g^j \e \D^\nu X^8 \D_\nu X^8 \D_\mu X^j - \lambda_2 \g^\mu \g^j \e \D^\nu X^8  \D_\mu X^8 \D_\nu X^j \nn \\
&+ \lambda_2 \g^\mu \g^j \e \D_\mu X^j \D_\nu X^8 \D^\nu X^8 - \lambda_2 \g^\mu \g^j \e \D_\nu X^j \D_\mu X^8 \D^\nu X^8 -4 \lambda_1 \g_\mu \e \D_\nu X^j \D^\nu X^j \D^\mu X^8 \nn \\
& - \lambda_2 \g_\mu \e \D_\nu X^j \D^\mu X^j \D^\nu X^8 + \lambda_2 \g_\mu \e \D_\nu X^j \D^\nu X^j \D^\mu X^8 - \lambda_2 \g_\mu \e \D^\nu X^8 \D^\mu X^k \D_\nu X^k\nn \\
& + \lambda_2 \g_\mu \e \D^\mu X^8 \D^\nu X^k \D_\nu X^k - \lambda_2 \e^{\mu \nu \lambda} \g^{ij} \e \D_\mu X^i \D_\lambda X^8 \D_\nu X^j +4 \lambda_1 \g^\mu \g^j \e \D_\nu X^k \D^\nu X^k \D_\mu X^j \nn \\
& - \lambda_2 \g^\mu \g^j \e \D_\mu X^k \D^\rho X^k \D_\rho X^j - \lambda_2 \g^\nu \g^i \e \D_\rho X^i \D^\rho X^j \D_\nu X^j - 8 \lambda_3 \e^{\mu \nu \rho} \g^{ijk} \e \D_\mu X^i \D_\nu X^j \D_\rho X^k\nn \\
 &+\overbrace{g_{YM}( -\lambda_2 \g_{\mu \nu} \g^j \e \D^\nu X^8 \D^\mu X^k X^{kj} + \lambda_2 \g_{\mu \nu} \g^i \e X^{ij} \D^\mu X^j \D^\nu X^8 + 24 \lambda_3 \g^{ijk} \e \D^\mu X^8 \D_\mu X^i X^{jk}}^{\textrm{Two Derivative}} \nn \\
 & + 2 \lambda_1 \g^{ij} \e \D^\mu X^8 \D_\mu X^8 X^{ij} - 2 \lambda_1 \g^{ij} \e \D_\mu X^k \D^\mu X^k X^{ij} + \lambda_2 \g^{\mu \nu} \e \D_\mu X^j X^{jk} \D_\nu X^k  \nn \\
& + \lambda_2 \g^{ij} \e \D_\rho X^i \D^\rho X^k X^{kj} + \lambda_2 \g^{ij}\e X_{ik} \D^\rho X^k \D_\rho X_j + 12 \lambda_3 \g^{\mu \nu} \g^{ijkl} \e \D_\mu X^i D^\nu X^j X^{kl}) \nn \\
& +\overbrace { g^2_{YM}  (-2 \lambda_1 \g_\mu \e X^{ij} X^{ij} \D^\mu X^8 - 6 \lambda_3 \g_\mu \g^{ijkl} \e \D^\mu X^8 X^{ij} X^{kl} + 2 \lambda_1 \g^\mu \g^j \e X^{kl} X^{kl} \D_\mu X^j}^{\textrm{One Derivative}} \nn \\
&+ \lambda_2 \g^\nu \g^i \e X^{ij} X^{jk} \D_\nu X^k + \lambda_2 \g^\mu \g^j \e \D_\mu X^k X^{kl} X^{lj} + 6 \lambda_3 \g^\mu \g^{ijklm} \e \D_\mu X^i X^{jk} X^{lm}) \nn \\
&- \overbrace{ g^3_{YM} (\lambda_1 \g^{ij} \e X^{kl} X^{kl} X^{ij} + \lambda_2 \g^{ij} \e X^{ik} X^{kl} X^{lj} + \lambda_3 \g^{ijklmn} \e X^{ij} X^{kl} X^{mn} )}^{\textrm{Zero Derivative}}.  \label{massive}
\end{align}

\section{SO(8) supersymmetry transformations}
In the previous section we applied the dNS prescription to the non-abelian D2-brane supersymmetry transformations. We would now like to re-write these expressions in SO(8) form. We will see that this is only possible for the fermion supersymmetry transformation. The scalar transformation is plagued by the same problems we encountered in the abelian theory. We will end this section with a discussion of how one might go about determining the scalar supersymmetry transformation.

\subsection{$\delta \psi$}
Earlier in this chapter we were able to determine the abelian supersymmetry transformation of the fermion by using abelian duality in (2+1) dimensions. In the process we were able to fix the coefficients appearing in \eqref{fermi}. Looking at \eqref{hf} we see that the first three terms are exactly the same as the terms appearing in \eqref{fermi} but with partial derivatives replaced by covariant derivatives. As a result we find that the coefficients are related in exactly the same way. Knowledge of the relationship between $\lambda_1$, $\lambda_2$ and $\lambda_3$, namely
\begin{equation}
\lambda_2 = 4 \lambda_1 ; \quad \lambda_2 = - 24 \lambda_3
\end{equation} 
allows us to re-write all the coefficients in \eqref{massive} in terms of $\lambda_1$. Furthermore by looking at the invariance of the higher order abelian Lagrangian we were able to fix $\lambda_1 = \frac{1}{32}$. Making use of this information, as well as the SO(8) relations outlined in the appendix, it is possible to re-write the two-derivative, one-derivative and zero-derivative terms in \eqref{massive} in an SO(8) invariant form. The final answer for the $l_p^3$ correction to the fermion supersymmetry transformation in Bagger-Lambert theory is
\begin{align}
\delta \psi = l_p^3 [ &\frac{1}{8} \g_\mu \g^I  D_\nu X^J D^\nu X^J D^\mu X^I - \frac{1}{4} \g_\mu \g^I  D^\mu X^J D_\nu X^J D^\nu X^I - \frac{1}{24} \e^{\mu \nu \rho} \g^{IJK}  D_\mu X^I D_\nu X^J D_\rho X^K \nn \\
&+ \frac{1}{8} \g^{\mu \nu} \g^I  D_\mu X^J D_\nu X^K X^{JKI}  + \frac{1}{8} \g^{IJK}   D_\mu X^L D^\mu X^J X^{ILK} - \frac{1}{48} \g^{IJK}  D_\mu X^L D^\mu X^L X^{IJK} \nn \\
& + \frac{1}{48} \g^{\mu \nu} \g^{IJKLM}  D_\mu X^I D_\nu X^J X^{KLM} - \frac{1}{8} \g_\mu \g^J  D^\mu X^K X^{KLM} X^{LJM} \nn \\
&+ \frac{1}{32}\g_\mu \g^{IJKLM}  D^\mu X^M X^{IJN} X^{KLN} + \frac{1}{48} \g_\mu \g^J D_\mu X^J X^{KLM} X^{KLM}  \nn \\
&- \frac{1}{48} \g_\mu \g^{IJKLM} D^\mu X^N X^{IJM} X^{KLN} + \frac{1}{16} \g^{IJKLMNP}  X^{IJQ} X^{KLQ} X^{MNP}  \nn \\
&+ \frac{1}{32}\g^{IJM}  X^{IKN} X^{KLN} X^{LJM} +  \frac{1}{144} \g^{IJM} X^{KLN} X^{KLN} X^{IJM} ]\e.
\end{align}
It is pleasing to see that the dNS duality transformation has allowed us to uniquely determine the structure of the fermion variation. It would be nice to extend this analysis to include tri-linear fermion terms. 

\subsection{$\delta X^I$}
Given that the dNS prescription works for the fermion transformation one might hope that it would also work for the scalar transformation. However, as we observed for the abelian scalar transformation, this is not the case. The two-derivative terms appearing in \eqref{sc1} are of the same form as the abelian scalar terms, with covariant derivatives replacing partial derivatives. For this reason, the non-abelian scalar transformation inherits the same problems we encountered before. For the abelian theory we used a different approach to determine the scalar transformation. This involved checking the invariance of the abelian Lagrangian under a proposed set of supersymmetry transformations (determined by dimensional analysis). The same should be possible for the non-abelian theory. The $l_p^3$ corrected Bagger-Lambert Lagrangian was derived in \cite{Alishahiha:2008rs, Ezhuthachan:2009sr}. Checking that this Lagrangian is invariant (up to surface terms) under the transformations \eqref{hf} and \eqref{hj} should fix the coefficients. Not only would this determine the scalar transformation but would also provide an independent test of the fermion variation calculated using the dNS prescription. 

Ultimately one would like to know how to modify the dNS prescription in such a way that it is possible to derive the scalar variation. Toward this end it may prove useful to determine the scalar transformation by an independent method such that a comparison can be made between the known result and the dNS transformed result \eqref{sc1}. Another complication worth mentioning is related to the gauge field transformation \eqref{pfm}. For the lower order abelian supersymmetry transformations we observed that the eighth component of the scalar variation $\delta X^I$ arises after dualising $\delta F_{\mu\nu}$. More specifically, looking at \eqref{bt} we see that at lowest order $\partial^\lambda \delta X^8 = i \be \g^8 \partial^\lambda \psi$. In this case, since there is only one field and one derivative on the right-hand side, it is possible to simply `pull off' the derivative to determine $\delta X^8$. This is no longer true at higher order and determining $\delta X^8$ becomes a non-trivial task. 
\chapter{Conclusions and outlook}
The content of this thesis is testament to the power of supersymmetry. In Chapter 2 we began with a review of M-theory branes and their interactions from the perspective of spacetime and worldvolume supersymmetry alegebras. This was followed in Chapter 3 by a review of recent attempts to model multiple M2-branes. In Chapter 4 we calculated the extended worldvolume superalgebra of the $\mathcal{N}=6$ Bagger-Lambert Theory. With a particular choice of 3-bracket we were able to derive the ABJM superalgebra. We found that the charges ${\mathcal{Z}}_i$ and ${\mathcal{Z}}^{AB}_{EF,i}$ characterise the topological information corresponding to two sets of BPS equations. In particular we were able to identify the central charge corresponding to the half-BPS fuzzy funnel configuration of the ABJM theory. This was confirmed by deriving the corresponding  BPS equation by completing the square of the Hamiltonian. In order to derive the ABJM fuzzy-funnel BPS equation it was necessary to consider a configuration in which two of the four complex scalar fields were zero. It would be interesting to try and find solutions to generalised BPS equations in the case where more than half the scalar fields are active. Related to this is the question of whether its possible to write the $\mathcal{N}=6$ Bagger-Lambert scalar Hamiltonian as 
\begin{align}
 H &=  \int dx^1 ds \Tr ( \partial_s Z^A - g^{AB}_{\phantom{1} \phantom{1} \phantom{1}CD}\Upsilon^{CD}_{B} )^2  + T \label{upo}
\end{align}
with the condition that
\begin{equation}
g^{AB}_{\phantom{1} \phantom{1} \phantom{1}CD}g_{AE}^{\phantom{1} \phantom{1} \phantom{1}FG} \Tr (\Upsilon^{CD}_{B} , {\bar{\Upsilon}}^{E}_{FG} ) = \frac{2}{3}\Tr (\Upsilon^{CD}_{B} , {\bar{\Upsilon}}^{B}_{CD} ) \label{const}
\end{equation}
where T is a topological term and $\Upsilon$ is defined in \eqref{upsi}. If this constraint is satisfied then we have a set of BPS equations of the form
\begin{equation}
\partial_s Z^A - \kappa g^{AB}_{\phantom{1} \phantom{1} \phantom{1}CD}\Upsilon^{CD}_{B} = 0 \label{BPS1}
\end{equation}
where $A,B = 1, \ldots 4$. It is interesting to note that the constraint \eqref{const} is analogous to the situation encountered when considering M5-brane calibrations \cite{Berman:2005re, Krishnan:2008zm}. In the case of the $\mathcal{N}=8$ Bagger-Lambert theory the constraint takes the form 
\begin{equation}
\frac{1}{3!} g_{IJKL} g_{IPQR} \Tr ([X^J, X^K, X^L],[X^P, X^Q, X^R]) = \Tr([X^I, X^J, X^K],[X^I, X^J, X^K]).
\end{equation}
The $g_{IJKL}$ are related to the calibrating forms of the cycle on which the M5-brane wraps and are therefore completely antisymmetric in their indices. For the case in which only half the scalar fields are activated it is possible to solve the constraint by writing $ g_{IJKL} = \varepsilon_{IJKL} $. This choice corresponds to a fuzzy-funnel configuration in which multiple M2-branes expand into a single M5-brane, and is described by the standard Basu-Harvey equation. For the situation in which more scalars are activated, additional constraints arise which have to be imposed alongside the Basu-Harvey equation. It would be interesting to see whether the results of \cite{Berman:2005re} can be derived from the ABJM theory. It would also prove interesting to consider the enhancement of the superymmetry of the algebra (for Chern-Simons levels $k=1,2$) through the use of monopole operators \cite{Gustavsson:2009pm, Kwon:2009ar}. This would then allow for direct comparison with the $\mathcal{N} = 8$ Bagger-Lambert supersymmetry algebra.

In Chapter 5 we investigated the worldvolume superalgebra of the Nambu-Poisson M5-brane. This was achieved by re-expressing the Bagger-Lambert M2-brane worldvolume central charges in terms of infinite dimensional generators defined by the Nambu-Poisson bracket. This resulted in a new set of charges expressed in terms of six-dimensional fields. These we interpreted as charges corresponding to worldvolume solitons of the M5-brane worldvolume theory in the presence of a background 3-form gauge field. In particular we found that the charges of the M5-brane theory could be grouped according to the power of the coupling constant $g$. At $\mathcal{O} (g^0)$ we found charges corresponding to the self-dual string soliton which describes an M2-M5 intersection as well as a 3-brane soliton corresponding to an M5-M5 intersection. We also found that background C-field modifies the energy bound of the self-dual string charge by shifting the $\mH_{\dot{1} \dot{2} \dot{3}}$ component of the field strength. Our results therefore reinforce the results presented in \cite{Ho:2008nn, Ho:2008ve} in which the Bagger-Lambert theory based on Nambu-Poisson bracket was interpreted as a theory describing an M5-brane in a background of 3-form flux. Furthermore we investigated the double dimensional reduction of the M5-brane Nambu-Poisson superalgebra. This gave rise to the Poisson bracket terms of a non-commutative D4-brane superslagebra. We found charges corresponding to spacetime intersections of the D4-brane with D0-branes, D2-branes and D4-branes, as well as the endpoint of the fundamental string. We also found charges that bare a structural similarity to Nahm-type configurations involving multiple D2-branes ending on the D4-brane (with the matrix commutator replaced by Poisson bracket). Finally we found a charge reminiscent of the D4-brane charge found in Matrix theory. It would appear that the superalgebra of the Nambu-Poisson M5-brane contains a large amount of information about spacetime configurations. In this thesis we were only able to provide spacetime interpretations for some of the M5-brane charges. However, knowledge of the D4-brane spacetime configurations may provide a hint about how to interpret the remaining M5-brane charges. 

In Chapter 6 we began an investigation into the $l_p^3$ corrections to the Bagger-Lambert supersymmetry transformations. For the abelian theory we were able to determine the the fermion supersymmetry transformation by using an abelian duality transformation. For the scalar transformation we had to use a different approach in which invariance of the higher order abelian Lagrangian was used to fix the coefficients of the transformation. For the non-abelian theory we were able to use the dNS duality transformation to uniquely determine the fermion supersymmetry transformation at $\mathcal{O} (l_p^3)$. It would be interesting to establish the reason why the dNS duality fails to work for the scalar supersymmetry transformation. It should be possible to uniquely determine the form of the $\mathcal{O} (l_p^3)$ scalar transformation by checking the invariance of the higher order Lagrangian derived in \cite{Alishahiha:2008rs, Ezhuthachan:2009sr}. This would also provide an independent check on the fermion result derived using the dNS duality approach. It would also prove interesting to extend this analysis to the $\mathcal{N} = 6$ ABJM theory. Finding such an extension is of great interest as these theories have a clear spacetime interpretation in M-theory. One possibility for how to derive the $\mathcal{N}=6$ result would be to make use of the $\mathcal{N}=8$ result and $SO(8)$ triality. This should work in the same way that it works for the lowest order Bagger-Lambert theory. In \cite{Gustavsson:2009pm} it was shown that the Bagger-Lambert Lagrangian fields could be `triality rotated' in such a way that $({\textbf{8}}_V , {\textbf{8}}_S , {\textbf{8}}_C) \rightarrow ({\textbf{8}}_S , {\textbf{8}}_C , {\textbf{8}}_V)$, where ${\textbf{8}}_V , {\textbf{8}}_S , {\textbf{8}}_C$ are the vector, spinor and cospinor representations of SO(8) respectively. After performing this transformation it is possible to break $SO(8) \rightarrow SU(4)\times U(1)$ and decompose the $SO(8)$ spinor and cospinor fields (and gamma matrices) in order to rewrite the original $\mathcal{N} = 8$ expression in terms of ABJM fields and so-called `non-ABJM' fields. In \cite{Gustavsson:2009pm} the non-ABJM terms were shown to vanish as a result of certain algebraic constraints (deriving from the flatness condition of the gauge field strength). It would be interesting to see whether this analysis can be extended to higher order and if so, whether additional algebraic constraints would be necessary to eliminate the higher order `non-ABJM' terms.

\begin{appendix}
\chapter{}
\section{Conventions}
\subsection{$\mathcal{N}=6$ Bagger-Lambert Theory}
The supersymmetry parameters of the $\mathcal{N}=6$ ABJM theory transform in the $\textbf{6}$ representation of $SU(4)$. We can write the susy parameter $\varepsilon_{AB}$ in terms of a basis of  $4 \times 4$ gamma matrices as
\begin{equation}
\varepsilon_{AB} = \epsilon^I .(\G^I_{AB}),
\end{equation} 
with $I= 1, \ldots 6$. The gamma matrices are antisymmetric $(\G^I_{AB} = -\G^I_{BA})$ and satisfy the following relation
\begin{equation}
\G^I_{AB} \Gt^{JBC} + \G^J_{AB} \Gt^{IBC} = 2 \delta^{IJ} \delta^C_A \label{bew}
\end{equation}
where
\begin{equation}
\Gt^{IAB} = \frac{1}{2} \varepsilon^{ABCD} \G^I_{CD} = - (\G^I_{AB})^*. \label{up}
\end{equation}
We note that the $4 \times 4$ matrices $\G^I$ act on a different vector space to the $2 \times 2$ matrices $\gamma^\mu$ which are defined as world volume gamma matrices. These two types of gamma matrix commute with one another. It is also important to note the following relations
\begin{align}
\G^I_{AB} \Gt^{ICD} &= -2 \delta^{CD}_{AB} = -2 (\delta^C_A \delta^D_B - \delta^C_B \delta^D_A ) \label{qed}\\ 
\G^I_{AB} \Gt^{IBD} &= 6 \delta^D_A. \label{qed2}
\end{align}
Acting with $\varepsilon^{ABMN} \varepsilon_{CDPQ}$ on both sides of \eqref{qed} one can show that
\begin{equation}
\Gt^{ICD}\G^I_{AB} = - 2 \delta^{CD}_{AB}.
\end{equation}
It therefore follows that
\begin{align}
\G^I_{AB} \Gt^{ICD} + \Gt^{ICD} \G^I_{AB} &= -4\delta^{CD}_{AB} \label{nonzero}\\
\G^I_{AB} \Gt^{ICD} - \Gt^{ICD} \G^I_{AB} &= 0 \label{zero}
\end{align}
We will also need the following identity in what follows
\begin{align}
\G^I_{AB} \Gt^{AC} + \Gt^{IAC} \G^J_{AB}  &= \G^I_{AB} \Gt^{AC} + \frac{1}{4} \varepsilon^{ACDE} \varepsilon_{ABFG} \G^I_{DE} \Gt^{JFG} \nn \\
&= \frac{1}{2} \delta^C_B \G^I_{FG} \Gt^{JFG} = 2 \delta^{IJ} \delta^C_B. \label{apefarm}
\end{align}
and therefore
\begin{equation}
\G^I_{FG} \Gt^{JFG} = 4 \delta^{IJ}.
\end{equation}
Note that in obtaining the last line of \eqref{apefarm} we made use of \eqref{bew} and the epsilon tensor identity
\begin{align}
\varepsilon^{ACDE} \varepsilon_{ABFG} = &+ \delta^C_B \delta^D_F \delta^E_G + \delta^C_F \delta^D_G \delta^E_B + \delta^C_G \delta^D_B \delta^E_F \nn \\
&- \delta^C_B \delta^D_G \delta^E_F - \delta^C_F \delta^D_B \delta^E_G - \delta^C_G \delta^D_F \delta^E_B  
\end{align}
Similarly we have
\begin{equation}
\G^I_{AB} \Gt^{AC} - \Gt^{IAC} \G^J_{AB} = 2 \G^I_{AB} \Gt^{AC} - \frac{1}{2} \delta^C_B \G^I_{FG} \Gt^{JFG}.\label{1d2}
\end{equation}
It is possible to derive identities involving $\e_{AB}$ based on the relations between the basis gamma matrices $\Gamma^I$. In \cite{Bagger:2008se} Bagger and Lambert make use of the following identities
\begin{equation}
\frac{1}{2}\be^{CD}_1 \gamma_\nu \e_{2CD} \delta^A_B = \be^{AC}_1 \gamma_\nu \e_{2BC} - \be^{AC}_2 \gamma_\nu \e_{1BC} \label{iden}
\end{equation}
and
\begin{align}
2 \be^{AC}_1 \e_{2BD} - 2 \be^{AC}_2 \e_{1BD} = &+ \be^{CE}_1 \e_{2DE} \delta^A_B - \be^{CE}_2 \e_{1DE} \delta^A_B \nn \\
&- \be^{AE}_1 \e_{2DE} \delta^C_B + \be^{AE}_2 \e_{1DE} \delta^C_B \nn \\
&+ \be^{AE}_1 \e_{2BE} \delta^C_D - \be^{AE}_2 \e_{1BE} \delta^C_D  \label{swiss} \\
&- \be^{CE}_1 \e_{2BE} \delta^A_D + \be^{CE}_2 \e_{1BE} \delta^A_D. \nn
\end{align}
Both of these identities can be re-written in terms of identities involving the Majorana spinors $\epsilon^I$ and the gamma matrices $\Gamma^I$.

\subsection{Nambu-Poisson M5-brane}
In the BLG Nambu-Poisson model of the M5-brane, the worldvolume is taken to be the product manifold $\mathcal{M} \times \mathcal{N}$. We assume a Minkowski metric $\eta_{\mu \nu} = \textrm{diag} (-++)$ on $\mathcal{M}$ and a Euclidean metric $\delta_{\mud \nud} = \textrm{diag} (+++)$ on the internal space $\mathcal{N}$. The supersymmetry transformation parameter $\e$ and the fermion $\psi$ of the Bagger-Lambert theory belong to the ${\textbf{8}}_s$ and ${\textbf{8}}_c$ representations of the SO(8) R-symmetry and are 32-component spinors satisfying
\begin{equation}
\g^{\mu \nu \rho} \e = + \e^{\mu \nu \rho} \e , \quad \quad \g^{\mu \nu \rho} \psi = - \e^{\mu \nu \rho} \psi. \label{chirality}
\end{equation}
We assume that $\varepsilon_{012} = - \varepsilon^{012}$ and thus
\begin{align}
\e_{\mu \nu \lambda} \e^{\mu \rho \sigma} &= - (\delta^\rho_\nu \delta^\sigma_\lambda - \delta^\rho_\lambda \delta^\sigma_\nu ). \\
\e_{\mu \nu \lambda} \e^{\mu \nu \sigma} &= -2 \delta^\sigma_\lambda.
\end{align}
The following relations on $\mathcal{M}$ follow from the chirality constraint \eqref{chirality}
\begin{align}
\e^{\mu \nu \rho} \g_{\nu \rho} \e &= - 2 \g^\mu \e. \\
\e^{\mu \nu \rho} \g_\rho \e &= \g^{\mu \nu} \e .\\
\e^{\mu \nu \rho} \g_{\nu \rho} \psi &= + 2 \g^\mu \psi. \\
\e^{\mu \nu \rho} \g_\rho \psi &= - \g^{\mu \nu} \psi.
\end{align}
There is no analogue of the relations \eqref{chirality} on $\mathcal{N}$. However we do find the following information useful 
\begin{align}
\e_{\dot{1} \dot{2} \dot{3}} &= \e^{\dot{1} \dot{2} \dot{3}}. \\
\e_{\mud \nud \lambdad} \e_{\mud \rhod \sigmad} &= \delta_{\nud \rhod} \delta_{\lambdad \sigmad} - \delta_{\nud \sigmad} \delta_{\lambdad \rhod}. \\
\e_{\mud \nud \lambdad} \e_{\mud \nud \sigmad} &= 2 \delta_{\lambdad \sigmad}.\\
\g_{\mud \nud \rhod} &= \g_{\dot{1} \dot{2} \dot{3}} \e_{\mud \nud \rhod}. \\ 
(\g_{\dot{1} \dot{2} \dot{3}})^2 &= -1.
\end{align}

\section{$\mathcal{N}= 6$ Bagger-Lambert Superalgebra Calculations} \label{calc}
In this section we calculate the supersymmetric variation of $J^{0, I}$. Given the supercurrent expression \eqref{supercurrent} one finds
\begin{align}
\delta J^{0,I} = &+ \overbrace{\Tr (\G^I_{AB} \Gt^{JAC}\gamma^\nu \gamma^0 \gamma^\rho  D_\nu Z^B  , D_\rho \bz_C \varepsilon^J ) + \Tr (\Gt^{IAB} \G^J_{AC} \gamma^\nu \gamma^0 \gamma^\rho D_\nu \bz_B , D_\rho Z^C)}^{(a)} \nn \\
&- \overbrace{\Tr (\G^I_{AB} \gamma^\nu \gamma^0 D_\nu Z^B , N^{JA} \varepsilon^J ) - \Tr (N^I_A \Gt^{JAC} \gamma^0 \gamma^\rho,  D_\rho \bz_C \varepsilon^J )}^{(b)} \nn \\
&+ \Tr (\Gt^{IAB} \gamma^\nu \gamma^0 D_\nu \bz_B, N^J_A \varepsilon^J ) + \Tr(N^{IA} \Gamma^J_{AC} \gamma^0 \gamma^\rho D_\rho Z^C \varepsilon^J ) \\
&+ \overbrace{\Tr (N^I_A \gamma^0 , N^{JA} \varepsilon^J ) + \Tr (N^{IA} \gamma_0 , N^J_A \varepsilon^J)}^{(c)}. \nn
\end{align}
\subsection{(a) terms}
The $(a)$ terms may be written as
\begin{align}
(a) = &-\Tr ((\G^I_{AB} \Gt^{JAC} + \Gt^{IAC} \G^J_{AB})\gamma^0 D_0 Z^B , D_0 \bz_C \varepsilon^J ) \nn \\
&-\Tr ((\G^I_{AB} \Gt^{JAC} + \Gt^{IAC} \G^J_{AB})\gamma^i D_0 Z^B , D_i \bz_C \varepsilon^J ) \nn \\
&-\Tr ((\G^I_{AB} \Gt^{JAC} + \Gt^{IAC} \G^J_{AB})\gamma^i D_i Z^B , D_0 \bz_C \varepsilon^J ) \nn \\
&- \Tr ((\G^I_{AB} \Gt^{JAC} + \Gt^{IAC} \G^J_{AB}) \gamma^0 D_i Z^B , D^i \bz_C \varepsilon^J ) \\
&- \Tr ((\G^I_{AB} \Gt^{JAC} - \Gt^{IAC} \G^J_{AB})\gamma^{ij} \gamma^0 D_i Z^B , D_j \bz_C \varepsilon^J ) \nn .
\end{align}
The first four terms can be further simplified by using the relation \eqref{apefarm}.
\begin{align}
(a) = &-2\delta^{IJ} \Tr(\gamma^0 D_0 Z^B , D^0 \bz_B \varepsilon^J) -2 \delta^{IJ} \Tr( \gamma^0 D_i Z^B , D^i \bz_B \varepsilon^J ) \nn \\
&-2 \delta^{IJ} \Tr (\gamma^i D_0 Z^B , D_i \bz_B \varepsilon^J ) -2 \delta^{IJ} \Tr( \gamma^i D_i Z^B , D_0 \bz_B \varepsilon^J ) \\
&- \Tr ((\G^I_{AB} \Gt^{JAC} - \Gt^{IAC} \G^J_{AB})\gamma^{ij} \gamma^0 D_i Z^B , D_j \bz_C \varepsilon^J ) \nn.
\end{align}
\subsection{(b) terms}
The $(b)$ terms may be written as
\begin{align}
(b) = &-2 \delta^{IJ} \Tr (D^i \bz_B , [Z^D , Z^B ; \bz_D] \gamma^0 \gamma^i \varepsilon^J ) + 2 \delta^{IJ} \Tr (D^i Z^B , [\bz_D , \bz_B ; Z^D] \gamma^0 \gamma^i \varepsilon^J ) \nn \\
&+ \Tr((\G^I_{DE} \Gt^{JAC} + \Gt^{IAC} \G^J_{DE}) [Z^D , Z^E ; \bz_A ] \gamma^0 \gamma^i , D_i \bz_C \varepsilon^J ) \nn \\
&- \Tr ((\G^I_{AB} \Gt^{JCD} + \Gt^{ICD} \G^J_{AB})[\bz_C , \bz_D ; Z^A] \gamma^0 \gamma^i ,  D_i Z^B \varepsilon^J )\nn  \\
&+ \Tr ((\G^I_{AB} \Gt^{JAC} - \Gt^{IAC} \G^J_{AB}) [Z^D , Z^B ; \bz_D ] , D_0 \bz_C \varepsilon^J ) \nn \\
&+ \Tr ((\G^I_{AB} \Gt^{JAC} - \Gt^{IAC} \G^J_{AB}) [\bz_D , \bz_C ; Z^D] , D_0 Z^B \varepsilon^J ) \\
&- \Tr ((\G^I_{AB} \Gt^{JAC} - \Gt^{IAC} \G^J_{AB}) [Z^D , Z^E ; \bz_A ] , D_0 \bz_C \varepsilon^J )\nn \\
&- \Tr ((\G^I_{AB} \Gt^{JAC} - \Gt^{IAC} \G^J_{AB}) [\bz_C , \bz_D ; Z^A] , D_0 Z^B \varepsilon^J )\nn 
\end{align}
The terms involving $D_0$ can be greatly simplified by using \eqref{swiss}. After a bit of rearrangement and relabeling we can write the (b) terms as
\begin{align}
(b) = &+2 \delta^{IJ} \Tr (D^i \bz_B , [Z^D , Z^B ; \bz_D] \varepsilon^{ij} \varepsilon^J ) - 2 \delta^{IJ} \Tr (D^i Z^B , [\bz_D , \bz_B ; Z^D] \varepsilon^{ij} \varepsilon^J ) \nn \\
&+ \Tr (\G^{CD(IJ)}_{AB} D^i \bz^B , [\bz_C , \bz_D ; Z^A] \varepsilon^{ij} \gamma^j \varepsilon^J ) - \Tr (\G^{CD(IJ)}_{AB} D_i \bz_D [Z^A , Z^B ; \bz_C ] \varepsilon^{ij} \gamma^j \varepsilon^J ) \nn \\
&- \Tr (\Gt^{AE[IJ]}_{DE} D_0 \bz_C , [Z^D, Z^C ; \bz_A ] \varepsilon^J ) - \Tr (\Gt^{AE[IJ]}_{DE} D_0 Z^C [\bz_A , \bz_C ; Z^D] \varepsilon^J ),
\end{align}
where
\begin{align}
\G^{CD(IJ)}_{AB} &= \G^I_{AB} \Gt^{JCD} + \Gt^{ICD} \G^J_{AB} ; \\
\Gt^{A[IJ]}_{D} &= \G^I_{DE} \Gt^{JAE} - \Gt^{IAE} \G^J_{DE},
\end{align}
and we have used the fact that in 3 dimensions $\gamma^{ij} \propto \varepsilon^{ij}$. We have also used the fact that $\gamma^0 \gamma^i = - \varepsilon^{ij} \gamma^{012}$ and $\gamma^{012} \varepsilon^J = \varepsilon^J$.
\subsection{(c) terms}
The (c) terms may be written as
\begin{align}
(c) = &-2 \delta^{IJ} \Tr ([Z^C , Z^B ; \bz_C ],[\bz_F , \bz_B ; Z^F ])\varepsilon^J \nn \\
&- \G^{EF(IJ)}_{AB}\Tr ( [Z^C, Z^B ; \bz_C ], [\bz_E , \bz_F ; Z^A ])\varepsilon^J \nn \\
&- \G^{EF(IJ)}_{AB}\Tr ( [Z^A , Z^B ; \bz_E ],[\bz_C , \bz_F ; Z^C])\varepsilon^J \\
&+ \G^{EF(IJ)}_{AB}\Tr ( [Z^A, Z^B ; \bz_C ],[\bz_E, \bz_F ; Z^C])\varepsilon^J. \nn
\end{align}
We can make use of the fact that the potential is
\begin{equation}
V = \frac{2}{3} \Tr ([Z^C , Z^D ; \bz_B ],[\bz_C , \bz_D ; Z^B]) - \frac{1}{3} \Tr ([Z^B , Z^D ; \bz_B ],[\bz_F , \bz_D ; Z^F])
\end{equation}
to write (c) as
\begin{align}
(c) = &-2 \delta^{IJ} (V-V_1)\varepsilon^J \nn \\
&- \G^{EF(IJ)}_{AB}\Tr ( [Z^C, Z^B ; \bz_C ], [\bz_E , \bz_F ; Z^A ])\varepsilon^J \nn \\
&- \G^{EF(IJ)}_{AB}\Tr ( [Z^A , Z^B ; \bz_E ],[\bz_C , \bz_F ; Z^C])\varepsilon^J \\
&+ \G^{EF(IJ)}_{AB}\Tr ( [Z^A, Z^B ; \bz_C ],[\bz_E, \bz_F ; Z^C])\varepsilon^J, \nn
\end{align}
where
\begin{equation}
V_1 = \frac{2}{3} \Tr([Z^C, Z^D ; \bz_B ],[\bz_C , \bz_D ; Z^B]) - \frac{4}{3} \Tr ([Z^C, Z^B ; \bz_C ],[\bz_E , \bz_B ; Z^E])
\end{equation}
\subsection{$\delta J^0$}
We can combine (a), (b) and (c) terms 
\begin{align}
\delta J^{0,I} =  &-2\delta^{IJ} T^0_\mu \gamma^\mu \varepsilon^J  + 2\delta^{IJ} V_1 \gamma^0 \varepsilon^J \nn \\
&+2 \delta^{IJ} ( \Tr (D_i \bz_B , [Z^D , Z^B ; \bz_D] ) -  \Tr (D_i Z^B , [\bz_D , \bz_B ; Z^D] ) \varepsilon^{ij}  \gamma^j \varepsilon^J \nn \\
&- \G^{C[IJ]}_{B}\Tr ( D_i Z^B , D_j \bz_C  )\varepsilon^{ij} \gamma^0 \varepsilon^J \nn \\
&- \G^{C[IJ]}_{B}( \Tr (D_0 \bz_A , [Z^B, Z^A ; \bz_C ] ) + \Tr ( D_0 Z^A, [\bz_C , \bz_A ; Z^B] )) \varepsilon^J \nn \\
&+ \G^{CD(IJ)}_{AB}( \Tr ( D^i \bz^B , [\bz_C , \bz_D ; Z^A] ) - \Tr ( D_i \bz_D [Z^A , Z^B ; \bz_C ] ))  \varepsilon^{ij} \gamma^j \varepsilon^J \nn \\
&- \G^{EF(IJ)}_{AB}\Tr ( [Z^C, Z^B ; \bz_C ], [\bz_E , \bz_F ; Z^A ])\varepsilon^J \nn \\
&- \G^{EF(IJ)}_{AB}\Tr ( [Z^A , Z^B ; \bz_E ],[\bz_C , \bz_F ; Z^C])\varepsilon^J\nn  \\
&+ \G^{EF(IJ)}_{AB}\Tr ( [Z^A, Z^B ; \bz_C ],[\bz_E, \bz_F ; Z^C])\varepsilon^J, \nn
\end{align}
where we have used
\begin{align}
T_{00} &= \Tr (D_0 Z^B , D_0 \bz_B ) + \Tr (  D_i Z^B , D^i \bz_B ) + V ; \\
T_{0i} &= \Tr ( D_0 Z^B , D_i \bz_B  ) + \Tr (  D_i Z^B , D_0 \bz_B ). 
\end{align}
\subsection{Potential}
In this appendix we show the equivalence of the Bagger-Lambert and ABJM potential. The Bagger-Lambert potential is given by
\begin{align}
V = \frac{2}{3} \Tr (\Upsilon^{CD}_B , {\bar{\Upsilon}}^B_{CD})
\end{align}
where
\begin{align}
\Upsilon^{CD}_B = [Z^C, Z^D;\bz_{B} ] - \frac{1}{2} \delta^C_B [Z^E, Z^D; \bz_E ] + \frac{1}{2}\delta^D_B [Z^E, Z^C; \bz_E ].
\end{align}
We can define the inner product as
\begin{align}
\Tr (X,Y) = \tr (X^\dagger Y)
\end{align}
where $\dagger$ denotes the transpose conjugate and $\tr$ denotes the ordinary matrix trace. Thus
\begin{align}
(\Upsilon^{CD}_B)^\dagger &= [Z^{D\dagger} , Z^{C\dagger} ; \bz_B^\dagger ] - \frac{1}{2} \delta^C_B [Z^{E \dagger} , Z^{D \dagger}, \bz^\dagger_{E}] + \frac{1}{2} \delta^D_B [Z^{E \dagger} , Z^{C\dagger} ; \bz^\dagger_{E}] \\
{\bar{\Upsilon}}^B_{CD} &= [\bz_C , \bz_D ; Z^B] - \frac{1}{2} \delta^B_C [\bz_E , \bz_D ; Z^E ] + \frac{1}{2} \delta^B_D [\bz_E , \bz_C ; Z^E ].
\end{align}
Making use of the above information one finds that
\begin{align}
V &= \frac{2}{3} \Tr (\Upsilon^{CD}_B , {\bar{\Upsilon}}^B_{CD}) \\
&=\frac{2}{3} \tr ((\Upsilon^{CD}_B)^\dagger {\bar{\Upsilon}}^B_{CD}) \\
&= \frac{2}{3} \tr \left([Z^{D \dagger} , Z^{C \dagger} ; \bz^\dagger_B ][\bz_C , \bz_D ; Z^B ] + \frac{1}{2} [Z^{E \dagger} , Z^{C \dagger} ; \bz^\dagger_E ][\bz_B , \bz_C ; Z^B ] \right). \label{bn}
\end{align}
For the particular choice
\begin{align}
[X, Y; Z] = \lambda (X Z^\dagger Y - Y Z^\dagger X) \label{vb}
\end{align}
it was shown by Bagger and Lambert that the $N=6$ ABJM potential is recovered. Inserting \eqref{vb} into \eqref{bn} one finds
\begin{align}
V = \lambda^2 \tr  ( &- \frac{1}{3} Z^{E \dagger} \bz_E Z^{C \dagger} \bz_C Z^{B \dagger} \bz_B - \frac{1}{3} Z^{C \dagger} \bz_E Z^{E \dagger} \bz_B Z^{B \dagger} \bz_C \nn \\
&- \frac{4}{3} Z^{D \dagger} \bz_B Z^{C \dagger} \bz_D Z^{B \dagger} \bz_C + 2 Z^{D\dagger} \bz_B Z^{C \dagger} \bz_{C} Z^{B \dagger} \bz_D ).
\end{align}
Comparing with
\begin{align}
V = \frac{4 \pi^2}{k^2}\tr (&- \frac{1}{3} X^A X_A X^B X_B X^C X_C - \frac{1}{3} X_A X^A X_B X^B X_C X^C  \nn\\
 &- \frac{4}{3} X_A X^B X_C X^A X_B X^C + 2 X^A X_B X^B X_A X^C X_C )
\end{align}
we see that the two expressions are equivalent given the redefinitions $Z^{A \dagger} \rightarrow X^A$ and $\bz_A \rightarrow X_A$, as well as $\lambda = 2 \pi/k$.
\section{Higher order $SO(8)$ invariant objects}
In this section we list $SO(8)$ invariant combinations which give rise to terms appearing in the dNS transformed superymmetry transformations of the previous section. Note that we have suppressed the symmetrised trace in all the expressions that follow.
\subsection{$\delta \psi$}
\subsubsection{Zero Derivative}
\begin{align}
\g^{IJM} X^{KLN} X^{KLN} X^{IJM} &\rightarrow 9g^3_{YM}\g^{ij} X^{kl} X^{kl} X^{ij}  \nn \\
\g^{IJM} X^{IKN} X^{KLN} X^{LJM} &\rightarrow g^3_{YM}\g^{ij} (4X^{ik} X^{kl} X^{lj} - X^{kl} X^{kl} X^{ij}) \nn \\
\g^{IJKLMNP} X^{IJQ} X^{KLQ} X^{MNP} &\rightarrow 3 g^3_{YM} \g^{ijklmn} X^{ij} X^{kl} X^{mn}  \nn 
\end{align}
\subsubsection{One Derivative}
\begin{align}
\g_\mu \g^{J} X^{KLM} X^{KLM} D^\mu X^J &\rightarrow 3 g^2_{YM}(\g_\mu \g^8 X^{kl} X^{kl} D^\mu X^8 +  \g_\mu \g^j X^{kl} X^{kl} D^\mu X^j )\nn \\
\g_\mu \g^J D_\mu X^K X^{KLM} X^{LJM} &\rightarrow g^2_{YM}(2 \g_\mu \g^j D^\mu X^k X^{kl} X^{lj} - \g_\mu \g^8 D^\mu X^8 X^{lm} X^{lm}) \nn \\
\g^\mu \g^J X^{JKM} X^{KLM} D_\mu X^L &\rightarrow g^2_{YM}( 2 \g^\mu \g^j X^{jk} X^{kl} D_\mu X^l - \g^\mu \g^8 X^{km} X^{km} D_\mu X^8) \nn \\
\g_\mu \g^{IJKLM} &\rightarrow g^2_{YM} (\g_\mu \g^{ijklm} D^\mu X^m X^{ij} X^{kl} + \g_\mu \g^{ijkl} D^\mu X^8 X^{ij} X^{kl}) \nn\\
\g_\mu \g^{IJKLM} D^\mu X^N X^{IJM} X^{KLN} &\rightarrow 3 g^2_{YM} \g_\mu \g^{ijkl} D^\mu X^8 X^{ij} X^{kl} \nn \\
\g_\mu \g^{ijklm} D^\mu X^m X^{ij} X^{kl} &\leftarrow g^2_{YM}\g_\mu \g^{IJKLM} (D^\mu X^M X^{IJN} X^{KLN} - \frac{1}{3} D^\mu X^N X^{IJM} X^{KLN}) \nn
\end{align}
\subsubsection{Two Derivative}
\begin{align}
\g^{IJK} D_\mu X^L D^\mu X^L X^{IJK} &\rightarrow 3g_{YM}( \g^{ij} D_\mu X^k D^\mu X^k X^{ij} +  \g^{ij} D_\mu X^8 D^\mu X^8 X^{ij} )\nn \\
\g^{\mu \nu} \g^{M} D_\mu X^K X^{KLM} D_\nu X^L &\rightarrow g_{YM}\g^{\mu \nu}( \g^i D_\mu X^k X^{ik} D_\nu X^8 \nn  +  \g^i D_\mu X^8 X^{li} D_\nu X^l \nn \\
 &+ \g^8 D_\mu X^k X^{kl} D_\nu X^l) \nn \\
\g^{IJK} D_\mu X^I D^\mu X^L X^{LJK} &\rightarrow g_{YM} (\g^{ijk} D_\mu X^i D^\mu X^8 X^{jk} + 2 \g^{ij} D_\mu X^i D^\mu X^l X^{lj} \nn \\
&\quad + \g^{ij} D_\mu X^8 D^\mu X^8 X^{ij} ) \nn \\
\g^{IJK} X^{ILK} D_\mu X^L D^\mu X^J &\rightarrow  g_{YM} (\g^{ijk} X^{jk} D_\mu X^8 D^\mu X^i + 2 \g^{ij} X^{ik} D_\mu X^k D^\mu X^j \nn \\
&\quad + \g^{ij} X^{ij} D_\mu X^8 D^\mu X^8 ) \nn \\
\g^{\mu \nu} \g^{IJKLM} D_\mu X^I D_\nu X^J X^{KLM} &\rightarrow 3 g_{YM}\g^{\mu \nu} \g^{ijkl} D_\mu X^i D_\nu X^j X^{kl} \nn .
\end{align}
\subsubsection{Three Derivative}
\begin{align}
\g_\mu \g^I D_\nu X^J D^\nu X^J D^\mu X^I &\rightarrow  \g_\mu \g^i D_\nu X^j D^\nu X^j D^\mu X^i + \g_\mu \g^i D_\nu X^8 D^\nu X^8 D^\mu X^i \nn \\
& \quad + \g_\mu \g^8 D_\nu X^j D^\nu X^j D^\mu X^8 + \g_\mu \g^8 D_\nu X^8 D^\nu X^8 D^\mu X^8 \nn \\
\g_\mu \g^I D^\mu X^J D_\nu X^J D^\nu X^I &\rightarrow \g_\mu \g^i D^\mu X^j D_\nu X^j D^\nu X^i + \g_\mu \g^i D^\mu X^8 D_\nu X^8 D^\nu X^i \nn \\
& \quad + \g_\mu D^\mu X^j D_\nu X^j D^\nu X^8 + \g_\mu D^\mu X^8 D_\nu X^8 D^\nu X^8 \nn  \\
\e^{\mu \nu \rho} \g^{IJK} D_\mu X^I D_\nu X^J D_\rho X^K &\rightarrow \e^{\mu \nu \rho} \g^{ijk} D_\mu X^i D_\nu X^j D_\rho X^k + 3 \e^{\mu \nu \rho} \g^{ij} D_\mu X^i D_\nu X^j D_\rho X^8. \nn
\end{align}

\subsection{$\delta X_i$}
\subsubsection{Zero Derivative}
\begin{align}
\be \g^J \psi X^{IKL} X^{JKL} &\ra 2 g^2_{YM}\be \g^j \psi X^{ik} X^{jk} \nn \\
\be \g^{JKL} \psi X^{IJN} X^{KLN} &\ra g^2_{YM}\be \g^{jkl} \psi X^{ij} X^{kl} \nn \\
\be \g^{IJKLM} \psi X^{JKN} X^{LMN} &\ra g^{2}_{YM} \be \g^{ijklm} \psi X^{jk} X^{lm} \nn \\
\end{align}
\subsubsection{One Derivative}
\begin{align}
\be \g_\mu \g^{JKL} \psi D^\mu X^I X^{JKL} &\ra 3g_{YM} \be \g_\mu \g^{jk} \psi D^\mu X^i X^{jk} \nn \\
\be \g_\mu \g^{IKL} \psi D^\mu X^J X^{JKL} &\ra 2g_{YM} \be \g_\mu \g^{ij} \psi D^\mu X^k X^{kj} + g_{YM}\be \g_\mu \g^{ijk} \psi D^\mu X^8 X^{jk} \nn \\
\be \g_\mu \g^K \psi D_\mu X^J X^{JKI} &\ra g_{YM}\be \g_\mu \g^8 \psi D_\mu X^j X^{ij} + g_{YM}\be \g_\mu \g^k \psi D_\mu X^8 X^{ki} \nn \\
\be \g^\mu \g^{IJKLM} \psi D_\mu X^J X^{KLM} &\ra 3g_{YM} \be\g^\mu \g^{ijkl} \psi D_\mu X^j X^{kl} \nn \\
\be \g_\mu \g^{JKL} \psi D^\mu X^K X^{IJL} &\ra 2 g_{YM} \be \g_\mu \g^{jk} \psi D^\mu X^k X^{ij} \nn
\end{align}
\subsubsection{Two Derivative}
\begin{align}
\be \g^I \psi D_\mu X^J D^\mu X^J &\ra \be \g^i \psi D_\mu X^8 D^\mu X^8 + \be \g^i \psi D_\mu X^j D^\mu X^j \nn \\
\be \g^I \g^{\mu \nu} \psi D_\mu X^J D_\nu X^J &\ra  0\nn \\
\be \g^{\mu \nu}\g^{IJK} \psi D_\mu X^J D_\nu X^K &\ra  2\be \g^{\mu \nu} \g^{ij} \psi D_\mu X^j D_\nu X^8 + \be \g^{\mu \nu} \g^{ijk} \psi D_\mu X^j D_\nu X^k \nn \\
\be \g^J \psi D_\mu X^I D^\mu X^J &\ra \be \g^8 \psi D_\mu X^i D^\mu X^8 + \be \g^j \psi D_\mu X^i D^\mu X^j \nn \\
\be \g^J \g^{\mu \nu} \psi D_\mu X^I D_\nu X^J &\ra \be \g^8 \g^{\mu \nu} \psi D_\mu X^i D_\nu X^8 + \be \g^j \g^{\mu \nu} \psi D_\mu X^i D_\nu X^j \nn \\
\be \g^{IJK} \psi D_\mu X^J D^\mu X^K &\ra 0
\end{align}
\subsection{$\delta A_\mu$}
\subsubsection{Zero Derivative}
\begin{align}
\be \g_\mu \chi X^{IJK} X^{IJK} &\ra 3 g^2_{YM} \be \g_\mu \chi X^{ij} X^{ij} \nn \\
\be \g_\mu \g^{IJ} \chi X^{ILM} X^{JLM} &\ra 2 g^2_{YM} \be \g_\mu \g^{ij} \chi X^{il} X^{jl} \nn \\
\be \g_\mu \g_{IJKL} \chi X^{IJN} X^{KLN} &\ra  g^2_{YM}\be \g_\mu \g^{ijkl} \chi X^{ij} X^{kl} \nn \\
\be \g_\mu  \g^{IJKLMN} \chi X^{IJK} X^{LMN} &\ra 0 \nn 
\end{align}
\subsubsection{One Derivative}
\begin{align}
\be \g^{JK} \chi D_\mu X^I X^{IJK} &\ra g_{YM}(\be \g^{jk} \chi D_\mu X^8 X^{jk} + 2 \be \g^j \chi D_\mu X^i X^{ij} )\nn \\
\be \g^{IJKL} \chi D_\mu X^I X^{JKL} &\ra 3 g_{YM} \be \g^{ijk} \chi D_\mu X^i X^{jk} \nn \\
\be \g_{\mu \nu} \g^{JK} D^\nu X^I X^{IJK} &\ra g_{YM}(\be \g_{\mu \nu} \g^{jk} \chi D^\nu X^8 X^{jk} + 2 \be \g_{\mu \nu} \g^j \chi D^\nu X^i X^{ij}) \nn \\
\be \g_{\mu \nu} \g^{IJKL} D^\nu X^I X^{JKL} &\ra 3 g_{YM }\be \g_{\mu \nu} \g^{ijk} \chi D_\mu X^i X^{jk} \nn 
\end{align}
\subsubsection{Two Derivative}
\begin{align}
\be \g_\mu \chi D_\nu X^K D^\nu X_K &\ra \be \g_\mu \chi D_\nu X^8 D^\nu X^8 + \be \g_\mu \chi D_\nu X^i D^\nu X^i \nn \\
\be \g_\mu \g^{IJ} \chi D_\nu X^I D^\nu X^J &\ra \be \g_\mu \g^{ij} \chi D^\nu X^i D^\nu X^j + \be \g_\mu \g^i \chi D_\nu X^i D^\nu X^8 - \be \g_\mu \g^i \chi D_\nu X^8 D^\nu X^i =0\nn \\
\be \g_\nu \chi D^\nu X^K D_\mu X^K &\ra \be \g_\nu \chi D^\nu X^i D_\mu X^i + \be \g_\nu \chi D^\nu X^8 D_\mu X^8 \nn \\
\be \g_\nu \g^{IJ} \chi D^\nu X^I D_\mu X^J &\ra \be \g_\nu \g^{ij} \chi D^\nu X^i D_\nu X^j + \be \g_\nu \g^i \chi D^\nu X^i D_\mu X^8 - \be \g_\nu \g^i \chi D^\nu X^8 D_\mu X^i \nn \\
\be \g_{\mu \nu \lambda} \chi D^\nu X^J D^\lambda X^J &\ra 0 \nn \\
\be \g_{\mu \nu \lambda} \g^{IJ} D^\nu X^I D^\lambda X^J &\ra \be \g_{\mu \nu \lambda} \g^{ij} D^\nu X^i D^\lambda X^j + \be \g_{\mu \nu \lambda} \g^i D^\nu X^i D^\lambda X^8 - \be \g_{\mu \nu \lambda} \g^i D^\nu X^8 D^\lambda X^i \nn 
\end{align}

\end{appendix}

\bibliographystyle{JHEP}
\bibliography{DansBib}
\end{document}